\documentclass[pre]{revtex4}
\usepackage{psfrag}
\usepackage{amssymb,amsmath,amsthm}
\usepackage[dvips]{graphicx}
\usepackage{verbatim}

\usepackage{bm}

\usepackage{color}

\definecolor{pink}{rgb}{1,1,0} 
\definecolor{red}{rgb}{1,0,0}
\definecolor{yellow}{rgb}{1,1,0}
\definecolor{orange}{rgb}{1,0.5,0}
\definecolor{rose}{rgb}{1,0.5,1}
\definecolor{purple}{rgb}{1,0,1}
\definecolor{white}{rgb}{1,1,1}
\definecolor{blue}{rgb}{0,0,1}
\definecolor{green}{rgb}{0,1,0}

\newcommand{\red}{\color{red}}  
 
\newcommand{\blue}{\color{blue}} 
\newcommand{\gr}{\color{green}}

\newcommand{\purple}{\color{purple}}

\begin{document}

\title{Theory of real supersolids}
\author{
Gustavo During$^1$, Christophe Josserand$^{2}$, Yves Pomeau$^3$ and Sergio Rica$^{4,5}$\\
$^1$ Center of Soft Matter Research, New York University, 4 Washington Place, New York, NY, 10003, USA.\\
$^2$ Institut D'Alembert, CNRS \& UPMC (Univ. Paris VI) UMR 7190, 4 place Jussieu, 75005 Paris, France.\\
$^3$ Univ. Arizona, Dept Math, Tucson, AZ 85721 USA.\\
$^4$ Facultad de Ingenier\'ia y Ciencias, Universidad Adolfo Ib\'a\~nez, Avda. Diagonal las Torres 2640, Pe\~nalol\'en, Santiago, Chile.\\
$^5$ INLN, Institut Non Lin\'eaire de Nice, CNRS \& Univ. Nice-Sophia-Antipolis, 1361 Route des lucioles, Sophia Antipolis, F-06560 Valbonne, France} 
\date{\today}

\begin{abstract}
We review the main properties of a supersolid. We describe first the macroscopic equation that satisfies a supersolid based on general arguments and symmetries and show 
that such solids might exhibit simultaneously or independently both elastic behavior and superfluidity. We then
explain why a supersolid state should exist for solids at very low temperature but with a very small superfluid fraction. Finally, we propose a mean-field model, based on the
Gross-Pitaevski\u{\i} equation, which presents the general properties expected for a supersolid and should therefore provide a consistent framework to study its dynamical 
properties.

\end{abstract}

\maketitle

\tableofcontents

\section{Introduction}\label{intro}
The striking observation by Kim and Chan in 2004 of a rapid drop in the rotational inertial of solid Helium below $0.1$Kelvin~\cite{chan04a,chan04b} has revived the question whether the solids could exhibit a new state of the matter at low temperature, the so-called ``supersolidity". In these experiments, the supersolidity appears as a decoupling between a ``superfluid" and a ``normal" part of the solid: everything happens as if a small fraction of the total mass of solid Helium (of the order of a few percent) is at rest while a rotation is imposed to the 
sample!
In fact, since the liquefaction of Helium at $T=4.22$ Kelvin in 1908 by Kammerlingh Onnes, the physics of Helium has accompanied many important scientific discoveries in the last century. Among them, the 
remarkable property of superfluidity was observed in 1938 for Helium 4 (the bosonic and most abundant isotope of Helium, noted also $^4$He) by P.L. Kapitza~\cite{kapit38,AlMi38}. 
Below the $\lambda$-point separating the two Helium liquid phases noted Helium I (above $2.2$ Kelvin) and Helium II (below)~\cite{rollin35,keesom36}, the resistance to flow of Helium 
through thin capillaries drops suddenly so that it can be considered a zero-viscosity liquid. Kapitza named this effect superfluidity by analogy with the superconductivity although Helium 
4 is a boson while superconductivity concerns fermions, the electrons. A slightly later Allen and Jones~\cite{AlJo38} exhibited the fountain effect, a macroscopic striking consequence of 
the superfluidity. It is generally believed that superfluidity is related to the Bose-Einstein (B-E) condensation~\cite{bose24,einstein24} and both the superconductivity and the superfluidity 
of Helium 3 (a fermion) are explained by the formation of Cooper pairs which are bosons. However, no direct connection between superfluidity and Bose-Einstein condensation is invoked in the classical 
theory of L.D.  Landau~\cite{landau41,landau46} where the dynamics of a so-called ``coherent state" is deduced within the framework of general quantum many body theory, 
with no reference to the quantum statistics, fermionic or bosonic. Let us remark that in this respect the Bose-Einstein condensate is {\emph{not}} superfluid, while it becomes superfluid 
when interactions are taken into account~\cite{bogo47}. Landau is led to introduce then the concept of a two-fluids model: the liquid Helium below the $\lambda$-point  can be decomposed 
between a superfluid part which has zero viscosity and a normal part which consists of the thermal excitations of the liquid. 

On the other hand, the solidification of
Helium being achieved in 1926 by Keesom, the ability for solid Helium to exhibit B-E condensation has been naturally questioned~\cite{Bookeesom}.
Later on, in a seminal paper where they link Bose-Einstein 
condensation (and thus in their spirit superfluidity) to off diagonal long range order (ODLRO), Penrose  and Onsager~\cite{po} have investigated the possibility of a 
superfluid-like behavior in a solid at low temperature, finally for concluding to the inconsistency for a perfect crystal of such a ``supersolid" state: the short range solid order is incompatible with a long range order.
Further works by Andreev and Lifshitz  \cite{andreev}, Reatto~\cite{reatto} and Chester~\cite{chester} have  suggested that supersolids, if they exist, might be a kind of Bose-Einstein condensation of defects, vacancies, interstitials, etc. that display a coherent state and can realize a matter flow through the crystal. 
Following the theory by Andreev and Lifshitz~\cite{andreev} such a superflow was expected under forces due to an imposed pressure gradient, likewise what had been seen by Kapitza in liquid Helium. The very first experiment by Andreev and collaborators in~\cite{andreev2} did not show superflow generated in this way: they put  a small sphere inside a bucket filled of solid helium, expecting it to fall down across the solid under its own weight in the field of gravity. Similar experiments have shown as well that solid Helium does not flow under an imposed pressure difference~\cite{Grey77,Castaing89}.

In likely  the most important theoretical paper on supersolids since Penrose and Onsager's first publication, Leggett suggested~\cite{leggett}  that non classical rotational inertia (NCRI) may be a test for supersolid.  NCRI consists in an observable drop of the effective moment of inertia of a substance in a container of given geometry, like annular, cylindrical, cubic or porous vycor.  This moment of inertia is measured sensitively by the frequency of resonance of a torsion pendulum, where the mass is a piece of supersolid. Such a drop is caused by the fact that the superfluid component of the supersolid does not follow the rotational motion of the rest of the solid, although the details depend on the geometry of the solid. Such effect is of course inspired by the classical measurement of superfluid density of liquid Helium by Androniksahvili~\cite{Andro,AndroRMP}. 
In a NCRI experiment the effective moment of inertia (which is always less than the one of a rigid body, the difference being due to the nonconvected superfluid fraction) is determined  for different temperatures thanks to a torsional resonator. Once geometrical effects are taken into account, measurements of NCRI give access to the superfluid density which should be a universal function of the thermodynamical parameters, pressure and temperature. It happens that, experimentally, and contrary to what happens with superfluid liquid helium, the superfluid fraction of a supersolid is {\emph{not}} a function of the thermodynamic parameters only, a point we discuss at length below.

Although non classical rotational inertia was expected to be a fundamental characteristic of supersolids, experiments by Reppy and collaborators~\cite{reppy81} in the early 80's failed to exhibit any measurable NCRI in solid helium at cryogenic temperatures. 
In fact, over the last 40 years a more or less ongoing research activity has failed to put in evidence in a convincing way 
the supersolid state of matter~\cite{meisel}, either by NCRI or through eventual changes in the modes of propagation of disturbances like sound waves brought by the superfluid component~\cite{goodkind}.  But recently Chan and Kim~\cite{chan04a,chan04b,chan06,chan07ncri} have claimed the observation of NCRI in solid Helium 4 below $0.1-0.2 K$, something which gave new life to the subject. Measuring the resonant frequency of a torsional oscillator filled of solid Helium, they observe a rapid drop of the rotational moment of inertia of the order of a few percent when decreasing the temperature below $0.1$K.
The successful difference between Chan and Kim's experiments and the previous attempts  lied in particular in the very small amplitude of the oscillations achieved to measure the momentum of inertia. To date, at least five other groups have confirmed Kim and Chan findings~\cite{Beamish05,RiRe06,kojima,kubota,shirihama,davis} but showing important variations in the superfluid fraction with the experimental configuration as already noticed by Kim and Chan. In particular, crystal annealing is shown to lower dramatically the superfluid fraction with a strong dependence on the ratio between the surface and the bulk of the sample (something quantified by the disorder in the solid crystal) up to a variation of almost three orders of magnitude for the NCRI fraction (NCRIF)~\cite{RiRe06,RiRe07}. 
In addition, no evidence of superflow has been noticed in solid He$^4$ submitted to a localized pressure jump~\cite{Beamish05,Beamish06,Rittner09} in the conditions where NCRI exists. For these reasons, the status of this NCRI effect as a proof of a new ``supersolid" state of matter has been often questioned and remains a subject of scientific debates.
However, we would like to emphasize here that they are some important experimental facts in favor of the existence of this supersolid state at low temperature in He$^4$, in particular: no NCRI is seen for He$^3$~\cite{chanHe3};  a (small) thermodynamical signature of a phase transition has been measured for the heat capacity at $T \sim 0.1 K$~(\cite{chan07} and see also \cite{goodkind07} where it is shown that no clear conclusions can be drawn yet). Amazingly, an increase of the shear modulus for solid helium is also observed with a temperature dependence similar to the observed NCRI~\cite{Beamish07,Beamish10,Rojas10}. Notice that this change in the shear modulus cannot explain mechanically the drop in NCRI but could witness the occurence of phase transition in solid Helium although it has been observed in solid Helium3 as well! Finally, in what could be considered as an {\it ``experimentum crucis''} Kim and Chan~\cite{chan04b} have shown that the NCRI signal was suppressed when changing the topology of the rotating container in order to block a macroscopic superflow. Regarding this so-called ``blocked annulus experiment'', the only interpretation of the observed NCRI otherwise is by the existence of macroscopic superflow, and, to the best of our knowledge, no other interpretation of this crucial experiment has been published.

Given the present experimental context it would be interesting to investigate theoretically the following questions:
\begin{quote}
{ \it  I -- Why does solid helium yield a coherent superflow in a  non classical rotational inertia experiment but responds as an ordinary elastic solid in pressure/external force driven experiments~?}
 \end{quote}
  \begin{quote}
{ \it II -- Why is the superfluid or supersolid fraction so small~?}
 \end{quote}
  \begin{quote}
{ \it III -- Why does the superflow density change so widely from sample to sample? and what is the role of the He$^3$ impurities in the observed superflow density?  }
 \end{quote}
  \begin{quote}
{ \it IV -- Why is there an anomaly of heat capacity~? Is this excess of heat capacity related to the superfluid fraction~?}
 \end{quote}
  \begin{quote}
{ \it V --  Why is there an anomaly of shear modulus~?  and how is this anomaly related to the supersolid fraction, if it is~?  }
 \end{quote}
 
 The first question,  { \it  I}, suggests that, indeed, supersolidity is a rather complex phenomenon:  there are two different large scale collective motions.  A bit like in Landau's two fluid model for superfluid helium, one is the lattice deformation as in ordinary solids while the second is a superflow, the lattice and the superfluid motion may be realized independently. In some situations, depending on the boundary conditions, the system reacts as an ordinary solid, and with a superfluid type of motion in other occasions. In this sense, because of rotation the superfluid mode is excited or imposed by the boundary conditions, while the solid under gradient pressure does not require superflow to reach equilibrium.

Concerning the second point, { \it  II}, the zero temperature limit gives a ``superfluid fraction'' or ``supersolid fraction'' about only 1 \%, while in superfluids the superfluid fraction is always 100\% at zero temperature. In his seminal paper Leggett \cite{leggett} indicates that the crystalline structure provides a natural lower value for the superfluid fraction at $T=0K$. However theoretical understanding of this point remains a major, if not the major, challenge.

Although the points { \it  III}, { \it  IV} and {\it V} deserve careful investigations, there is yet no fair understanding of their possible connection with the phenomenon of supersolidity and we will leave them for future research. 

\medskip
 
The theory of supersolids presents also some outstanding issues. Using an argument already presented by Penrose and Onsager~\cite{po},  Prokof'ev and Svistunov~\cite{ProSvi05} claimed that a defect-free supersolid cannot exist. Despite the presentation by these authors of their
claim as a theorem (meaning etymologically a statement that must be
respected) there has been an ongoing research solving numerically by one way or another the Schr\"{o}dinger equation to prove or disprove
the existence of superfluidity in a regular lattice, with widely varying
not to say contradictory results. Like many others in the field, they assume a fixed lattice without quantum fluctuations
induced by zero point phonons in the lattice, although those fluctuations are crucial (section II-C) to yield number fluctuations in the system and
ultimately a long range phase coherence.
On the numerical side, based on path integral Montecarlo,  there is no crystal clear conclusion. For instance, Ceperley and Bernu~\cite{CeBu04} conclude to the absence of superfluid density in some numerical model while Cozorla and Boronat~\cite{boronat}  obtain a finite superfluid fraction or NCRI. Clark and Ceperley~\cite{ClarkCe06} and Boniensegui {\it et al.}~\cite{Boninsegni06} predict that a perfect crystal cannot exhibit off diagonal long range order, in contradiction with the conclusions of  Reatto {\it et al.}~\cite{Rea05}. 
It is necessary to be cautious with such numerical estimates: path integral Montecarlo estimations are difficult in the limit $T\rightarrow  0$. They rely on the winding number that seems to be valid only in one (or quasi one) dimensional multiconnected system, known not to be crystal in the ground state. Variational approaches  and the conclusions drawn depend sensitively on the class of trial functions used. Moreover the analysis based on the notion of winding number assumes, although somewhat implicitly, a fixed network, although this network experiences quantum fluctuations as well. This type of fluctuations yields a well defined quantum phase to the ground state, even if there is one particle per site, an essential remark for the quantum coherence of this state which is ultimately responsible of NCRI. Finally, let us mention that the disorder itself has been also invoked to explain the NCRI signature observed in experiments, through the concept of quantum glass~\cite{biroli,zamponi} and/or the motion of dislocation network within the solid~\cite{pollet07,Aleini10,Beamish10,Rojas10}. However, to date, neutron diffraction experiments on solid helium 4 show
sharp Bragg peaks typical of a regular lattice structure in the
thermodynamical conditions where NCRI is observed with no evidence at all
of a glass like structure.

\medskip
In this review, we present a general description of the theory of supersolids. In particular, one of the aim of this paper is to reach a clear and coherent understanding  of the macroscopic properties of the supersolid state, so that it could be tested in experiments. Although mainly  of a phenomenological character, we expect that our predictions should be pertinent to solid Helium four at low temperatures (below 0.1 K). Next section (section \ref{macro}) discusses the macroscopic properties of a supersolid, based on general arguments and on a Landau type approach. Then, in section \ref{sec:dense} we show how a quantum coherent phase can be present in ordinary solids at low enough temperature. In particular, we explain that the non-zero exchange term between particles in a crystal leads to supersolidity. Finally, before the general conclusion, section \ref{Approach1} describes a consistent mean-field model of a supersolid that exhibits the main features observed in experiments and which satisfies naturally the macroscopic properties drawn in section \ref{macro}.

\section{Macroscopic description of a supersolid.}
\label{macro}

In this section, we show how macroscopic approaches and models can explain the remarkable properties of supersolidity. Contrary to microscopic approaches that would describe the dynamics at the atom level, they are based on thermodynamic variables such as the mean density, the displacement field and the quantum phase. Indeed, an implicit assumption beside is that a kind of quantum coherence is present like in superfluidity. Such hypothesis leans on the crucial experiment of the blocked annulus which cannot be explained by an anomaly in the elastic properties of this solid. We will first rapidly review the Andreev-Lifshitz two fluids model which was the first macroscopic approach for supersolidty. Then, we
present how a Lagrangian approach can be deduced from general principles giving a similar set of macroscopic equations that in the Andreev-Lifshitz model. Moreover we obtain from this model the consistent set of boundary conditions that apply to these macroscopic approaches.
It explains in particular the counterintuitive fact that rotation induces a superflow but not a pressure difference. Finally, we sketch the relevant properties and some specific solutions of this macroscopic description of a supersolid.

Let us first emphasize that the present theories are based on the mismatch in the the local number density of particles $\rho$ due to quantum fluctuations: in ordinary (classical) solids an elastic deformation induces a change of density $\delta \rho$ given by the constitutive relation:
$$\frac{\delta \rho}{\rho} =  -{\bm \nabla}\cdot {\bm u} \mathrm{,} $$
with $ {\bm u}$ the displacement field. However in a quantum solid, quantum fluctuations give rise to additional elastic deformation so that one can expect:
$$\frac{\delta \rho}{\rho} + {\bm \nabla}\cdot {\bm u} = {\mathcal O} (\varrho^{ss}) $$
where $\varrho^{ss}$ is a  quantity (given for the moment) of quantum origin that represents the superfluid or supersolid fraction. It is a dimensionless number found to be of the order of $10^{-2}$ or $10^{-3}$ in the experiements done on solid Helium.

 \subsection{Andreev-Lifshitz two fluids model}
 
In the late 60's, Andreev and Lifshitz~\cite{andreev}, following Landau's approach for the description of superfluidity, developed a set of macroscopic equations that would describe the dynamics of a supersolid. These equations provide the dynamical evolution of the mass density, the superfluid velocity, the elastic deformation and the entropy transport thanks to the total (that is the superfluid plus the ordinary solid part) energy and momentum densities:
\begin{eqnarray}
{\mathcal E } &= &\frac{1}{2} m  \varrho^{ss} {\bm v^s}^2 + \frac{m }{2} (\rho - \varrho^{ss}) \dot {\bm u}^2
+ {\mathcal E}_0(\rho, s , \epsilon_{ik})  , \label{energy}\\
 {\bm j} &=&  m  \varrho^{ss } {\bm v}^s  +  m (\rho - \varrho^{ss})  \dot {\bm u} 
\label{current}\end{eqnarray}
Here ${\mathcal E}_0(\rho, s , \epsilon_{ik})  $ is the internal energy of the body that depends explicitly on the number density $\rho$, the entropy  density $s$, and the elastic strain $\epsilon_{ik} = (\partial_i u_k + \partial_k u_i )/2$. The supersolid density, $\varrho^{ss}$, is simply a parameter here. 
As usual,  the conservation equations for the total mass, momentum and entropy give~\cite{andreev,kalath}:
\begin{eqnarray}
 \partial_t \rho + {\bm{\nabla}}\cdot \left(  {\bm j}/m \right) &=& 0
  \label{massEq}\\
 \partial_t j_i   + \partial_k {\cal T}_{ik} &= &0 \label{momentumEq}\\
 \partial_t s + {\bm{\nabla}}\cdot \left( s  \dot{\bm u}  \right) &=&0,
\label{entropy}
\end{eqnarray}	 
where ${\cal T}_{ik}$ is the momentum tensor that can be deduced from the energy balance (see below).
Additionnally, the superfluid velocity satisfies the Landau equation \cite{landau41}:
\begin{eqnarray}
\partial_t {\bm v}^s + {\bm \nabla} \left(\frac{1}{2} ({\bm v}^s)^2 + \varphi \right)=0.
\label{landauEqn}
\end{eqnarray}
Thus the unknown quantities ${\cal T}_{ik} $ and $\varphi$ in previous equations are derived from the energy conservation:
\begin{eqnarray}\partial_t{\mathcal E} + {\bm{\nabla}}\cdot {\bm Q} \equiv 0.\label{ConsEnergy} \end{eqnarray} 

After a straightforward calculation, one readily obtains a well defined closed system of partial differential equations, using the Einstein's convention of summation for the indices:
\begin{eqnarray}
\varphi &=&   \frac{1}{m}  \mu - \frac{1}{2} ({\bm v}^s - \dot {\bm u} )^2   , \label{varphi} \\
{\cal T}_{ik}   &=& m \varrho^{ss}  v^s_iv^s_k +  m  (\rho- \varrho^{ss}) \dot{u}_i  \dot{u}_k  -\left[ {\mathcal E}_0 - Ts - \mu \rho +  \frac{m}{2} \varrho^{ss} ({\bm v}^s - \dot {\bm u} )^2 \right] \,\delta_{ik} - \sigma_{ik} ,\label{Tik}\\
{\bm Q}_i &=& m \varrho^{ss} \left[ \frac{\mu}{m}+ {\bm v}^s\cdot \dot{\bm u} - \frac{1}{2}  \dot {\bm u} ^2 \right] v^s_i +  \left(   s T + (\rho-\varrho^{ss} )  \left( \frac{m}{2} \dot {\bm u} ^2+ \mu \right)\right)\dot {u}_i + \sigma_{ik}\dot {u}_k ,  \label{energyflux}
\end{eqnarray}	  
where the temperature is $T =  \frac{\partial {\mathcal E}_0 }{\partial  s}$, the chemical potential is $\mu = \frac{\partial {\mathcal E}_0 }{\partial  \rho}$ and the stress tensor  is $\sigma_{ik} = \frac{\partial {\mathcal E}_0 }{\partial  \epsilon_{ik}}$.
Therefore the macroscopic set of equations finally reads:

\begin{eqnarray}
 \partial_t \rho + {\bm{\nabla}}\cdot \left( \varrho^{ss } {\bm v}^s  +  (\rho - \varrho^{ss})  \dot {\bm u}  \right) = 0,
  \label{massEq2}\\
  \partial_t {\bm v}^s + {\bm \nabla} \left(\frac{1}{2} ({\bm v}^s)^2 + \varphi \right)=0,\label{landauEqn2}\\
 m \partial_t ( (\rho - \varrho^{ss})  \dot {\bm u} )+ m\varrho^{ss} v_k^s \partial_i v_i^s  + \partial_i (m (\rho-\varrho^{ss}) \dot u_i \dot u_k)    -  \partial_k [{\mathcal E}_0 - Ts - \mu (\rho -\varrho^{ss}) ] +\partial_i \sigma_{ik} = 0,  \label{momentumEq2}\\
 \partial_t s + {\bm{\nabla}}\cdot \left( s  \dot{\bm u}  \right) =0.
\label{entropy2}
\end{eqnarray}	  


The first equation (\ref{massEq2}) is the usual equation of mass conservation, while equation (\ref{landauEqn2}) is an Euler like equation for the superfluid velocity ${\bm v}^s$. In particular, if the initial superflow is not of a vortex type then the superfluid velocity displays a potential flow, ${\bm v}^s=\frac{\hbar}{m} {\bm \nabla}\Phi$, with $\Phi$ a dimensionless  field which is eventually linked to the global phase of the $N$ body wave function of the system (see below). Equation (\ref{momentumEq2}) is the usual equation for the elastic response of the system:  if no superfluid is present ($\varrho^{ss}=0$), then one recovers the usual equation for elastic dynamics. Again, the last equation (\ref{entropy2}) stands for the 
entropy conservation.
Boundary conditions have to be specified for solving this set of equations: surprisingly, such conditions are lacking in Andreev and Lifshitz approach and we will deduce them from the Lagrangian approach described below. 
Let us emphasize at this stage that this model opens the way to a simple explanation of the apparent paradoxical behavior observed experimentally, that is a nonclassical rotational inertia fraction in the limit of small rotation speed but a solid-like elastic response to small stress or external force field: the implicit coupling between the superfluid dynamics and the elasticity present both in the equations (\ref{massEq2} and \ref{landauEqn2}) might be so that the solid can react very differently depending on the external constraints applied, as we shall see in Sections \ref{Examples} and \ref{Utube}.

 \subsection{Lagrangian Continuum description of a supersolid}
 
In this section we present a derivation of the equations of motion for a supersolid in the general framework of continuum mechanics and based on general principles. It is based on a Lagrangian formalism, neglecting dissipation. The final result includes the equations of elasticity together with the equation for the superfluid component. This is valid under the general conditions of applicability of continuum mechanics: in particular the perturbations under consideration are long wave (practically much longer than the lattice size). The form of this Lagrangian is constrained by various symmetries, the Galilean invariance and the invariance under rotation and translation. This leaves only one possible form, where the microphysics enters only in the values of various coefficients like the Lam\'e coefficients of elasticity and the superfluid density $\varrho^{ss}$. The power of this Lagrangian formulation, proposed also by Son~\cite{son}, is that it does not require any explicit microscopic model but only laws of symmetry. One of the best related example is the theory of superfluidity by Landau, free of any explicit connection between the phenomenon itself and its detailed microscopic explanation. Besides the general symmetries, Landau made use of  the very deep remark that the chemical potential is conjugate (in the quantum sense) to the number of particles. A similar property is used below for the supersolid, but it has to be changed in order to take into account that the number density of particles may change either by compression/dilation of the lattice or by changing the density of the superfluid component. 
 
This derivation of the equations of motion from a Lagrange principle provides a natural derivation of the boundary conditions which are crucial here because they explain why superflow cannot be induced by a pressure gradient in a supersolid (contrary to the case of a superfluid) although one is generated by rotation, as explained above. Although the model is very much inspired by the deduction by Landau of the equations of motion for a superfluid as a mixture of normal fluid and of its superfluid component, it differs from Andreev and Lifshitz approach because of the Lagrangian formalism. Here the ``normal'' fluid is somehow the crystal lattice, although the superfluid is very much like the superfluid component of the ``mixture'' superfluid/normal fluid.  
Thus, the dynamical equations are obtained through the variation of the action $S =\int {\mathcal L}_{eff}\, {\mathrm{d}}t {\mathrm{d}}^3 {\bm r}$, where the effective Lagrangian density reads \cite{ss1,ss2}:
 
 \begin{equation}
 {\mathcal L}_{eff} =-  \rho\,\left[  \hbar \frac{\partial \Phi}{\partial t} + \frac{\hbar^2}{2m} \left({\bm \nabla}\Phi\right)^2\right]  +   \frac{\hbar^2}{2m}(\rho\delta_{ik} -  \varrho^{ss}_{ik}) \left({\bm \nabla}\Phi - \frac{m}{\hbar}  \frac{{\mathrm D} {\bm u}}{{\mathrm D}t} \right)_i  \left({\bm \nabla}\Phi - \frac{m}{\hbar}  \frac{{\mathrm D} {\bm u}}{{\mathrm D}t} \right)_k     -{\mathcal E}(\rho) - \frac{1}{2}\lambda_{iklm}  \epsilon_{ik}\epsilon_{lm}  \label{effLagrangian}
      \end{equation}
In the above expression we have used the so-called material derivative defined naturally as 
$$ \frac{{\mathrm{D}}{\bm{u}}}{\mathrm{D}t} = 
	  \frac{\partial{\bm{u}}}{\partial t} + 
	  \frac{\hbar}{m}{\bm{\nabla}}\Phi\cdot \bm {\nabla} \bm{u}
  \mathrm{.}$$ It is the rate of change of a quantity along the flow lines defined by the velocity potential $\Phi$. This derivative makes the rate of change Galilean invariant. 
As before, 
$$ \epsilon_{ik} =  \frac{1}{2} \left(\frac{\partial u_i}{\partial x_k} + \frac{\partial u_i}{\partial x_k}  \right)$$ 
is the strain tensor of the Hookean elasticity theory, $u_i$ being the displacement field. The quantity $\varrho^{ss}_{ik}$ is the supersolid density tensor, which is in general a symmetric matrix. We shall keep it aside for a while, but for more specific applications and also for the sake of simplicity, we shall often reduce it to an isotropic form  $\varrho^{ss}_{ik} = \varrho^{ss}\delta_{ik}$. Everywhere the indices are for Cartesian coordinates. These equations of motion involve the displacement field $u_i$, as in usual solids, the number density $\rho$ and the quantum phase $\Phi$, related to a ground state wave function. All these three quantities are functions of the position $\bm{r}$ and the time $t$.  
  
 This expression of the Lagrangian density given in equation (\ref{effLagrangian}) is explained as follows: 

\begin{quote}
1) The part representing the change of energy due to the elastic deformations of the crystal  is the same as in the standard mechanics of solids. In the limit of small deformations,  the final result is minus the (positive) quadratic form $- \frac{1}{2}\lambda_{iklm}  \epsilon_{ik}\epsilon_{lm}$, therefore the elastic stresses  are quantities proportional to the gradient of the displacement field $u_i$, with coefficients of proportionality called the Lam\'e coefficients. The number of independent Lam\'e coefficients depend on the symmetries of the crystal and is at least two. 
\end{quote}

\begin{quote}
2) From the general principles of quantum mechanics, ${\bm v}^s = \frac{\hbar}{m} {\bm \nabla}\Phi$ is the superfluid velocity of the system, due to a Galilean boost of the ground state wave function. But in the present case the crystal lattice may have a different velocity than the superfluid component. This velocity difference is  proportional to $ \left({\bm \nabla}\Phi - \frac{m}{\hbar}  \frac{{\mathrm D} {\bm u}}{{\mathrm D}t} \right)$. If this difference vanishes (that is if the lattice and the superfluid component move at the same speed) the kinetic energy of the motion is just the usual $\frac{\hbar^2}{2m}\rho( {\bm \nabla}\Phi)^2$ where $\rho$ is the total number density. If only the crystal lattice moves, the velocity $\frac{\hbar}{m} {\bm \nabla}\Phi$ is zero and the kinetic energy is just half the mass density of the crystal lattice times the square of the lattice speed. Those two possibilities (global uniform speed or displacement of the lattice only) are well taken into account by adding the two contributions, as done in $$-\frac{\hbar^2}{2m}\left[\rho \left({\bm \nabla}\Phi\right)^2 - (\rho\delta_{ik} -  \varrho^{ss}_{ik}) \left({\bm \nabla}\Phi - \frac{m}{\hbar}  \frac{{\mathrm D} {\bm u}}{{\mathrm D}t} \right)_i  \left({\bm \nabla}\Phi - \frac{m}{\hbar}  \frac{{\mathrm D} {\bm u}}{{\mathrm D}t} \right)_k \right] \mathrm{.}$$ 
This is because the difference between the total density $\rho$ and the superfluid density tensor, $\varrho^{ss}_{ik}$, $(\rho \delta_{ik} -  \varrho^{ss}_{ik}) $, is the mass density of the crystal lattice, a rank two tensor for a crystal in general. This density is multiplied by the square of the difference of the superfluid velocity minus the velocity of the lattice in order to make this term exactly zero if the two velocities are equal, as it happens when the full system (superfluid part and lattice) moves at the same speed: in this case, all the kinetic energy is given by the contribution  $\frac{\hbar^2}{2m}\rho \left({\bm \nabla}\Phi\right)^2$ to $ {\mathcal L}_{eff}$ so that the other contribution to the kinetic energy must cancel if ${\bm \nabla}\Phi = \frac{m}{\hbar} \frac{{\mathrm D} {\bm u}}{{\mathrm D}t}$. Notice that this expression of the Lagrangian density can be derived rigorously for the mean field model of supersolid~\cite{ss1,ss2}, as introduced in section \ref{Approach1}. 
\end{quote}
\begin{quote}
3) The macroscopic (or averaged) quantities ${\mathcal E}$, $\lambda_{iklm} $ and $ \varrho^{ss}_{ik}$ are respectively the internal energy of the solid, the Lam\'e elastic constants and, last but not least, the ``superfluid'' density tensor. These  expressions do depend, in principle, on thermodynamical variables, crystal structure, etc, and these functions should be given by experimental measurements. In particular, we shall explicitly express that the superfluid density is a function of the local density number $\rho$ and it will be often considered further on within an isotropic symmetry $ \varrho^{ss}_{ik}= \varrho^{ss}(\rho) \delta_{ik}$
\end{quote}

Finally, notice that this Lagrangian is Galilean invariant and that the Hamiltonian of the system writes:
\begin{eqnarray}
{\mathcal H } & = & \Phi_t \frac{\delta {\mathcal L}_{eff}}{\delta \Phi_t } + {\bm u}_t \cdot \frac{\delta {\mathcal L}_{eff}}{\delta {\bm u} _t } -{\mathcal L}_{eff},
\nonumber\\
&= & \frac{\hbar^2}{2m} \varrho^{ss}_{ik}\partial_i\Phi \partial_k\Phi + \frac{m}{2}  (\rho\delta_{ik} -  \varrho^{ss}_{ik})
	 \frac{{\mathrm D} {u_i}}{{\mathrm D}t}  \frac{{\mathrm D} {u_k}}{{\mathrm D}t}  + {\mathcal E} + \frac{1}{2}\lambda_{iklm}  \epsilon_{ik}\epsilon_{lm}   .
\label{hamilton}
\end{eqnarray}	

Taking the variations of $\mathcal L$ as a functional of $\rho$, $\Phi$ and $\bm{u}$, the final result is a set of coupled partial differential 
equations for the those fields~\cite{ss1,ss2}:

\begin{eqnarray}
 \frac{\partial \rho}{\partial t} +  {\bm{\nabla}}\cdot\left( \rho\, \frac{\hbar}{m}   {\bm{\nabla}}{\Phi }\right) + 
\frac{\partial}{\partial x_{k}} \left( (\rho-\varrho^{ss}) (\delta_{ki} - \partial_k u_i)\left(\dot u_i  -  \frac{\hbar}{m} \partial_i {\Phi} \right)\right) &=& 0
   \label{eq:lagrangtimedeptotalmass}\\
m \frac{\partial}{\partial t}\left[
   (\rho-\varrho^{ss})\left(  \dot {u_k}  - \frac{\hbar}{m} \partial_k \Phi  \right) \right]  + \hbar   \frac{\partial  }{\partial x_{k}} \left[(\rho- \varrho^{ss}) \left(   \dot {u}_i   - \frac{\hbar}{m} {\partial_i} \Phi  \right)\partial_k \Phi  \right]   + \frac{\partial}{\partial x_{k}}\left(\lambda_{iklm} \epsilon_{lm}\right)      &= &0
\label{eq:lagrangtimedeptotalcauchy}\\
\hbar \frac{\partial \Phi}{\partial t} + 
\frac{\hbar^2}{2m} \left({\bm{\nabla}}\Phi\right)^2  + 
  {\cal E} '(\rho)  & = & 0.
\label{eq:lagrangtimedeptotalBernoulli} 
\end{eqnarray}	
Here, for the sake of simplicity and further on (except when explicitly stated), we have assumed isotropic macroscopic properties for the supersolid tensor $ \varrho^{ss}_{ik}=\varrho^{ss} \delta_{ik}$.
To these equations the least action or Hamilton principle provides also the boundary conditions, which were lacking for the Andreev-Lifshitz model. In the  stationary case for free boundary condition it writes:
\begin{eqnarray}
  \left[   \hbar (\rho-\varrho^{ss}) \left( \frac{{\mathrm D}u_i}{{\mathrm D}t} -
      \frac{\hbar}{m}\frac{\partial 
      \Phi}{\partial x_{i}} \right)  \frac{\partial 
      \Phi}{\partial x_{k}} + \lambda_{iklm} \epsilon_{lm}  \right] \hat e_k
  & = &0  \quad {\rm on} \quad \partial\Omega 
\label{BCelastic3}\\
\left[ \rho\partial_i {\Phi} - (\rho - \varrho^{ss}) (\delta_{ik} - \partial_i u_k)\left(\partial_k {\Phi} - \frac{m}{\hbar}\frac{\mathrm{D}u_k}{\mathrm{D}t}\right) \right] \hat e_i &=& 0 \quad {\rm on} \, \partial\Omega,
   \label{BCmass3}
\end{eqnarray}	
where $\hat e_i$ is a normal vector to the surface. Similarly in the case of rigid boundary conditions:
\begin{eqnarray}
 u_i & = &0  \quad {\rm on}\quad  \partial\Omega 
\label{BCelastic4}\\
 \hat e_i \, \partial_i {\Phi} &=& 0 \quad {\rm on} \quad \partial\Omega,
   \label{BCmass4}
\end{eqnarray}	

Equation (\ref{eq:lagrangtimedeptotalcauchy}) is the usual equation for the elastic response of the system. Equation (\ref{eq:lagrangtimedeptotalBernoulli}) is a Bernoulli like equation for the superfluid velocity ${\bm v}^s = \frac{\hbar}{m}{\bm \nabla}\Phi$ that gives back the usual Landau equation for the superfluid component $\partial_t {\bm v}^s= - {\bm \nabla}\Psi$, $\Psi$ being a potential defined directly from  (\ref{eq:lagrangtimedeptotalBernoulli}). Notice that equation (\ref{eq:lagrangtimedeptotalmass}) reduces to the familiar equation of mass conservation 
for potential superflows whenever $\varrho^{ss}_{ik} \rightarrow \rho \delta_{ik} $, corresponding namely to a $T=0$ superfluid. 

Finally, we can deduce from these boundary conditions the set of boundary conditions for the Andreev-Lifshitz, first for free surface (FS) boundary where all fluxes vanish:
\begin{eqnarray}
  \left[    m \varrho^{ss}  v^s_iv^s_k +  m  (\rho- \varrho^{ss}) \dot{u}_i  \dot{u}_k  -\left( {\mathcal E}_0 - Ts - \mu \rho +  \frac{m}{2} \varrho^{ss} ({\bm v}^s - \dot {\bm u} )^2 \right) \,\delta_{ik} -  \sigma_{ik} \right] \hat e_k
     &=&0  \quad {\rm on} \quad  \partial\Omega 
\label{BCelastic1}\\
 \left( \varrho^{ss } {\bm v}^s  +  (\rho - \varrho^{ss})  \dot {\bm u}  \right) \cdot \hat {\bm e} &=& 0 \quad {\rm on}\quad\partial\Omega,
   \label{BCmass1}\\
    s  \dot{\bm u}  \cdot \hat {\bm e}  &=&0   \quad {\rm on}\quad  \partial\Omega   \label{BCentropy1}
\end{eqnarray}	
where $\hat {\bm e}$ is a normal vector to the surface. On the other hand, for rigid boundaries (RB), it gives 
\begin{eqnarray}
u_i & = &0  \quad {\rm on} \quad \partial\Omega 
\label{BCelastic2}\\
 {\bm v}^s   \cdot \hat {\bm e} &=& 0 \quad {\rm on}\quad \partial\Omega,
   \label{BCmass2}\\
    s & = &0  \quad {\rm on}\quad  \partial\Omega   \label{BCentropy2}
\end{eqnarray}	

The free surface (FS) boundary condition (\ref{BCelastic1}) expresses the balance between the normal stress with a kinetic (superfluid) pressure. In particular the zero normal stress is recovered when no superfluid phase is present.  For rigid boundary (RB) condition (\ref{BCelastic2}) no displacement is allowed at the boundary (obviously, in the case of moving boundaries, the condition adapts such that the crystal moves together with the boundaries). The boundary conditions for the phase (\ref{BCmass1} and \ref{BCmass2}) correspond to a zero flux of matter condition at the boundary (again easily extendable to moving boundaries). Conditions (\ref{BCentropy1} and \ref{BCentropy2}) account for the entropy fluxes at the boundaries. 

 \subsection{Superfluid Properties: Sound Waves, Non-classical Rotational of Inertia}
\label{sec4}

Given the general model obtained above, where the superfluid tensor $\varrho^{ss}$ and the elasticity parameters have to be derived from a microscopic model or from experimental
measurements, we will present below its characteristics properties. In particular, we will explain how the supersolidity appears in the NCRI, we will investigate the long-wave perturbations dynamics and the existence of quantum vortices and permanent currents.

\subsubsection{Superfluid density}
The mass conservation  equation (\ref{eq:lagrangtimedeptotalmass}) defines a current reminiscent of Landau's ``two-fluid'' expression for the mass current, at least up to linear order
$$
 j_i =  \varrho^{ss}_{ik} v^s_k  +   (\rho \delta_{ik}  -   \varrho^{ss}_{ik}) \dot u_k
$$
with  ${\bm v}^s=  \frac{\hbar}{m}{\bm \nabla}\Phi$.
This quantity plays an important role in the NCRI that we shall describe below. Being a fundamental quantity for the experimental observations we shall consider in more details how it is possible to compute it starting from a microscopic model. 

\subsubsection{Non-classical Rotational of Inertia}

As suggested by Leggett \cite{leggett}, supersolidity should clearly show up {\it via} an Andronikashvili kind of experiment of non-classical rotational inertia (NCRI), namely a difference between the mass carried under rotational motion and the total mass. Experimentally, this is measured with great accuracy by the frequency of a torsion pendulum. Indeed let us suppose that the wall of the container of volume $\Omega$ rotates with an uniform angular speed $\omega$. For low angular speed the crystal moves rigidly with the container $\dot {\bm u} = {\bm \omega}\times {\bm r}$ without any elastic deformation. The densities $\rho$ and $\varrho^{ss}$ (where $\rho$ is the number density) being constant in space, equation (\ref{eq:lagrangtimedeptotalmass}) simplifies into 
\begin{equation}
\nabla^2\Phi=0 \quad {\rm in} \quad \Omega \quad {\rm with} \quad {\bm \nabla} \Phi \cdot\hat e = (m/\hbar) ({\bm \omega}\times {\bm r}) \cdot \hat e \quad {\rm on} \quad \partial\Omega .\label{perfect} \end{equation}
It corresponds to the classical problem of the flow of an inviscid and incompressible (perfect) fluid inside a rotating container. It has a unique solution, and exact solutions can be computed for various geometries in 2D~\cite{milne}. For instance, as was shown by a student of Kirchhoff, a rotating ellipsoid yields a potential flow that is a linear function of the coordinates, a solution given in the book by Lamb~\cite{lamb}. 
The effective or measurable moment of inertia comes directly from the energy per unit volume of the system (\ref{hamilton}).
In the rotating case $\dot {\bm u} = {\bm \omega}\times {\bm r}$ the energy (\ref{hamilton}) can be written as
$$E = \frac{1}{2} I_{ss} \omega^2$$ where $I_{ss}$ is the $zz$ component (the angular velocity ${\bm \omega}$ being along the $z$-axis) of the inertia moment (we use here an isotropic supersolid density tensor):
$$
I_{ss} = m \varrho^{ss}  {\mathcal I}_{pf} + m(\rho - \varrho^{ss}) {\mathcal I}_{rb}$$
with 
$${\mathcal I}_{pf} = \int_\Omega (\bm \nabla \Phi)^2 d{\bm r},$$ with $\Phi$ solution of \begin{equation}
\nabla^2\Phi=0 \quad {\rm in} \quad \Omega \quad {\rm with} \quad {\bm \nabla} \Phi \cdot\hat e = (\hat z \times {\bm r} ) \cdot \hat e \quad {\rm on} \quad \partial\Omega .\nonumber \end{equation}
This number depends on the geometry only \cite{milne,fetter}. The constant  ${\mathcal I}_{rb} $ is also a geometrical factor corresponding  to rigid body rotational inertia ($x\& y$ orthogonal to the axis of rotation)  $${\mathcal I}_{rb}= \int_\Omega(x^2+y^2) d{\bm r}.$$ 
The relative change of the moment of inertia whenever the supersolid phase appears is (here ${I}_{rb}=m \rho  \, {\mathcal I}_{rb}$)
\begin{equation}\frac{(I_{ss}-I_{rb})}{I_{rb} } = - \frac{\varrho^{ss}}{\rho} \left(1-\frac{{\mathcal I}_{pf}}{{\mathcal I}_{rb}}\right)\label{ncri}\end{equation}
Because ${\mathcal I}_{pf}<{\mathcal I}_{rb}$, one has $ (I_{ss}-I_{rb})/I_{rb} \leq 0$ as expected and observed experimentally \cite{chan04a,chan04b,chan06}. The NCRI fraction disappears when the solid returns from supersolid to the ordinary solid phase ($ \varrho^{ss} \rightarrow 0$).

\subsubsection{Sound waves of isotropic lattices}
To characterize the dynamics of this system, we will investigate the small perturbations around a non-deformed ($\bm u=0$) and steady ($\bm \nabla \Phi =0$) state of average density $\bar\rho$ using the linearized version of (\ref{eq:lagrangtimedeptotalmass},\ref{eq:lagrangtimedeptotalcauchy},\ref{eq:lagrangtimedeptotalBernoulli}). Before carrying the usual small amplitude perturbation analysis, we need to account that  an ordinary compression in a solid changes the number density of particle in an obvious way, because of the deformation itself, {\it i. e.} $\rho =  \bar\rho +  {\bar\rho}  {\bm \nabla}\cdot {\bm u} + {\delta \rho} $. Finally, the linearized system of equations for ${\bm u}$, $\delta\Phi$ and $\delta \rho$ reads:

\begin{eqnarray}
 \frac{\partial\, \delta \rho }{\partial t}+     \varrho^{ss}\,  \left( \frac{\hbar}{m} \nabla^2 \delta\Phi - ({\bm \nabla}\cdot \dot {\bm u} )\right)
  &=& 0
   \label{sound1}\\
\hbar \frac{\partial \delta\Phi}{\partial t}  + {\cal E} ''(\bar\rho) \delta\rho & = & 0
\label{sound2} \\
 m  (\bar\rho-\varrho^{ss}) \frac{\partial}{\partial t}
   \left(  \partial_t {\bm u}   - \frac{\hbar}{m}{\bm \nabla }   \delta\Phi  \right)   - (\lambda+\mu_s) {\bm \nabla} ( {\bm \nabla }\cdot {\bm u} ) - \mu_s \nabla^2 {\bm u}   
     &= &0.
\label{sound3}
\end{eqnarray}	
Here we have assumed an isotropic solid so that the elastic term in (\ref{eq:lagrangtimedeptotalcauchy}) simplifies into:
$$  \frac{\partial}{\partial x_{k}}\left(\lambda_{iklm} \frac{\partial u_l}{\partial x_{m}}\right)  =  (\lambda+\mu_s)  \partial_{ik} u_k + \mu_s \nabla^2 u_i ,$$
where $ \lambda=K-\frac{2}{3} \mu_s$ is the second Lam\'e coefficient, and $K$ and  $\mu_s$ are the compressibility and shear modulus of the solid. 

Firstly, taking curl of equation (\ref{sound3}), we obtain that the shear waves are decoupled from other modes and propagate following ( $\varpi = {\bm \nabla }\times {\bm u}$)
$$ m  (\bar\rho-\varrho^{ss}) \frac{\partial^2}{\partial t^2}
 \varpi    - \mu_s \nabla^2 \varpi  =0.$$
 As fist noticed by Andreev-Lifshitz the shear mode velocity depends on the supersolid density, namely
 
\begin{equation}
c_{shear} = \sqrt{ \frac{\mu_s}{m(\bar\rho - \varrho^{ss})}}.
\label{shearwaves}
\end{equation}
It is thus tempting to deduce from the dependence of the shear mode velocity an alternative measure of $\varrho^{ss}$ and also to explain qualitatively the increase of the
effective shear modulus (extracted experimentally from the shear mode velocity) observed at low temperature~\cite{Beamish06} simply through the growth of the superfluid fraction.
However, although the increase of the shear mode velocity and the trend of its dependence are correct, such comparison fails quantitatively by about one order of magnitude since
the shear modulus is observed to vary of more than $10$ \% for solid Helium samples for which the NCRIF is expected to be of the order of one \%. Recent results~\cite{Beamish10,Rojas10} suggest that such shear modulus variation could be also due to dislocation motion. However, up to now, let us remark that NCRIF and shear modulus have never been measured on the same sample simultaneously. Given the large
variations of NCRIF between experimental conditions, such simultaneous measurements would be important. 

In addition to this shear mode, taking divergence of equation (\ref{sound3}), we obtain for the elastic compressibility $\delta\psi = {\bm \nabla }\cdot{\bm u}$
\begin{eqnarray}
 m  (\bar\rho-\varrho^{ss}) \frac{\partial}{\partial t}
   \left( \frac{\partial}{\partial t} \delta\psi  - \frac{\hbar}{m}{\bm \nabla }^2   \delta\Phi  \right)   - (\lambda+2 \mu_s) {\bm \nabla} ^2  \delta\psi  =0.
\label{sound4}
\end{eqnarray}	
Together with (\ref{sound1}) and (\ref{sound2}), it form a coupled set of equations for the compression and for the phase (Bogoliubov-like) modes. 
The dispersion relation for these modes follows from a linear eigenvalue problem for the variables $\delta\rho$, $\delta\Phi$, and $\delta\psi $ in the Fourier space (with ${\mathcal E}''(\bar\rho)=\frac{d^2{\mathcal E}}{d\bar\rho^2}$) :
$$
\left( \begin{array}{ccc}
i \omega  & -\frac{\hbar}{m} k^2 \varrho^{ss}  &-i \varrho^{ss} \omega \\
{\mathcal E}''(\bar\rho)& i\hbar \omega &0 \\
0 & (i \hbar \omega (\bar\rho - \varrho^{ss})  k^2 )&   \left((\lambda+ 2 \mu_s)  k^2  -m(\bar\rho  - \varrho^{ss} ) \omega^2 \right) \\
\end{array}\right)  \left( \begin{array}{c}
 \delta\, \hat \rho_{\bm k}\\
 \delta\hat \Phi_{\bm k}\\
 \delta\hat\psi_{\bm k}
 \end{array}\right) =0.
$$
Solving this linear system provides a simple algebraic equation for the dispersion relation, from which we can easily see that it is linear, so $\omega = v k$, where the wave speed $v$ is given by the roots of the determinant of the matrix, so that:

$$ v^2=\frac{K+\frac43 \mu_s}{2m (\bar\rho- \varrho^{ss})} \left( 1 \pm \sqrt{1-\frac{4\varrho^{ss} {\mathcal E}''(\bar\rho)(\bar\rho- \varrho^{ss})}{K+\frac43 \mu_s}} \right) $$

In the limit $\varrho^{ss} \rightarrow 0$, which is the most realistic situations, the two propagation speeds are:

$$v_1^2  = \frac{K+\frac43\mu_s}{m\bar\rho} \left( 1 +  \frac{\varrho^{ss}}{\bar\rho}  \left( 1 -\frac{\bar\rho^2 {\mathcal E}''(\bar\rho)}{K+\frac43\mu_s} \right) \right)+ {\mathcal O}( \varrho^{ss} )^2 $$ 
and 
$$  v_2^2 = \frac{ \varrho^{ss} {\mathcal E}''(\bar\rho)}{m}   + {\mathcal O}( \varrho^{ss} )^2.$$
Introducing the classical longitudinal elastic wave speed $c_K = \sqrt{\frac{K+\frac43\mu_s}{m\bar\rho}}$ and noting that $v_2$ is the Bogoliubov speed for a weakly interacting Bose gas of density $\varrho^{ss}$, we remark that:
$$ v_1^2= c_K^2\left( 1 + \frac{\varrho^{ss}}{\bar\rho}\right)  -v_2^2. $$
Therefore, the second mode, $v_2$ (related to the phase mode), disappears at the supersolid-ordinary solid transition while the first mode describes simply the ordinary compression 
waves.

\subsection{Simple steady {\it elastic} solutions of the macroscopic equations of a supersolid}
\label{Examples}

On the other side, we will show here that pure elastic behavior can also be observed in such solid under external constraint, with no macroscopic quantum phase and thus no superflow. Indeed, this is quite natural since classical elasticity of solid is retrieved in the dynamics when no superflow is present. 
 
For the sake of simplicity we shall assume an isotropic solid so that the elastic tensor term in (\ref{eq:lagrangtimedeptotalcauchy}) reads $$ \partial_k\sigma_{ik}   =  (\lambda+\mu_s)  \partial_{ik} u_k + \mu_s \nabla^2 u_i ,$$ where $ \lambda=K-\frac{2}{3} \mu_s$ is the second Lam\'e coefficient, and $K$ and  $\mu_s$ are the compressibility and the shear strees of solid helium. 

\subsubsection{Gravity driven flow}

\begin{figure}[hc]
\begin{center}
\centerline{a) \includegraphics[width=5cm]{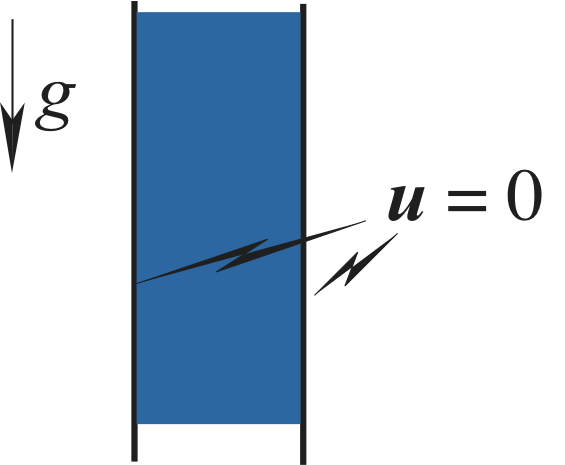} \quad \quad \quad b)  \includegraphics[width=5cm]{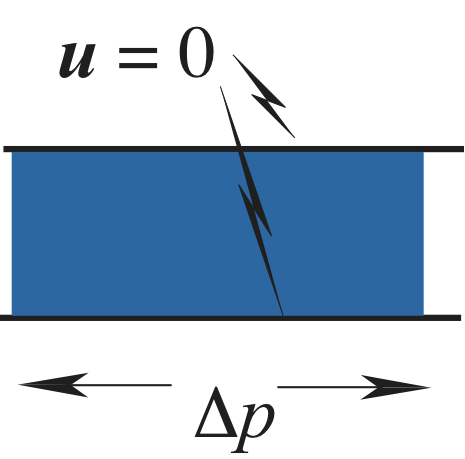} } 
\centerline{c) \includegraphics[width=5cm]{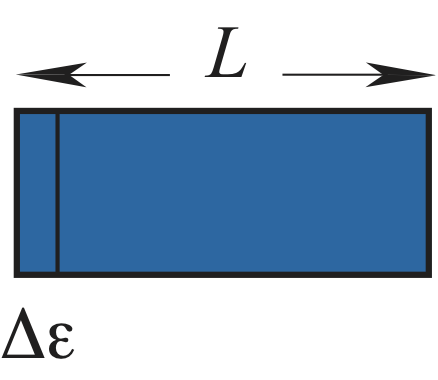}   \quad \quad \quad d) \includegraphics[width=5cm]{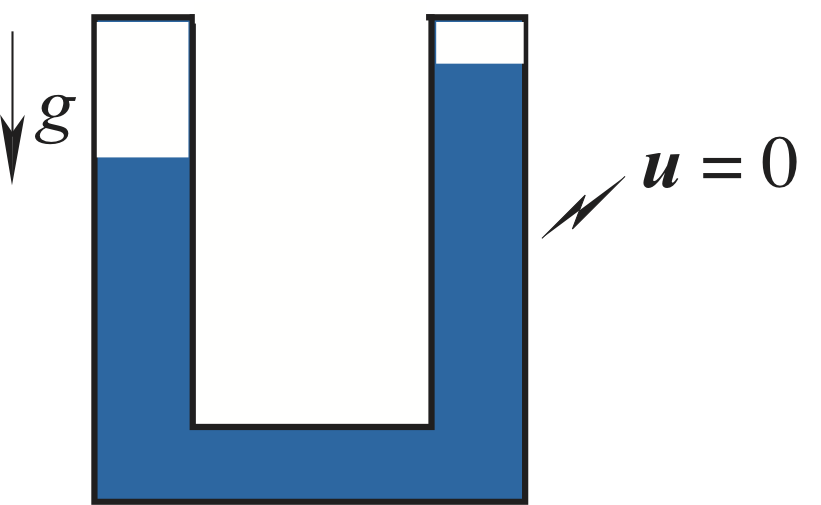}  } 
\caption{\label{steadyeg} 
}
\end{center}
\end{figure}

Let us see if a flow is driven by gravity inside a simple domain as shown in Fig. \ref{steadyeg}-a. One has to solve the equations  (\ref{eq:lagrangtimedeptotalmass},\ref{eq:lagrangtimedeptotalcauchy},\ref{eq:lagrangtimedeptotalBernoulli}) adding a bulk force per unit mass $-g \hat {\bm z} $ to  (\ref{eq:lagrangtimedeptotalcauchy}) and satisfying the boundary conditions that ${\bm u} =0$.
It is straightforward to observe that steady solutions are satisfied with an uniform phase, that is a zero superfluid velocity: $${\bm v}^s =0,$$
with a steady displacement of the lattice: $$\dot {\bm u}=0.$$ 
The equilibrium condition is the elastic equilibrium of a body in gravity for the vertical displacement ($u_z$) in terms of the transverse variable ($x$):

\begin{eqnarray}
  -\mu_s   \frac{\partial^2 u_z}{\partial x^2}   
     &= &- m \rho g
\label{gravity1bis}\\
  \mu/m  +g z & = & Ct.
\label{gravity2bis} 
\end{eqnarray}	
where the second condition follows directly (\ref{eq:lagrangtimedeptotalBernoulli}), assuming that $\rho$ is constant (or that its variations can be neglected) and introducing the chemical potential $\mu$

Solving the equation (\ref{gravity1bis}) for the displacement with fixed boundary conditions, the second equation leads to the variation of a chemical potential: $ \mu = \mu_0 -m g z$. 
\begin{eqnarray}
 u_z &=&\frac{m \rho g }{2\mu_s} x (x-\ell) ,
\label{gravity1} \\
 \mu& =& \mu_0 - m g z
\label{gravity2}
\end{eqnarray}	
Where $\ell$ is the width of the channel. In conclusion, no superflow is formed in this configuration: the response to imposed gravity is a strained lattice without superflow, while, on the other hand, the chemical potential varies with the potential energy.

\subsubsection{Uniform stress}
We shall focus now on the case of an uniform stress over a supersolid (Fig. \ref{steadyeg}-b). Looking for steady state solutions of  (\ref{eq:lagrangtimedeptotalmass},\ref{eq:lagrangtimedeptotalcauchy},\ref{eq:lagrangtimedeptotalBernoulli}) one notices that 

\begin{eqnarray}
  \varrho^{ss}{\bm v}^s  +   (\rho   -   \varrho^{ss}) \dot {\bm  u} & =&Ct
\label{uniformstress1}\\
 m (\rho   -   \varrho^{ss})(  \dot { u}_i -  v^s_i) v^s_i - \sigma_{ik} &=&Ct
\label{uniformstress2} 
\end{eqnarray}	
 Where the first relation (\ref{uniformstress1}) is the conservation of mass, and the second one (\ref{uniformstress2}) tells us about the conservation of momentum and relates in a quite interesting way the elastic stresses $ \sigma_{ik}$ with the superfluid velocity.  Because the symmetries one has that ${\bm v}^s =0 \quad \& \quad \dot {\bm u}=0.$ So that at the end $$\sigma_{ik}= Ct.$$

\subsubsection{Uniform strain}
Similarly, the case of an uniform strain (Fig. \ref{steadyeg}-c) presents a no superflow solution because  ${\bm v}^s =0 \quad \& \quad \dot {\bm u}=0$ by symmetry. The deformation is simply a uniform compression: $$u_x = \frac{\Delta\epsilon}{\ell} x.$$

\subsubsection{U-tube}
The U-tube experiment \cite{Castaing89,Bali06} (Fig. \ref{steadyeg}-d) combines all the previous cases in a single one: indeed, the left and right vertical channels under an imposed gravity while the horizontal tube in between is not submitted to any stress.

In the vertical tubes, the equilibrium state is given by (\ref{gravity1bis}) and (\ref{gravity2bis}), while $\epsilon_{ll} \neq 0$ in the horizontal tube and the equations of equilibrium read:
\begin{eqnarray}
\varrho^{ss}  {\bm v}^s + 
 (\rho-\varrho^{ss}) \dot {\bm u}  &=& 0
   \label{MassUtubechannel2}\\
 \hbar   \frac{\partial  }{\partial x_{k}} \left[(\rho- \varrho^{ss}) \left(   \dot {u}_i   - \frac{\hbar}{m} {\partial_i} \Phi  \right)\partial_k \Phi  \right]   + \frac{\partial}{\partial x_{k}}\left(\lambda_{iklm} \epsilon_{lm}\right) 
     &= &0
\label{ForceUtubechannel2}\\
  {\cal E} '(\rho)  = \mu & = & Ct.
\label{SuperUtubechannel2} 
\end{eqnarray}	
Remark that the chemical potential is constant in the horizontal channel. One can easily see that again
a pure ``elastic" solution with no superflow can be found by combining the solutions obtained above for the three different ``bricks" of the U-tube. We thus emphasize that 
the lack of superflow for a pressure or gravity driven U-tube experiment cannot prove the relevance of disorder to the observed supersolidity of solid Helium: indeed, no superflow can be induced in a good quality crystal by pressure gradient, with or without defects, at least at small stress.

\section{A possible analytical approach of the real dense supersolid}
\label{sec:dense}

As already noticed above in section \ref{intro}, the blocked annulus experiment~\cite{chan04b} represents a strong argument in favor of the existence of a supersolid state for Helium 4 at low enough temperature. Let us emphasize here that this experiment rules out explanation(s) of the observed Non-classical inertia (NCRI) by anomalies in the behavior of the torsion pendulum used to put in evidence this phenomenon.  It rules out also explanations relying on anomalies in the elastic behavior of solid Helium. In superfluid {\emph{liquid}} Helium 4 NCRI was observed long ago by Andronikashvili, but with major differences with super{\emph{solid}} Helium 4 : 

{\it{i)}} In the liquid, the density of the superfluid component is a function of the thermodynamic parameters: pressure and temperature only \footnote{This is true at low speed. At larger speeds, the velocity difference between normal and superfluid becomes an additional thermodynamic parameter.}. On the contrary in solid  Helium 4 the  {\it observed} density of the superfluid component depends on non-thermodynamical  parameters, that may be roughly categorized as ``crystal defects" because they change from one experiment to the other under the same nominal values of the thermodynamic parameters.  

{\it{ii)}} In the zero temperature limit the whole liquid becomes superfluid. On the contrary the superfluid component of solid Helium 4 remains relatively small at the lowest accessible temperature. 

{\it{iii)}} The density of the superfluid component, always quite small, tends to zero as the pressure increases in the crystal.

Each of the points made above requires explanation, which we shall attempt to give below. To set the stage, we shall outline first some basic ideas about the supersolid state, following the general schema initiated by Onsager \cite{po}.

\subsection{Phase coherence in a quantum crystal}
\label{phasecoherence}

We shall introduce first the idea that superfluidity in a crystal is related to the exchange by quantum tunneling of atoms from one lattice site to its neighbour. We show that tunneling, along with the quantum fluctuations of the phonons, is essential to ensure quantum coherence and so superfluidity, at least in the ground state. 
 
A standard result of quantum mechanics (a straighforward extension of a classical theorem by Liouville for eigenvalues and eigenfunctions of real symmetric linear differential operator) is that the  wave-function of the ground state of a system of Bosons has a constant phase in the absence of rotation. Therefore, it seems obvious that this system, whatever is its quantum ground state, liquid or crystal, has the property of off-diagonal long range order (ODLRO), understood as equivalent to infinite range correlations of the phase of the wavefunction. Nevertheless, things are not that simple because an argument presented below shows that  no such order shows-up  in a perfect quantum crystal (to be defined precisely) if one limits oneself to a particular entry of the density matrix. This raises several issues that we address. 

We define first what we mean by perfect quantum crystal, that is a crystal of identical atoms where each atom occupies a lattice site with small quantum fluctuations off this site. This smallness of the quantum fluctuations gives the idea of a small parameter, ultimately associated to the site-to-site exchange energy and the Debye theory is valid at leading order when this exchange energy is small. In a second stage, we consider the property of ODLRO and show that it shows-up for some components of the density matrix. 

The idea of studying dense quantum crystals by expansion in a small parameter goes back to a 1960 paper by Yang ~\cite{yang}. Yang looked at the ground state of an assembly of identical quantum hard spheres near close packing. In this limit, particles have, by definition, little free space. This increases their quantum kinetic energy which diverges at close packing with the power law $(\rho_c - \rho)^{-2}$, $\rho$ actual number density, less than but close to $\rho_c$, density at close packing. The inverse square law is explained by the property that the quantum kinetic energy is like the inverse square of a distance, and that this length is like the size of the available space for fluctuations of position of the hard spheres, a length scale tending to zero like $(\rho_c - \rho)$ as $\rho$ tends to $\rho_c$ by inferior values.  For arbitrary potentials one can try to extend the approach by Yang and to build up a rational theory of dense quantum solids, outside of a problematic numerical solution of the Schr\"odinger equation for a sizable number of particles\footnote{As well-known, it is {\emph{practically}} impossible to represent accurately in a computer the wave function of $N$ interacting particles as soon as $N$ is bigger than a rather small number. Assume that every particle needs, say $3Q$ points in its position space ($Q$ per coordinate), $N$ particles require $Q^{3N}$ points of collocation in the position space. Taking $Q =2$, which is not very large (this amounts to define a function by its value at two points only on the real line!), one gets a enormous number of collocation points as soon as $N$ is larger than a few units: for instance $ N = 8$ and $Q =2$ yields $2^{24} = 1.68\times  10^{7}$, meaning that the value of the wavefunction should be known at this number of points, even though the  the wave function is still quite poorly known (two points per coordinate) and the number of particles is not terribly large (eight). This is the well-known problem of computing by brute force properties of quantum systems with more than a few interacting particles. In this respect there is a significant difference between the computation of properties like the ground-state energy, never very wrong because of the Rayleigh minimum principle behind it, although it is much harder to get good representation of functions, like pair correlation for instance, which carry far more information than a number, the ground-state energy, without being related to a variational principle.} 
 
Given the {\emph{practical}} closure of direct numerical access to the properties of many-body quantum system with interactions, one is left with an analytical approach relying on the existence of a small parameter, following the line of thinking started by Yang. In the (generic) case of a smooth potential of interaction, this requires first to find such a small relevant dimensionless number, other than the density difference $(\rho_c - \rho)$ of Yang and to expand the relevant quantities in the limit where it is small.  The zeroth-order theory in this approach is the Debye theory of quantum lattices. The quantum Debye theory of solids accounts well for the thermodynamics of crystals, at low temperature at least. One expects it to become invalid if the thermal and/or the quantum fluctuations become large enough to make the interaction between phonons non negligible. 
In Debye's theory the zero-point fluctuations are absent. This is equivalent to assume that the width of the wave packet of each atom at its site is much smaller than the period of the lattice. As well-known such fluctuations preclude the existence of  infinite range positional order in 1D. In higher dimensions the quantum fluctuations do not destroy the long range order, but make possible the overlap of the wave functions of neighbors on the lattice. This leads to a coherent ground-state wave-function extended all over the lattice that is ultimately responsible of ODLRO. Let us put those remarks in an analytic framework. 

The small parameter we are going to introduce is the ratio of the width of the quantum wave packet near each site to the lattice size. This small parameter (the $(-1/4)$ power of the quantity $\Lambda$ introduced below) tends to zero as the number density $\rho$  increases. As shown below this makes the classical (that is, {\bf non-quantum}) theory more and more exact in this high density limit, and explains that the superfluid density tends to zero when the  density of the supersolid increases (something hitherto unexplained to the best of our knowledge).    
When $1/\Lambda$  is small one can expand the energy of the ground state explicitly (see equation (\ref{eq:expenergy}) below). We assume a smooth  (except at zero inter-particle distance) two-body potential. Let us estimate the width of the wave-packet of an atom near one lattice site in the field of its neighbors. We do it first for a two-body potential depending with a power law from the distance, like $V(r) = \frac{A}{r^s}$, $A$ and $s$ are positive, the crystal being held by outside pressure. This potential $V(r)$ needs to be computed at distances shorter than usual in physical applications, typically  fraction of the radius at the minimum of a Lennard-Jones potential. Physically, the Lennard-Jones potential increases too much at short distance (like $1/r^{12}$) : there the dominant interaction is the Coulombian repulsion between the nuclei, far less singular than $1/r^{12}$. This could affect significantly estimates of the supersolid density relying on tunneling effects (see \ref{subsec:tunnelingam}). The present approach does not work for the hard-core interaction considered by Yang~\cite{yang}.  Lastly, the true soft-core interaction at short distance could explain why the observed transition temperature to the supersolid state does not change much with pressure.  


According to the rules of quantum mechanics the position of each atom fluctuates near the minimum of the classical potential at each lattice node. This neglects correlations between fluctuations at different lattice sites, a kind of correlation taken into account in Debye's theory through a potential energy term involving cross products of displacements at neighboring sites. This does not change fundamentally (at least in space dimensions higher than 1) the order of magnitude estimates. At a given lattice site the potential energy due to the neighbors has a minimum and varies near this minimum like $V_{site} \approx \frac{s^2}{2}\frac{A}{a^s}  \frac{q^2}{a^2} $, where $a$ is the lattice spacing and $q$ the magnitude of the displacement near the minimum. This estimate should include a numerical coefficient of order one depending on the structure of the lattice. The first term in this expansion is quadratic in $q$, the departure from the classical equilibrium position. 

Balancing the potential energy $V_{site}(q)$  and the quantum kinetic energy that scales like $\frac{\hbar^2}{2 m q^2}$, $m$ mass of the He-4 atom, one finds  that the order of magnitude of the zero-point  fluctuations of $q$ is $ q_{Q} \sim  a\Lambda^{-1/4}  \mathrm{,}$ where $\Lambda$, a kind of dimensionless de Boer parameter, is such that 
$\Lambda = \frac{m A}{\hbar^2 a^{s-2}}$. In the high density limit and if the potential is steeper than quadratic ($s>2$, as we shall assume), $q_{Q}$ is much smaller than $a$, the limit we shall consider here. This is relevant for most, if not all, real solids, including solid Helium.  The dimensionless quantity $\Lambda_E \sim \frac{m a^4 }{\hbar^2} V''_{site}(r=a)$ is a natural extension of the parameter $\Lambda$ for an arbitrary two-body potential. The second derivative $V''_{site}(r=a)  $ and Young's modulus $E$ of the crystal are linked in such a way that  $V''_{site}(r=a)  \sim \frac{E a}{\rho a^3}$. Finally, the parameter 
$\Lambda_E =  \frac{m a^2 }{\hbar^2} \frac{E}{\rho} $, proportional to the quantity denoted as $\Lambda$ for a power law potential, measures the relative magnitude of the quantum effects, independently on any assumption on the details of the crystal (two or more bodies interaction, etc.). Although $\sqrt{\Lambda_E}$ is the smallest, about  $7.4$, for Helium 4, it is still a not too large  for  solid Hydrogen ($12.2$) and for solid Neon ($14.9$). 

We shall continue our developments for the case of a power law dependent interaction. The parameter $\Lambda$ can be used to scale the various contributions to the energy of a dense $N$-particle system. Let  $ \psi({\bm {r}}_1,{\bm{r}}_2,\dots, {\bm{r}}_N) $ be the symmetric $N$-particle wave-function. 
Let us scale ${\bm r}_k$ as $a \tilde{\bm r}_k$ (tilde being for dimensionless quantities of order 1, $k$ particle index), the potential energy as $V(|{\bm r}|) =  V(a) \tilde V(|\tilde{\bm r}|) $ and introduce  the wavefunction $ \psi({\bm r}_1,{\bm r}_2,\dots, {\bm r}_N)  = a^{3N/2}\tilde  \psi(  \tilde {\bm r}_1,  \tilde{\bm r}_2,\dots, \tilde {\bm r}_N)$. After the tildes are dropped  the energy  reads: 
$$E =  \frac{\hbar^2}{ ma^2}  \int \left( \frac{1}{2}  \sum_{i=1}^N{|{\bm \nabla}_i \psi |^2}   + \Lambda \sum_{i<k}^N V(|{\bm r}_i- {\bm r}_k|) | \psi |^2  \right) d^{3N}{\bm r},  $$ where
 $d^{3N}{\bm r} = d{\bm r}_1 d{\bm r}_2 \dots  d{\bm r}_N$. 
 
In the large $\Lambda$ limit the leading contribution to be minimized in the ground state is the potential energy. If one assumes one particle per site, this minimization is exactly equivalent to the one for the {\it classical} ground state, something done rather easily if one assumes a simple crystal structure and short range forces.
The quantum and classical problems differ because one does not have to put exactly one particle in each site in the quantum case. The wave-function of the full system is a product of normalized one particle wave-functions at a prescribed subset of lattice sites among the $N_{\mathrm{s}}$ possible ones. Assuming a proportion $p$ of occupied sites, one adds all contributions obtained in this way for all possible combinations of $p N_{\mathrm{s}}$ sites. Lastly the wave-function is symetrized under exchange. The density of potential energy to be minimized is $ \frac{p^2}{2} V(a )$, the lattice size $a$ being such that $p = \rho a^3$, $\rho $ imposed number density. The minimization of the energy is done by changing $p$ in the interval  $0 < p \leq 1$, $\rho $ fixed. Decreasing $p$ at constant $\rho$ obliges to put the particles closer to each other. For a realistic $V(r)$ this increases the potential energy if the potential increases rapidly at short distance.  Therefore $ p =1$ is the optimal choice.  This defines what we mean by ``perfect quantum crystal": $\Lambda$ should be large and so the optimal configuration is for one atom per lattice site\footnote{Note that in our ``perfect crystal" there is one atom per lattice site in the ground state, but that is true {\it only} in the limit $\Lambda$ infinite. Otherwise the assumption  ``one atom per lattice site" is ambiguous because of quantum effects, see below.}.  Notice also that the expansion of any quantity, like the energy for instance, in inverse powers of $\Lambda$ is insensitive to the quantum statistics (Bose or Fermi) of the atoms. It is at transcendentally small order only that quantum statistics become relevant. It is also at an order which is non-algebraic with respect to $1/\Lambda$ that non-classical rotational inertia shows up \cite{ss1}. 

 Let us sketch the principles of an expansion of the ground-state energy in the limit $\Lambda$ large. As just said the leading order term is the classical potential energy. At the next order, one has to take into account the small amplitude fluctuations of the atoms near their equilibrium position. Since the fluctuations are small one can expand the potential energy to second order only in the excursion of each atom from its classical equilibrium position. Adding now the quantum kinetic energy, one gets a problem of coupled linear oscillators.
The diagonalization of the energy operator can be done explicitly by Fourier transform and the final result is that the first quantum correction to the energy of the ground state is the zero-point energy of the phonons, a quantity of relative order $\Lambda^{-1/2}$, therefore the beginning of the expansion of 
energy per particle reads  \cite{bender}: 
\begin{equation}
E_0 = \Lambda  \hbar^2/ma^2 \left( U(\rho) + \Lambda^{-1/2} E_{0,ph} +  \dots \right) 
\mathrm{.}
\label{eq:expenergy}
\end{equation}
The leading order term $U (\rho) $ is the energy in the classical limit, following a law of corresponding states,  and $E_{0,ph}$ is the zero-point energy of the phonons, etc. The next order term takes into account the interaction between the phonons. This expansion goes on to infinity with integer powers of  $\Lambda^{-1/2}$, although logarithms are likely to occur as always in expansions generated by quantum theory of interacting fields. However, some physical effects are missing in this expansion, most notably the site-to-site tunneling which yields a contribution to $E_0$ transcendentally small with respect to the expansion parameter as shown in the next  subsection. 

\subsection{Order of magnitude of the tunneling amplitude in the limit $\Lambda$ large}
\label{subsec:tunnelingam}

As shown in section \ref{macro}, the ``superfluid'' density $\varrho^{ss}$ is defined through the
superfluid fraction of a rotating quantum solid~\cite{leggett}. This quantity is thus formally obtained through the
computation of the variation of the ground state of a system of $N$ particles under a small non-uniform
change of phase~\cite{onsager}. 

Finally we shall provide a relation between the ground-state wave function of the solid and $\varrho^{ss}$ in this large $\Lambda$ limit. This follows from a chain of arguments already used by Onsager in his derivation of the quantization of circulation in quantum fluids \cite{onsager}.
Consider a real and positive ground state wavefunction: $\psi_0({\bm r}_1,{\bm r}_2,\dots, {\bm r}_N) $ with energy $E_0$. We shall deal with steady states, so that the energy can be set to zero and one may deal with time independent problems only. 
  
Then one computes the change of kinetic energy by a coherent motion under a small variation of the ground state because of a non-uniform change of phase:

 \begin{equation} \psi({\bm r}_1,{\bm r}_2,\dots, {\bm r}_N,t) = e^{i( \phi({\bm r}_1) + \phi({\bm r}_2) + \dots + \phi({\bm r}_N))} \psi_0({\bm r}_1,{\bm r}_2,\dots, {\bm r}_N)
  \label{GSWavefunction}
       \end{equation} 
 where the phase $\phi({\bm{r}}_j)$ is taken the same for all particles. Notice that the specific form of the phase of the wave-function (\ref{GSWavefunction}) assumes {\it ab-initio} phase coherence, thus superfluidity. Introducing this Ansatz into the energy (\ref{energy}) one gets:

 \begin{equation}
 \delta  {\mathcal E } = E - E_0=  \frac{\hbar^2}{2 m}   \int \rho_0({\bm r}) ({\bm \nabla}\phi)^2 {\mathrm{d}}{\bm r}
  \label{deltaEnergy}
       \end{equation} 
where $\rho_0({\bm r})$ is the diagonal part of the one particle ground state density matrix $$\rho_0({\bm r}) \equiv N R_1({\bm r}, {\bm r}) $$
with 

\begin{equation}
 R_1({\bm{r}}_1, {\bm{r}}_{1}') = \int {\mathrm{d}}{\bm{r}}_2 {\mathrm{d}}{\bm{r}}_3...{\mathrm{d}}{\bm{r}}_N \psi({\bm{r}}_1, {\bm {r}}_2, ...{\bm{r}}_N) \overline{\psi}({\bm{r}}_{1}', {\bm{r}}_2, \dots {\bm{r}}_N)
 \label{eq:defR1}
 \mathrm{,}
 \end{equation}
 
where the overline is for the complex conjugation. Notice that 
the number density $\rho_0({\bm{r}}) > 0$ is the true density in the system, a non constant periodic function of ${\bm{r}}$ if the ground state is a crystal.  Because the ground state wave function $ \psi_0$ does not vanish then $\rho_0({\bm{r}})$ does not, something true for the ground state of Bosons by a standard result of Sturm-Liouville theory. 

According  to Ref.~\cite{ss1} the calculation 
of the NCRI amounts to estimate the average phase gradient making the smallest the change of energy:
$ \delta{\mathcal{E}} =  \frac{\hbar^2}{2 m}   \int \rho_0({\bm r}) ({\bm \nabla}\phi)^2 {\mathrm{d}}{\bm r},$
where ${\bm \nabla}\phi$ is the gradient of the phase and $\rho_0({\bm r})$ is the density distribution in the solid obtained by taking the diagonal part of the one-particle density matrix of the 
ground state. We sketch below the proof that, whenever there are wide variations of $\rho_0({\bm r})$ in the lattice cell (namely when the ratio $\frac{Max [\rho_0({\bm r})]}{Min [\rho_0({\bm r})]}$ is large), the equivalent superfluid density is the number density at the saddle point in the lattice cell.

The computation of  $\delta  {\mathcal{E}}$ is done under the condition that the phase $\phi$ changes at large scale at a constant rate, and that, given this condition, and a periodic $\rho_0({\bm r})$, the energy  $\delta  {\mathcal{E}}$ is minimum. For the writing of this energy, it is obvious that all the variation of $\phi$ must be concentrated in places where $\rho_0({\bm r})$ is mimimum. This is even more constraining when $\rho_0({\bm r})$ changes widely within a lattice cell. On the other hand, $\phi$, when it changes from one cell to the next must extrapolate between zones of almost constant value to another zone of constant value in the next cell, the variation being concentrated in the transition between the two zones of constant phase. Moreover, the contribution to $\delta  {\mathcal{E}}$ of this transition domain is kept to the minimum by making the transition in the saddle between the two domains where the phase is constant because $\rho_0({\bm r})$ is much larger than at the saddle point. According to those considerations, the dominant contribution to $\delta  {\mathcal{E}}$ comes from values of $\rho_0({\bm r})$ near saddle points. This can be made more quantitative by considering the local expansion of the Euler-Lagrange equation for the optimal phase distribution with a constant gradient at large scales.  We shall consider now the specific case where the phase gradient is linked to a global rotation. 

Under an uniform rotation the kinetic part of the energy becomes proportional to the square of  the angular speed of the imposed rotational motion  with angular velocity ${\bm{\omega}}$. The goal of the calculation is to find the coefficient of ${\bm{\omega}}^2$ in the expansion of the full energy near ${\bm{\omega}}=0$, this coefficient being the observed momentum of rotational inertia. Formally this is a standard problem of perturbation. It has been shown that the NCRI can be reduced, for a crystal-like ground state, to the calculation of a homogeneized response function. Let $\rho_0({\bm{r}})$ be the space dependent coherent density distribution in the solid. For a perfect crystal, this is a periodic and non-constant function of $\bm{r}$. The rotation of this crystal induces boundary conditions for the slowly varying part of the phase that change the energy.



 Using similar bounds than the one obtained by Leggett~\cite{leggett,bounds} one can  show that for the given {\it Ansatz}  used above the ``supersolid'' density satisfy the inequalities 
\begin{equation}
 \frac{1}{V} \int_V  {\rm min}_{x}\,  \rho_0(x,{\bm \zeta} ) \,  d {\bm r}  \leq \varrho^{ss}_{xx} \leq  \left( \frac{1}{V} \int_V  \frac{1}{ {\rm max}_{\bm \zeta} \, \rho_0(x,{\bm \zeta} ) } \, d {\bm r}  \right)^{-1 }
\label{bounds}\end{equation}
where $x$ is the direction of the boost and ${\bm \zeta}$ stands for the transversal coordinates. 
Although these bounds are less narrow than those found by Leggett \cite{leggett,bounds} both may be estimated in the large $\Lambda$ limit via steepest descent. Both sides are bounded by the value at the saddle point of $\rho_0({\bm r})$ times a constant.
Thus a non-zero $\varrho^{ss}$ is fundamentally linked  the  coherence which comes from the small overlap between the quantum fluctuations at different sites.

According to the considerations developed above, in the large $\Lambda$ limit one expects the density to concentrate in narrow domains near the minima of the potential. This yields large density contrasts in the unit cell, because most of the cell is filled with low to very low density wave-function.  Therefore, as we just explained, the integrals in equation  (\ref{bounds}) is dominated by the value of $\rho_0({\bm{r}})$ at saddle point values in the unit cell. This estimate could be sharpened by considering various rigorous bounds of the minimization problem \cite{bounds,mamandine}.This value of the density at the saddle in the unit cell is given by the low density part of the wave-function, off the minimum of the potential, related itself to the quasi-classical approximation in the Euclidean case, because this concerns classically forbidden region of the trajectories. The general result of this kind of theory is the estimate
 
$$ \frac{\varrho^{ss}}{\rho}  \sim  e^{- 2 \sqrt{ \Lambda }  \frac{S_0}{\hbar}} $$
where $S_0$, a Euclidean action ( ie. the action derived from the classical action by
changing the kinetic energy from $p^2/2m$ to $-p^2/2m$ to reach  quantum
corrections transcendentally small with respect to $\hbar$), depends on the lattice and of the interactions.  Notice that the significant quantity in the exponential of the amplitude of quantum tunneling is an {\emph{action}}, not an {\emph{energy}}, contrary to the case of thermally assisted tunneling.

The action  $S_0$ is found by solving the $N$-particle Schr\"odinger equation in the WKB limit for a well-defined process of exchange of atoms. We shall show first that the saddle point of the density is on a heteroclinic trajectory joining two equilibria (unstable in the Euclidean dynamics). In this WKB limit the wave-function  $\psi ({\bm{r}}_1, \dots,  {\bm{r}}_N)\sim e^{-\sqrt{\Lambda} S^{(N)}_0({\bm{r}}_1, \dots,  {\bm{r}}_N)/\hbar}$, where $S^{(N)}_0({\bm{r}}_1, \dots,  {\bm{r}}_N)$ is the classical action associated to a trajectory leaving a point of equilibrium to reach the point where the action is computed. The long time needed for this exchanges yields a cut-off for the frequencies such that the wave-function can be considered as globally coherent. 

If one takes an arbitrary point in the position space, many trajectories contribute to the amplitude of the wave-function there. 
Along each Euclidean trajectory the action increases with time and the contribution to the density decreases with the distance from the equilibrium point. Therefore a general trajectory does not contribute to the saddle point  density. The dominant contribution is reached when the trajectory is a heteroclinic connection between two equilibria exchanging a certain number of atoms between different sites. This
corresponds to a stationary action and the contributions add to each other to yield a saddle point of the sum of the two sides at the mid-trajectory. This saddle point is precisely what we are looking after. 
The quantity we need to know is the one particle density matrix as a function of the position. This is found by squaring the sum of the exponentials and integrating the result over all the ${\bm r}_2\dots{\bm r}_N$ positions.  We consider below a slightly more detailed calculation of   $S_0$ in a particular case.
 
\subsubsection{ A more detailed calculation in a special case.}   

We consider the action associated to an exchange trajectory in the case of a potential $V(r) = \frac{A}{r^s}$. We shall restrict ourselves to a slightly simpler case, the one of two atoms exchanging their equilibrium position at neighboring sites, all the other atoms being assumed to be classical points at rest during this exchange, all sites occupied. Let the atoms carry indices $1$ and $2$ at positions ${\bm{r}}_1$ and ${\bm{r}}_2$, the site themselves have index $\alpha$ and $\beta$. The two particles have a large repulsive interaction potential $ \Lambda V({\bm{r}}_1 - {\bm{r}}_2)$. The interaction with the rest of the lattice is represented as follows: near two neighboring sites $\alpha$ and $\beta$ the potential seen by particle $1$ and $2$ is a function  $W({\bm{r}}) = \Lambda  \Sigma_{\gamma \neq \alpha \mathrm{,}\beta} V({\bm{r}}  - {\bm{r}}_{\gamma}) \mathrm{,}$, with all the  ${\bm{r}}_{\gamma}$'s fixed for $\gamma \neq (\alpha, \beta)$ in a first approximation, with two sharp minima at ${\bm{r}} = {\bm{r}} _{\alpha}$ and  ${\bm{r}}= {\bm{r}} _{\beta}$. This neglects possible displacements of the particles other than the one located at ${\bm{r}} _{\alpha}$ and  ${\bm{r}} _{\beta}$. 

The wave-function  for this problem is a solution of the dimensionless eigen-equation (the energy is $E = \frac{\hbar^2}{ma^2}  \varepsilon$):
\begin{equation} 
\varepsilon \psi({\bm{r}}_1,  {\bm{r}}_2) =   -  \frac{1}{2} \left[{{\nabla}}^2_1 + {{\nabla}}^2_2 \right]\psi + \Lambda V({\bm{r}}_1 - {\bm{r}}_2) \psi + \Lambda( W({\bm{r}}_1) + W({\bm{r}}_2))\psi 
\mathrm{.}
\label{eq:shrod}
\end{equation}

In the WKB limit, the amplitude of the wave-function under the barrier, that is the tunneling region, is proportional to $\psi ({\bm{r}}_1 \mathrm{,} {\bm{r}}_2)\sim e^{-\sqrt{\Lambda} S^{(2)}_0({\bm{r}}_1 \mathrm{,} {\bm{r}}_2)}$, where $S^{(2)}_0({\bm{r}}_1 \mathrm{,} {\bm{r}}_2)$ is the dimensionless classical action associated to the Euclidean dynamical problem with the original kinetic energy $-\frac{\nabla^2}{2m} $  replaced by $\frac{\nabla^2}{2m}$. In this Euclidean dynamics  the potential wells of the Lagrangian dynamics are replaced by maxima of the external potential. The tunnel factor is found by imposing the trajectory  to start with particle $1$ in well $\alpha$ and particle $2$ in well $\beta$ and to end up with $2$ at $\alpha$ and $1$ at $\beta$ \footnote{The potential $ W({\bm{r}})$ is considered as fixed in a first approximation, the atoms others than $1$ and $2$ staying at rest whilst the atoms 1 and 2 are doing their tunneling trajectories. This is a first order approach to the tunneling problem, since one expects that the other particles do move less and less during the tunneling of $1$ and $2$ when they are farther and farther away. Including this motion of the other particles does not bring any fundamental difficulty, it just makes things more cumbersome. The other particles should follow each a homoclinic orbit by starting from  and coming back to the same site whilst the pair $(1,2)$ does an exchange trajectory. The tunneling region is everywhere outside of the bottom of the potential wells.}. In space dimensions higher than 1, such an Euclidean trajectory  exists with particles ``turning around'' each other if their interaction potential is repulsive (attractive in the Euclidean dynamics). There is an Euclidean trajectory joining  states where the two particles interchange their equilibrium positions \footnote{This is shown as follows: if there is no interaction between the two particles they run independent straight trajectories moving from one equilibrium to the other and crossing in the middle. Once the interaction $V({\bm{r}}_1 - {\bm{r}}_2)$ is turned on, one considers all trajectories leaving simultaneously the two equilibria. They are indexed by an angle, if this angle is zero or close to zero, the two particles will meet in their mid-course ({\it i.e.} near the saddle of the external potential) and fall on each other because of the strong attraction at short distance.  If on the contrary the angle is large, each particle will fall down to infinite depth of the potential well $(-\Lambda W)$ without interacting with the other one. In-between, there should be a value of the angle such that the two particles make it to the other equilibrium point.} and in practice the pair  $({\bm{r}}_1 \mathrm{,} \alpha  \mathrm{,} {\bm{r}}_2 \mathrm{,}  \beta)$  define the initial condition for the Euclidean trajectory. Fix now a value of $({\bm{r}}_1 \mathrm{,} {\bm{r}}_2)$, the action is $S^{(2)}_0({\bm{r}}_1 \mathrm{,} {\bm{r}}_2)= \int \left({\bm{\nabla}_1 S^{(2)}_0 }\cdot {\mathrm{d}}{\bm{r}}_1 + {\bm{\nabla}_2S^{(2)}_0 }\cdot {\mathrm{d}}{\bm{r}}_2\right) = \int \left({\bm{p}_1}\cdot {\mathrm{d}}{\bm{r}}_1 + {\bm{p}_2}\cdot {\mathrm{d}}{\bm{r}}_2\right)$ where the integral is  parametrized by the Euclidean time, although ${\bm{p_i}}$ is the momentum associated to the position ${\bm{r}}_i$. 
Because there are two starting points for the Euclidean trajectory, one has to add the two contributions, the overlap being negligible \footnote{The function $\psi ({\bm{r}}_1 \mathrm{,} {\bm{r}}_2)$ yields in a consistent way the full wave-function of the ground state. One might wonder if, in the full crystal, one should not take also into account the tunneling effects between three particles in neighboring sites, etc. Although exchange of many particles yields certainly  larger actions (and small contributions to the density) exchanges between few particles (larger than 2) may contribute significantly to $\varrho^{ss}$. We plan to investigate this.}.

Let us sketch the calculation of $S_0^{(2)}$ for a triangular lattice in 2D with a repulsive interaction $V(r) =1/ r^{s}$ in the limit $s$ large, quite realistic for a Lennard-Jones interaction increasing like an inverse twelfth power. The lattice size is taken as $1$ because every other dimensional quantity is absorbed into $\Lambda$ \footnote{Note that with the standard Lennard-Jones parameters of Helium ,  $\sigma$ and $\epsilon$, one has $\Lambda= \frac{4 m a^2 \epsilon}{\hbar^2} \left(\sigma/a\right)^{12}\approx 0.6$. The two body $V(r)$ is needed at shorter distances than usual in physical applications, a fraction of the minimum of a Lennard-Jones potential. The $1/r^{12}$ repulsion is too strong there because at close distances the dominant interaction is the Coulombian repulsion between the nuclei, far less singular than $1/r^{12}$. This could affect significantly the estimates of the tunneling contribution to the superfluid density. }. 
Therefore during the exchange only the interaction between the two exchanging particles and with the nearest neighbors is relevant. The trajectories must have finite curvature, which imposes that the superposition of all very large forces involved must be such that there is no force normal to the trajectories, which would induce a very large curvature. This yields a geometrical equation for the trajectory. For a triangular lattice one gets:
\begin{equation}
r(\theta) =  \left(\sqrt{3 + \sin^2\theta} - \sin\theta\right)/(2 \sqrt{3} ) \mathrm{,}
 \label{eq:traject}
 \end{equation}
where $r(\theta)$ is the radial equation with the center at the mid-point between the two equilibria the particles start from, those equilibria being located at $ r =1/2$ and $\theta = 0$ and $\pi$. The equation (\ref{eq:traject}) is valid for $0<\theta<\pi$.   
Using this knowledge of the geometrical shape of the trajectory, one finds the analytical expression for the dimensionless action $S_0^{(2)} = \frac{ \sqrt{\pi}  3 ^{s/4} }{\sqrt{s/2-1}}$ at the saddle-point.

\subsection{Off-Diagonal long range order or not in a perfect crystal?}
\label{subsec:ODLRO}
 
Having shown in the previous subsection the general existence of a non-zero exchange between nearest-neighbours sites, we consider now the existence or not of ODLRO in the large $\Lambda$ limit for a perfect quantum crystal. 

Let us consider first with the one-body density matrix and outline the arguments used to ``prove" sometime the lack of ODLRO in a ``perfect crystal". Let $\psi({\bm{r}}_1, {\bm {r}}_2, ...{\bm{r}}_N)$ be the full ground state wave function of the N atoms in the lattice with one atom per site. The one-body density matrix is given (\ref{eq:defR1}). The formal existence of ODLRO may be understood by saying that, when $|{\bm{r}}_1 -{\bm{r}}_{1}'|$ tends to infinity, $R_1({\bm {r}}_1, {\bm {r}}_{1}') $ tends to the product $f({\bm{r}}_1) \overline{f}({\bm{r}}_1')$ with a function $f \ne 0$. This amounts to have that, fixing the position ${\bm{r}}_1$, the phase of $R_1({\bm {r}}_1, {\bm {r}}_{1}') $ does not vary much with ${\bm{r}}_1'$ at whatever distance from ${\bm{r}}_1$. Such ODLRO exists for Bose-Einstein condensate, $f({\bm{r}}_1)$ being then the mean value of the creation (/annihilation) operator at ${\bm{r}}_1$. This property of ODLRO, if it is satisfied, is a statistical property, and its connection with macroscopic properties like superfluidity is not clear at all: remember for instance that an ideal Bose-Einstein condensate (without interaction) is {\it not} a superfluid. In this sense, it seems more relevant to deal directly with the phase itself, as we shall do further on, and particularly with the relationship between this phase and the number of particles~\cite{josephson,anderson}. 

Coming back to 
the one body density matrix $R_1({\bm {r}}_1, {\bm {r}}_{1}') $, one can argue that it does not show ODLRO ({\it i.e.} long range correlations between primed and unprimed positions) in a crystal with one atom per site because ${\bm{r}}_1$ and ${\bm{r}}_{1}'$ have to lay in the same vacant site of the lattice left once the other sites are occupied by particle number 2, 3, etc so that $R_1({\bm {r}}_1, {\bm {r}}_{1}') $ tends to zero as $|{\bm{r}}_1 -{\bm{r}}_{1}'|$ tends to infinity. Recent claims~\cite{ProSvi05} on the lack of ODLRO in lattices with exactly one atom per site rely on the property of the one-body density matrix we just showed. 
 
There is an obvious weakness in this argument because it considers only one specific entry of the density matrix (the one body), and so cannot exclude that other elements of this matrix show ODLRO, what we are going to show. Moreover it considers that the lattice as physically immobile, something incorrect, because it neglects the zero-point fluctuations of the phonons. Let us consider first another claim of reference \cite{ProSvi05} according to which the phase of a quantum crystal is not well defined because it fluctuates without bound.  Specifically this claim is based upon the assumption that, in such a perfect lattice, there is no quantum fluctuation of the number of particles. The
 relation~\cite{josephson,anderson} $$\delta n \; \delta \Phi = 1$$
 between the quantum fluctuations of the number of particles, $\delta n$, and of the phase $\delta\Phi$ shows that the phase fluctuations $\delta\Phi$ are unbounded if $\delta n =0$, implying that there is no ODLRO. This argument seemingly confirm that a perfect quantum crystal cannot be a supersolid but it does not apply to real crystals. Actually it omits two types of quantum fluctuations of $n$, some due to the phonons and others to site-to-site tunneling. Even in a quantum crystal with one atom per site, the zero-point fluctuations of the phonons, always present,  induce quantum fluctuations of the number of particles. In 1D this kind of fluctuation is so strong  that it forbids long range {\it positional} order at zero temperature. The only meaningful fluctuation in the number of atoms is found by considering what happens in a {\it fixed} box of volume of order $L^3$ with a size of order $L \gg a$,  $a$ lattice size. This size $L$ must be also much less than the external dimension of the crystal, since otherwise the density inside this full crystal does not fluctuate at all. This (obvious) remark is not without consequences: practically, to be sensitive to  those crucial fluctuations, a numerical simulation of whatever kind must be for a system big enough to show finite quantum fluctuations in its subpart. Practically, not only $N$, number of atoms, must be large but also $N^{1/3}$ (see below for the exponent $1/3$), which is, practically, much harder to reach.

The number of particles  $n$ laying in a finite box drawn on an infinite lattice fluctuates for two reasons: first because of the zero-point motion of the phonons, then, because once tunneling between sites exists (and it does, however small it is), there is always an indeterminacy whether a site on the border of the domain is filled or not because particles there have a small quantum probability of being on one side of the border or on the other. Consider first the contributions of the phonons to $\delta n$, then the one of tunneling. 
 
 \subsubsection{ Number fluctuations due to the phonons.} 
 
 Phonons changing  $n$ in the box of size $L$ have a wavelength of order $L$. The effect of the zero-point fluctuations on $n$ is found by balancing the zero-point energy of a phonon of wavelength $L$ with the fluctuation of elastic energy due to a variation $\delta n$ of $n$ in the box of size $L$ (note that the box should be of irregular shape, not a cube or a sphere,  where the fluctuations in the number of particles depend on the precise way the fixed box is placed with respect to the lattice). 
 
This yields 
 \begin{equation}
\frac{1}{2}  \hbar c_s \frac{2\pi}{L} \sim E \int_{L^3} {\mathrm{d}}^3{\bm{r}} \left(\frac{\partial u}{\partial r}\right)^2
  \label{eq:defR3}
 \mathrm{.}
 \end{equation}
In this equation, the left-hand side is the zero-point energy of the phonon of wavelength $L$ with $c_s$ speed of sound in the solid. The right-hand side gives the order of magnitude of the elastic energy of this phonon, with $E$ Young's modulus, $\left(\frac{\partial u}{\partial r}\right)$ strain in the solid, $u$ being the displacement field associated to the phonon. The strain $\left(\frac{\partial u}{\partial r}\right)$ induces a variation of $n$ of order $\delta n = \rho   \frac{\partial u}{\partial r} \, L^3$, $\rho_0$ mean number density. Once put into the right-hand side of the equation (\ref{eq:defR3}) this yields the order of magnitude of the mean-square value of zero-point fluctuations of $n$:
 \begin{equation}
 \left(\delta n\right)^2  \sim L^2 \frac{\hbar c_s \rho^2}{E } \sim \frac{1}{\sqrt{\Lambda_E} }\rho a L^2 \sim  \frac{n }{\sqrt{\Lambda_E} }  \frac{a}{L}
  \label{eq:defR4}
 \mathrm{.}
 \end{equation}
 Where $\Lambda_E = \frac{ma^2}{\hbar^2} \frac{E}{\rho}$ was defined previously.
Notice that, in 3D,  the fluctuation of $n$ inside the box are of order $N^{1/3} \sim L$. The equation (\ref{eq:defR4}) shows our point, namely that the zero-point fluctuations of the number density do {\it not} vanish in the perfect quantum crystal. Because they diverge as $L$ tends to infinity, the phase fluctuates less and less in this limit.  Moreover the scaling $ \left(\delta n\right)^2  \sim L^2$ yields  $ \left(\delta \varphi \right)^2  \sim L^{-2}$, fully consistent with the scaling derived by balancing an energy proportional to $\frac{\varrho^{ss}}{2} \int_{L^3} {\mathrm{d}}^3{\bm{r}} \left(\nabla\varphi\right)^2 \sim L  \left(\nabla\varphi\right)^2 $ with the left-hand side of equation (\ref{eq:defR3}) that is of order $1/L$ at large $L$ \footnote{This  kind of argument can be extended to show a fluctuation theorem relating the coefficients of response of the supersolid, including the superfluid density matrix $\varrho^{ss}_{ik}$ to fluctuations in the ground state, via the formal quantization of the macroscopic equations of motion. This can be done by using the fact that the phase is the conjugate of the number of particles, giving it a well defined meaning. To define also precisely what is meant by the lattice fluctuations, one has   define what is meant by lattice ordering in the ground-state, which can be done by using a method similar to the one leading to the supersolid equation in the mean field limit, as done in section \ref{Approach1}.
The starting point of the derivation of the quantum fluctuation theorem is in simple algebraic relations for the quantum harmonic oscillator. Let $\mathcal{H} = a \frac{p^2}{2} + b \frac{q^2}{2}$ be the energy operator of this oscillator, with $a$ and $b$ positive parameters and $p$ and $q$ non-commuting operators such that $[p,q] = i\hbar$. Let furthermore $\left<G\right>_0$ be the mean value of any observable $G(p,q)$ in the ground-state. The eigenvalue of $\mathcal{H}$ in the ground-state is $e_0 = \frac{\hbar (ab)^{1/2}}{2}$. As well-known there is equipartition of energy in this state, so that $\left<a \frac{p^2}{2}\right>_0  = \left<b \frac{q^2}{2}\right>_0 =\frac{e_0}{2}$, a relation allowing to compute $a$ and $b$ as functions of $\left<p^2\right>_0$ and $\left<q^2\right>_0$, which yields a kind of fluctuation theorem for the harmonic oscillator. To apply this to find the response functions of the supersolid (i.e. the Lam\'e coefficients and $\varrho^{ss}_{ik}$), one decomposes the possible fluctuations of the macroscopic equations for the supersolid (see section \ref{Approach1}) in normal modes. This yields a set of harmonic oscillators, four for each wavenumber (four = three elastic modes + one phase mode, all being linearly coupled), a set to which one can apply the fluctuation ``theorem" just derived. The final result is rather complex, because of the coupling effects between the different modes.} . Extending equation (\ref{eq:defR4}) to an arbitrary space dimension $d$ yields $ \left(\delta n\right)^2 \sim L^{d-1}$. In 1D the zero exponent is for a logarithmic dependance $\delta n \sim \log(L) $ showing  that the end atoms of a large fixed segment have no well defined position on average, a standard result. 

 \subsubsection{Number fluctuations due to tunneling.}
 
  Let consider, as before, a large box of size $L$ drawn on the 3D lattice. The fluctuations by tunneling are due to the fluctuations at the sites near the border of this volume. Because those fluctuations are random and not correlated at large distances, a fluctuation $\delta n$ results from the random addition or subtraction of $L^2$ variables, because there are $L^2$ lattice sites near the border. Therefore the magnitude of the fluctuations is like the square root of the number of sites near the border, namely like $L$. For a small probability of tunneling, as expected for real solid Helium 4, this small fluctuation is to be multiplied by the (small) square root of the probability of site-to-site tunneling.  Notice that this source of number fluctuations by tunneling continues to exist in a non fluctuating lattice as well, i.e. in a lattice fixed from the outside as the optical lattices for atomic vapors. Therefore, because of those quantum fluctuations, such a system should exhibit long range phase order, at least in its ground state. This could be tested perhaps for an atomic vapor with one atom per site of an optical lattice. Therefore, the fluctuation of the number of particles in a fixed box of size $L$ is of order $L$, either because of the phonon zero-point motion or of the site-to-site tunneling. We shall use now this estimate to discuss ODLRO.

 The above discussion gives the idea to introduce the quantum amplitude for a state with a given number $n$ of particles in a given volume, without specifying which particle is there, contrary to the one-body density matrix $R_1({\bm{r}}_1, {\bm{r}}_{1}')$ that relies on the specification of a particle, something that is not obviously compatible with the quantum non-discernability. Let $\chi_{\Omega}(\bm{r})$ be the function equal to $1$ if its argument $\bm{r}$ is inside a fixed box $\Omega$ included in the whole crystal and zero otherwise, let $\delta(\alpha,\beta)$ be the Kronecker discrete function  equal to 1 if $\alpha = \beta$ and zero otherwise, $\alpha$ and $ \beta$  integers. Introduce now the function $\mathrm{C}_{\Omega}(n | {\bm{r}}_1, {\bm{r}}_2,\dots {\bm{r}}_N) = \delta(\sum_{i=1}^{N} \chi_{\Omega}({\bm{r}}_i), n) $ that is equal to $1$ if there are $n$ particles of any index inside a given volume $\Omega$ and zero otherwise. Thanks to this counting function one can derive from the $N$-body wave-function $\psi({\bm{r}}_1, {\bm{r}}_2, \dots, {\bm{r}}_N)$ the quantum amplitude (a complex number) associated to $n$ particles inside $\Omega$ as  
 $$\psi(n) = \int d^{3N}{\bm r} \mathrm{C}_{\Omega}(n) \psi({\bm{r}}_1, {\bm{r}}_2 \dots {\bm{r}}_N)\mathrm{.}$$ The element of the density matrix between states of given number of particles in two volumes $\Omega$ and $\Omega'$ is:
 $$ R_{\Omega, \Omega'} (n, n')=  \int d^{3N}{\bm r} \mathrm{C}_{\Omega}(n) \mathrm{C}_{\Omega'}(n' ) R_N ({\bm{r}}_1, {\bm{r}}_2 \dots {\bm{r}}_N; {\bm{r}}_{1}, {\bm{r}}_{2} \dots {\bm{r}}_{N}) \mathrm{,}$$ where $R_N$ is the N-body density matrix.  
 The  relevance of the fluctuations of $n$ in a given volume $\Omega$ for the fluctuations of the phase shows-up when considering the Fourier transform of $\mathrm{C}_{\Omega}(n)$, namely the function $\mathrm{D}_{\Omega}(\varphi)$ such that:
 \begin{equation}
\mathrm{D}_{\Omega}(\varphi) = \sum_{n=0}^{\infty} \mathrm{C}_{\Omega}(n) e^{in\varphi}
  \label{eq:defR5.1}
 \mathrm{.}
 \end{equation}
This function $\mathrm{D}_{\Omega}(\varphi)$ is the quantum amplitude associated to the phase in the volume $\Omega$. To get rid of the arbitrariness in the phase reference, it is a slightly more convenient to introduce the phase density matrix 
$$\mathrm{D}_{\Omega, \Omega}(\varphi, \varphi') = \sum_{n=0}^{\infty} \sum_{n'=0}^{\infty}\mathrm{C}_{\Omega, \Omega}(n, n') e^{in\varphi + in'\varphi'}\mathrm{.}$$ Consider now the function $\mathrm{D}_{\Omega, \Omega}(\varphi, \varphi') = \mathrm{\Delta}_{\Omega}(\varphi- \varphi')$.
If $n$ does not fluctuate, $\mathrm{\Delta}_{\Omega}(\varphi -\varphi') $ is just a smooth function $ e^{in_0 (\varphi - \varphi')}$, with $n_0$ fixed value of $n$. We have shown that, including in a {\it perfect} quantum crystal, $n$ fluctuates in such a way that $\left< (\delta n )^2\right>^{1/2}$ is much bigger than 1, therefore $\mathrm{\Delta}_{\Omega}(\varphi- \varphi') $ is non-zero in a narrow range of values of $(\varphi - \varphi')$ only. 

This is enough to prove the existence of ODLRO in this system. Take a large volume for $\Omega$. From the argument just presented, the phase of  $\mathrm{C}_{\Omega}(n)$ is a well defined constant. Therefore it does not fluctuate.  Take now $n = n'$ and $\Omega'$ derived from $\Omega$ by a space translation in $R_{\Omega, \Omega'} (n, n')$. 
The lack of ODLRO amounts to the fact $ R_{\Omega, \Omega'} (\varphi, \varphi')$ tends to zero as the distance between $\Omega$ and $\Omega'$ tends to infinity. 
This is true if and only if the phase of $\mathrm{C}_{\Omega}(n)$ and $\mathrm{C}_{\Omega'}(n)$ are independent, which is clearly not the case because both functions have to have a constant phase and because these phases are not independent: they have to be the same  phase of a bigger volume encompassing both $\Omega$ and $\Omega'$. There is a slightly less direct argument in favor of ODLRO, that assumes the existence of macroscopic equations of motion of the supersolid, as written in section \ref{macro}. The phase $\varphi$ of $\mathrm{D}_{\Omega'}(\varphi)$ is actually the same phase (up to an arbitrary additive constant) as the phase entering the macroscopic equations of motion of the supersolid. This is because the phase $\varphi$ is conjugate to the particle number, both as a result of its definition by the Fourier transform as done here or by the derivation of the macroscopic dynamical equations by a Lagrange formalism (\cite{ss1} and section \ref{macro}-B). Furthermore, the equilibrium state of the system, as it follows from the macroscopic dynamical equations, is with an uniform phase, the property equivalent to ODLRO in its microscopic formulation.  
 
Let us come back to the meaning of the statement ``one particle per lattice site". Classically it is well defined because the counting of particles is unambiguous. Things are different in the quantum case. The number density $\rho_0({\bm{r}})$ in such a crystal is a periodic function of the position and, by definition, the integral of $\rho_0({\bm{r}})$ over one cell (denoted $\Omega_{C}$) , namely $\int _{\Omega_{C}}{\mathrm{d}}{\bm{r}}  \rho_0({\bm{r}})$, is the average number of particles in this cell. In general there is no reason for this to be exactly an integer in the ground state at a given mean density. The only constraint is that, given the number $N$ of particles and the box volume $|\Omega_{tot}|$, the average number density $\frac{N}{|\Omega_{tot}|}$ should also be $\frac{1}{{\Omega_{C}}} \int _{\Omega_{C}}{\mathrm{d}}{\bm{r}} \rho_0({\bm{r}})$. Furthermore the {\it mean} number of particles in a box of volume $|\Omega|$ is $\int _{\Omega}{\mathrm{d}}{\bm {r}} \rho_0({\bm{r}}) = \sum_{n = 1}^{\infty}  n R_{\Omega, \Omega} (n, n)$. The number of particles in any box fluctuates because of tunneling and the zero-point fluctuations and, because of that, it does not make sense to specify exactly the number of particles per lattice cell. In the ground state at least it does not make any sense to specify this number of particles per lattice cell, because, according to the general principles of quantum mechanics, this can be done if this number is an observable commuting exactly with the energy operator, something that is certainly not correct if exchange between neighboring cells is possible. However this exchange can be small, but non zero in the limit $\Lambda$ large. In the coming subsection we explore another approach to the same question, namely the modelization of a crystal by atoms staying at discrete sites of a fixed lattice. This approach does not take into account the vibration of the lattice itself which are essential, as we have just seen, to make the phase of the ground state wave function uniform in space. This question will be looked at afterwards. 
 
 \subsection{Representation of a supersolid by a discrete crystal: lattice without phonons}
 \label{subsec:discrete}
 The theory of solid state makes wide use of lattice models of various kinds, like the Ising model of ferromagnets for instance. Of course it is a natural idea to try to model a supersolid by a lattice with "particles" at its nodes. We explore below this possibility. 
 
We consider a model energy operator where particles are constrained to stay on a fixed regular lattice (although as we have just seen this assumption of a fixed lattice makes impossible some quantum fluctuations existing in a {\emph{real}} elastic lattice, see next section for a lattice with phonons). Let $I$ denote the generic lattice index (a set of $d$ indices at each node of a lattice of dimension $d$). The energy operator representing the energy at leading order is an operator $\sum_{I} {\mathcal{E}} f(n_{I})$ where ${\mathcal{E}}$ is an energy, large compared to the other energies we are going to introduce, and $f(n)$ is a function of the number of particles at the given site. It is a function with a minimum if the site is occupied by one particle ( $n =1$) and increasing very much if there is more than one particle at the site. It can be for instance a function $f(n) = n(n-1)^{\kappa}$ where ${\kappa}$ is a large integer. In terms of creation and annihilation operators at the site $I$, $n = a^{\dagger} a$ where $a^{\dagger}$ and $a$ are the usual Bosonic creation and annihilation operators such that $[a^{\dagger}, a ] =1$. 
 Indeed the energy operator $\sum_{I} {\mathcal{E}} f(n_{I})$ commutes exactly with all the $n_{I}$ and its ground state is exactly for the value of $n_I$ making $f(n)$ the smallest at each site. However this simple property disappears as soon as one adds a tunneling piece to the energy operator. Tunneling is described in this simple approach by  adding to the energy the operator $\sum_{(IJ)} {\mathcal{E}}_{tunn} (a^{\dagger}_I a_J + a^{\dagger}_J a_I)$ , the subscript  ${(IJ)}$ in $\sum_{(IJ)}$ being for a sum extended to all pairs of nearest neighbors. Define the energy  
 \begin{equation}  
 \mathcal{H} = \sum_{I} {\mathcal{E}} f(n_{I}) + \sum_{(IJ)} {\mathcal{E}}_{tunn} (a^{\dagger}_I a_J + a^{\dagger}_J a_I)
 \mathrm{,} 
 \label{eq:hamilton}
 \end{equation}
 It commutes with the {\it total} number of particles, but does not with the number of particles at each site, which cannot be considered anymore as specified in this ground sate.  Because the total number of particles commutes with the energy defined in equation (\ref{eq:hamilton}), an explicit calculation of this energy requires to impose the number of particles as well. The search of the optimal state of this system is slightly complicated. The physically meaningful quantity is the energy per particle, given the density of the system. The lattice size is {\emph{not}} fixed a priori. It results from the optimization of the total energy, including under variation of the lattice size, that is accounted for by changing the energy of the single particle energy, under changes of the average density of the crystal. We plan to consider this problem in the future. This simple model provides an example of quantum crystal where the number of particle per lattice site can be calculated in the (realistic) limit where $\mathcal{E}_{tunn}  \ll \mathcal{E}$. This can be done by standard algebraic perturbation method, considering the tunneling piece as a small perturbation. 
 
Another remark may be interesting at this stage: simple models like the energy (\ref{eq:hamilton}) relate at least in order of magnitude various physically measurable quantities. It is tempting to do this for the present model. The superfluid component is linked to the exchange energy ${\mathcal{E}}_{tunn} $. This energy yields also an order of magnitude of the transition temperature from normal solid to supersolid, this being a bit similar to what London did many years ago to show the connection between the Bose-Einstein transition and the observed anomaly in the thermodynamic of liquid Helium 4 at the lambda point. In the present case, this relates ${\mathcal{E}}_{tunn} $ to the transition temperature.
In turn this allow to estimate, by a London-like argument, the density of the superfluid component of the supersolid. 

 \subsection{Representation of a supersolid by a discrete crystal: lattice with phonons}
 \label{subsec:discretephonons}
The introduction of phonons in this discrete model requires to add one more (quantum) degree of freedom per lattice site. This degree of freedom represents the displacement of the atom, if it is there, This section is for the purpose of completeness, to show that a lattice model of supersolid may also include the phonons. Suppose we had only one atom in an harmonic potential. Its energy is:
 \begin{equation}  
 {\mathcal{H}}_{s} = \frac{p^2}{2m} + g  \frac{q^2}{2}
 \mathrm{,} 
 \label{eq:hamiltonsingle}
 \end{equation}
where $g$ is a positive stiffness coefficient, and where $p$ and $q$ are conjugate such that $[p,q] = i\hbar$. The extension of this to a lattice of interacting atoms is done as follows. The first term  of kinetic energy becomes in the full lattice $\sum_{I} n_I   \frac{p_I^2}{2m}$. The interaction term should respect the translation invariance of the system: if all atoms move by the same amount in the lattice, the energy remains the same. Then a rather straightforward extension of the potential energy part of $ {\mathcal{H}}_{s} $ to a full lattice is $g   \sum_{(IJ)} \frac{(q_I - q_J)^2}{2} n_I n_J$.  Because it depends only on the difference between $q_I$ and $q_J$, it is invariant under a general translation, as requested for the energy linked to the strain in  crystal. Therefore the full energy operator with the phonon part included reads:
 \begin{equation}  
 {\mathcal{H}}_{phonon} = \sum_{I} ({\mathcal{E}} f(n_{I}) +  \frac{p_I^2}{2m}n_{I}) + \sum_{(IJ)} \left( {\mathcal{E}}_{tunn} (a^{\dagger}_I a_J + a^{\dagger}_J a_I) + g  \frac{(q_I - q_J)^2}{2} n_I n_J\right)
 \mathrm{,} 
 \label{eq:hamiltonfull}
 \end{equation}

\subsection{How to reconcile theory and observations ?}
\label{susec:howto}
 As we have shown,  a defect-free crystal of Helium 4 can be a supersolid at low enough temperature if phonons and site-to-site tunneling are taken into account. This leads us to discuss how this can be reconciled with the experimental findings.   
 
A rather strong argument for the genuine existence of a supersolid state of bulk Helium 4 is that the observed transition temperature from normal to supersolid is fairly independent of the quality of the crystal and depends on the pressure only. This rules out any explanation of supersolidity relying on a superflow along a network of dislocations and/or grain boundaries, because it would yield a transition temperature depending strongly on the quality of the crystal, not what is seen. Another rather strong argument against this explanation by a superflow along a network of grain boundaries or in the core of a dislocation networks is that it would need a very high density of defects to explain even the small observed superfluid fraction, particularly at the highest pressure where supersolid behavior has been observed. We propose to explain the {\it observed} sensitivity of the superfluid density with respect to the quality of the crystal and to foreign He3 atoms as resulting from the polycrystalline structure of the real crystals. In such a structure the superfluid density $\varrho_{obs}^{ss}$ measured by nonclassical rotational inertia (NCRI) experiments, is a function of the superfluid density $\varrho^{ss}$ in each single crystal and of the way the quantum coherence tunnels from one single crystal to its neighbors across the grain boundaries separating the single crystals in the sample. The observed high sensitivity of   $\varrho_{obs}^{ss}$  to minute (bulk) concentrations of He3 can be explained by the concentration of He3 in grain boundaries (this concentration of impurities in defects like grain boundaries and others is a well-known general phenomenon in real crystals). This concentration of He3 atoms likely lowers the grain-to-grain tunneling of Helium 4, including after long time scales necessary to reach equilibrium density of He3 atoms in the grain boundaries. The general dependence of  $\varrho_{obs}^{ss}$  with respect to the He3 concentration and annealing time should be rather complex, and not easily describable by a single tendency. To check this picture one should measure NCRI in single crystals, to get rid of effects linked to the polycrystalline structure of the sample.
Somehow this explanation is in agreement with the general fact that it is very hard to predict accurately macroscopic elastic parameters of real crystals, because those properties are very much dependent upon grain boundaries, etc. The case of superconductors in this respect is quite different of the supersolid: in the usual case where the healing length of the superconducting wave functions is much larger than the thickness of the grain boundaries, the supraconducing properties are not dependent on the microstructure, polycrystals with many grain boundaries or single crystals.

\section{Mean-Field model of supersolidity.}\label{Approach1}
 Below we focus on a model of supersolid valid at $T=0$ that is fully explicit and has all the properties requested for this state of matter, and that is sufficiently simple to allow in depth calculation of its properties. This is a model in the true sense because it cannot be derived from the equations of atomic motion pertinent for solid Helium, it tries only to keep the most fundamental properties of this quantum solid to understand its behavior. This model has many advantages: the first one is that it is a model of supersolid in the sense of Leggett \cite{leggett}, {\it i.e.} it shows NCRI as well as an absence of superflow induced by pressure gradient; second, it may be easily implemented in numerical simulations  where superfluids as well as ordinary solid behavior may be tested; finally, some of predictions may be compared to some features observed in experiments.

Since the work of Kirzhnits and Nepomnyashchii \cite{nepo} and of Schneider and Enz \cite{enz} in the early seventies the transition from liquid to solid Helium has been regarded as a 
manifestation of an instability as the roton minimum in the energy-momentum spectrum touches the zero energy line. Although this idea has been circulating for many years, in 
Ref.~\cite{pomric}  we have proposed a mean field model consisting on the Gross-Pitaevskii equation \cite{pit,gross}\footnote{Recently P.W. Anderson [Science {\bf 324}, 631 (2009)] 
suggested a Gross-Pitaevskii treatment of supersolidity, this description is in a completely different frame. Anderson argues, after Andreev and Lifshitz, that, in solid Helium, vacancies 
undergo a Bose-Einstein transition and maybe considered as a dilute (because of the smallness of the superfluid fraction) interacting gas which is known to be ruled by the Gross-Pitaevskii  equation. Anderson assumes a repulsive interaction, which is true a short distance,  and neglects the attractive  long range elastic interactions between vacancies without 
rational justification. This is in sharp disagreement with the results of Montecarlo simulations showing that vacancies tend to aggregate in bubbles \cite{Boninsegni06b}. In this respect 
the ``ground-state" proposed by Anderson, assumed to be homogeneous in space, would be unstable consisting to an unstable condensate of vacancies. Indeed, Anderson dealing of 
the condensate of vacancies leads to the {\it focusing}-nonlinear Sch\"odinger equation. Is well known that this  ill-posed equation develops massless finite time singularities.  Physically it means that the ground state is made of two phases: a zero density phase with droplets of a high density phase where the growth of the long range instability is stopped by the short range repulsive forces, a state that is not describable by a weak interaction or mean field theory like the non linear Schr\"{o}dinger equation. 
} 
with a roton minimum in the dispersion relation, already introduced in 1993 \cite{roton}, that presents a first order phase transition to a crystalline state as the roton minimun decreases, but before it touches zero. The difference with the usual Gross-Pitaevskii or nonlinear Schr\"odinger equation (NLS) commonly used for dilute gases where it is a good approximation, is in the potential of 
interaction between atoms that includes now a non trivial dependence on their  distance. This was the way the original Gross-Pitaevskii equation was written, like:
\begin{equation}
    i\hbar \frac{\partial \psi}{\partial t} = 
    -\frac{\hbar^{2}}{2m}{\bm{\nabla}}^{2}\psi + \psi({\bm{r}})\int U(|{\bm{r}} - 
{\bm{r}}'|) |\psi({\bm{r}}')|^2 {{d}}{\bm{r}}'
    {,}
\label{nls.org}
\end{equation}
where $ U(|{\bm{r}} - 
{\bm{r}}'|)$ is the two-body interaction potential. At this stage, this model correspond to a superfluid-like system at $T=0$K. Remark that finite temperature can be mimicked for steady situations~\cite{giorgio,nestorT}. Defining the average number density as $ \rho = \frac{1}{V} \int_V   |\psi({\bm{r}})|^2 {{d}}{\bm{r}} $, and noting $a$ the characteristic length of the interaction potential, the model has a single dimensionless parameter that governs the type of solutions
$$ \Lambda =  \frac{m  a^2}{\hbar^2} \rho  \int U(|{\bm{r}}|) \, {d}{\bm{r}}, $$ 
a kind of inverse of the de Boer parameter. This equation is an accurate representation of the atomic reality in the mean field limit, namely when the quantum fluctuations inside the range of the potential $U(|{\bm{r}} - 
{\bm{r}}'|)$ are relatively small. This happens if the average number of particles inside this range is large. This is not the case for real atoms where the repulsive forces and attractive forces have about the same range. One could think of electric dipoles to enhance the range of the long range part of the interaction. This puts the problem of the ground state in a different setting, as the ground state then is made of chains of parallel dipoles, very different of the crystal structure.

 The main properties that may be predicted for this model are: 
 
 {\it{i)}} The ground state of the mean field model is a crystal, that is a periodic pattern (a hexagonal one in 2D and a {\it hcp} in 3D); 
 
{\it{ii)}}  The existence of NCRI; 

{\it{iii)}} Four sound modes instead of the usual three in solids; 

{\it{iv)}} Quantized vortex and persistent currents. 

Moreover, in Ref. \cite{ss1,ss2} we have developped a theory for the long wave perturbations of the ground state of this model of supersolid.   We discussed in detail the role of boundary conditions and how to handle steady rotation and pressure gradient in this model.  The mechanical equilibrium was studied under external constrains as steady rotation or external stress and the model displays a paradoxical behavior: 

\begin{quote}
{\it The existence of a non classical rotational inertia in the limit of small rotation speed  that do not require defects nor vacancies  (in full agreement with Leggett's ideas) and no superflow under small (but finite) stress nor external pressure gradient. The only matter flow for finite stress is due to plasticity being facilitated by the eventual presence 
of defects. }
\end{quote}
Our numerical simulations \cite{ss1} clearly show that, within the same model, nonclassical rotational inertia is observed as well a regular elastic response to external stress or forces without any flow of matter, as in experiments \cite{chan06,chan07ncri,Beamish05,Beamish06}.

In addition, a new propagating "sound" mode  appears besides the usual  longitudinal and transverse phonons in regular crystals. The speed of this mode is smaller that the usual elastic sound waves speed, it scales as $c \sim \sqrt{f^{ss}} $, with $f^{ss}=\varrho^{ss}/\rho$ the superfluid fraction  $\sim 10^{-4}$. This slow mode 
is partly a modulation of the coherent quantum phase, like the phonons in Bose-Einstein condensates. Indeed the $\sqrt{f^{ss}} $ dependence reminds us the Bogoliubov spectrum. 

The aim of the section below is to study the macroscopical properties of the model given by equation (\ref{nls.org}) and to compare them with experiments in order to achieve a coherent picture of supersolidity.

\subsection{The non local Gross-Pitaevski\u{\i} equation as a model of supersolid} 
\label{sec1}
 Our model is based upon the original form of the 
Gross-Pitaevski\u{\i} (G--P) equation, which is a non-linear and non-local partial differential equation,  for the wavefunction of a weakly 
interacting Bose-Einstein condensate (\ref{nls.org}). This is an equation for a 
complex valued function $\psi({\bm{r}},t)$ (this is a 
complex field, {\it not} an operator).

This equation can be derived from the full Schr\"odinger equation 
of many interacting bosons in the mean-field limit where the interaction 
potential is weak and the quantum fluctuations in the range of the potential are small. To have stability of long wave fluctuations, the two-body interaction potential, that depends on the relative distance, should satisfy 
 that  $\int U(|{\bm{r}}|) \, \mathrm{d}{\bm{r}} $ is finite and positive.
Moreover, we shall require also that  the Fourier transform \begin{equation}
\hat U_k= \int U(|{\bm{r}}|)\, e^{i {\bm k}\cdot {\bm r} } \mathrm{d}{\bm{r}}  \label{fourier}
\end{equation}
has to be bounded for all $\bm k$, and, as we will see later, $ \hat U_k$ should become negative in some band-width in the $k$ space.

The difference with the most 
commonly used form of the 
G--P equation is that the potential of 
interaction between atoms includes a non-trivial dependence on the 
distance although usually this interaction is taken as a 
$\delta $-function in such a way that the cubic term in equation 
(\ref{nls.org}) becomes simply $g 
\psi({\bm{r}})|\psi({\bm{r}})|^2$. For dilute vapours at low 
temperature, the latter 
model is a fair approximation of the dynamics, $g$ being proportional 
to the scattering length for $s$-waves. 
The authors does not know any experimental situation where this non local Gross-Pitaevskii equation may be applied, although it has been suggested recently that a Bose-Einstein condensate with such peculiar interactions could exhibit such property~\cite{Henk10,Bonin10}: whatever the context, such a system would be very interesting for the fundamental physics of many-body systems.

\subsubsection{Basic properties, invariances and conserved quantities}
\label{secc1}

The equation (\ref{nls.org}) possesses the following properties. 

Suppose $\psi(\bm{r},t)$ is a solution. Therefore

\begin{quote}
-translation invariance : $\psi(\bm{r}+\bm{d},t) $ is a solution as well, with $\bm{d}$ any constant vector.

\end{quote}
\begin{quote}
-phase invariance : similarly $\psi(\bm{r},t) e^{i\alpha}$ with $\alpha$ any real constant,  is a solution. 
\end{quote}
\begin{quote}
-galilean invariance : the ``boosted solution'' $\psi(\bm{r}- \bm{v}t ,t) \, e^{i (m \bm{v} \cdot \bm{r}/\hbar + \frac{1}{2}mv^2 t/\hbar)} $ with $\bm{v}$ any constant vector is a solution 
\end{quote}
Moreover 
\begin{quote}
- Equation (\ref{nls.org}) can be put into a Hamiltonian form:
$$
    i\hbar \frac{\partial \psi}{\partial t} = 
   \frac{ \delta H}{\delta \psi^*} 
    \mathrm{,}
$$
where $ \psi^*$ is the complex conjugate of  $\psi$ and where the Hamiltonian $H[.]$ is 
\begin{equation}
  H [\psi,\psi^*]  =  \frac{\hbar^{2}}{2m} \int  \left| {\bm \nabla} \psi({\bm{r}})  \right|^2  {\mathrm{d}}{\bm{r}} + \frac{1}{2}\int  \int U (|{\bm{r}}-{\bm{r}'}|) |\psi({\bm{r}})|^2    |\psi({\bm{r}'})|^2 \,    {\mathrm{d}}{\bm{r}} {\mathrm{d}}{\bm{r}}' 
    \mathrm{,}
\label{nls.hamilton}
\end{equation}
is positive if $U(s)\geq 0$ for all $s$.
\end{quote}
\begin{quote}
-The Hamiltonian $H$, the number of particles $N = \int    |\psi({\bm{r}})|^2 \,    {\mathrm{d}}{\bm{r}} $ and the linear momentum $ {\bm P} = -\frac{i \hbar}{2} \int  (\psi^*{\bm \nabla}\psi - \psi {\bm \nabla} \psi ^*)  {\mathrm{d}}x$ are conserved by the dynamics with adequate boundary conditions \footnote{Note that a Galilean boost with a speed ${\bm v}$ changes the energy $ H'= H + {\bm P}\cdot{\bm  v} + \frac{1}{2} m N v^2$ and the momentum ${\bm P}' = {\bm P} + mN {\bm v}$ as usual in classical mechanics.}.
\end{quote}
\begin{quote}
-The Hamiltonian is convex for real values of $\psi$, that is, if one defines  $E[\psi^2] \equiv   H [\psi,\psi]$ for a real wavefunction $\psi$, then \cite{rafa} 
$$ E[\psi^2 = \lambda \psi_1^2 + (1-\lambda) \psi_2^2] \leq \lambda E[ \psi_1^2] + (1-\lambda)E[ \psi_2^2].$$
\end{quote}
\begin{quote}
-After a change to polar variables (sometimes called Madelung transform in this case): $\psi = \sqrt{\rho} e^{i\phi}$ with $\rho$ and $\phi$ real fields,  
the Gross-Pitaevskii equation (\ref{nls.org}) is transformed into the equations of an inviscid compressible fluid. They split into the equation for the conservation of matter:
\begin{equation}
  \frac{\partial 
   \rho}{\partial t} + \frac{\hbar}{m} {\bm \nabla}  \cdot \left( \rho {\bm \nabla}   \phi  \right) = 0  
    \mathrm{,}
\label{eq:GPrho}
\end{equation}
and a Bernoulli-like  equation:
\begin{equation}
   \hbar \frac{\partial 
   \phi}{\partial t}+\frac{\hbar^{2}}{2m}\left( {\bm \nabla} \phi \right)^2 + 
\int    U (|{\bm{r}}-{\bm{r}'}|)    |\rho({\bm{r}})|^2 \,    {\mathrm{d}}{\bm{r}} '  + 
  \frac{\hbar^{2}}{4m}\left(
	   \frac{( {\bm \nabla} \rho)^2}{2\rho^2} - \frac{ \nabla^2 \rho}{\rho} \right)\,=\,0  
    \mathrm{.}
\label{eq:GPbernoulli}
\end{equation} 
The Hamiltonian takes the form in the polar variables:
\begin{equation}
  H[\rho,\phi] =  \frac{\hbar^{2}}{2m} \int   \left(  \frac{1}{4\rho(\bm{r}) } \left| {\bm \nabla} \rho({\bm{r}})  \right|^2 + \rho({\bm{r}})   \left| {\bm \nabla} \phi({\bm{r}})  \right|^2 \right){\mathrm{d}}{\bm{r}} + \frac{1}{2}\int    U (|{\bm{r}}-{\bm{r}'}|) \rho({\bm{r}})    \rho({\bm{r}}') \,    {\mathrm{d}}{\bm{r}} {\mathrm{d}}{\bm{r}}' 
    \mathrm{,}
\label{hamilton.polar}
\end{equation} 

\end{quote}

{\it Remark.} According to the energy (\ref{hamilton.polar}), the ground-state solution is real (up to a 
constant phase) and any non uniform phase increases the ground state energy. Moreover, in general 
the ground state cannot vanish. If, however, the  ground state vanishes at some point $r_*$ as 
$\rho_0({\bm{r}}) \sim |{\bm{r}}-{\bm{r}}_*|^\alpha$ with $\alpha>0$ one must have $\alpha >2-D$. Indeed a state with 
$0<\alpha \leq 2-D$ has an infinite (positive) energy, since the energy in (\ref{hamilton.polar}) diverges as: 
$$\int   \frac{1}{4\rho(\bm{r}) } \left| {\bm \nabla} \rho({\bm{r}})  \right|^2  {\mathrm{d}}{\bm{r}}  \approx {\rm finite \ term} +  \alpha^2 \int_{V({\bm{r}}_* )_\epsilon}   |{\bm{r}}-{\bm{r}}_*|^{\alpha-2}{\mathrm{d}}{\bm{r}}\approx  {\rm finite \ term} +  {\mathcal O}(\epsilon^{\alpha+D-2} ) ,$$ 
when $\epsilon\rightarrow 0$.

\subsubsection{Bogoliubov spectrum with rotons}

Given $\rho$ the mean density defined as the number of particles per unit length, surface or volume in one, two or three space dimensions respectively, the homogenous and 
steady (up to a phase frequency) function $\psi_0=\sqrt{\rho}  e^{-i\frac{E_0}{\hbar} t} $, where $E_0 =  \rho\hat U_0$, is a solution of the dynamics, where $\hat U_0 = \int   U (|{\bm{r}}|)   {\mathrm{d}}{\bm{r}}$ (more generally the Fourier transform of $U(\cdot)$ is $\hat U_k = \int   U (|{\bm{r}}|) e^{ i {\bm{r}}\cdot{\bm{k}}} {\mathrm{d}}{\bm{r}}$). 

As discussed in next section, one can show that this solution is 
stable and makes the ground state for small enough $n$.
Indeed, small perturbations around this uniform solution are dispersive waves:
$$ \psi({\bm{r}} ,t)=\psi_0+(u_{\bm k} e^{i({\bm{k}} \cdot {\bm{r}} -\omega_k t)}+v_{\bm k} e^{-i({\bm{k}} \cdot {\bm{r}} -\omega_k  t)})e^{-i\frac{E_0}{\hbar} t} $$
where ${\bm{k}} $ and $\omega_k$ satisfy the Bogoliubov dispersion relation or spectrum~\cite{bogo47} 
\begin{equation}
\hbar \omega_k = \sqrt{ \left( \frac{\hbar^2 k^2}{2 m} \right)^2  +    \frac{\hbar^2 k^2}{m}  \rho\hat U_k}.
\label{bog}
\end{equation}

We shall assume that the potential scales as $\hat U_0$ and possesses a single length scale $a$.  
As discussed in~\cite{pomric}, the spectrum depends then only on a single dimensionless parameter : $\Lambda\sim  (\rho  a^D) \frac{m   a^{2-D}}{\hbar^2}  \hat U_0$ that is the product of a de Boer kind of parameter (that measures the ratio between the particle interaction and a zero point energy) and of the dimensionless parameter $\rho a^D$. More precisely we define
\begin{equation}
\Lambda =  \frac{m   a^2}{\hbar^2}  \rho \hat U_0
\mathrm{.}
\label{Lambda}
\end{equation}
Notice that the existence of a single dimensionless parameter in the problem does not take fully into account the complex nature of real solids with, at least,  two independent  dimensionless numbers $\rho a^D$ and $  \frac{m   a^{2-D}}{\hbar^2}  \hat U_0$.

The dimensionless Bogoliubov dispersion relation becomes  

\begin{equation}
\hbar \omega_k =  \frac{\hbar^2 }{ ma^2}   \tilde\omega_\Lambda(k \, a ) ,\quad {\rm with } \quad  \tilde\omega_\Lambda(s ) =\sqrt{ \frac{s^4}{4}  +    \Lambda s^2 u_D(s)}
\mathrm{,}
\label{bogSansDim}
\end{equation}
where $u_D(s)=\hat U_{s/a} /\hat U_0 $ depends only on the interparticle potential and of space dimension.

For some analytical results and for the numerics  we will choose a soft core interaction \cite{roton}
\begin{equation}
U (|{\bm{r}}-{\bm{r}'}|) = U_0 \theta (a-|{\bm{r}}-{\bm{r}'}|)\label{U0}\end{equation} with $\theta(\cdot)$ Heaviside function:  $\theta(s)=1$ if $s>0$ and $\theta(s)=0$ if $s<0$.

The Fourier transform of this special interaction potential reads 
\begin{equation}
\hat U_k = \hat U_0 u_D(k a)\label{fourier1}
\end{equation} 
with 
$$
 \hat U_0  =  \left\{ 
		\begin{array}{ll}
		2 a U_0  & \quad 1-D; \\ \\
  		\pi a^2 U_0   & \quad 2-D; \\ \\
		\frac{4\pi}{3} a^3 U_0   & \quad 3-D. \\ \\
		\end{array}
\right.
$$
and
$$
u_D(s)  =  \left\{ 
		\begin{array}{ll}
		 \frac{ \sin s}{s} & \quad 1-D; \\ \\
  2\frac{J_1(s)}{s} & \quad 2-D; \\ \\
		\frac{3 }{s^2}  \left(
\frac{\sin s}{s} - \cos s \right) & \quad 3-D; \\ \\
		\end{array}
\right.
$$
where $J_1(x)$ is a  Bessel function. Note that $U_0$  is the amplitude of the interaction (with units of energy) while   $\hat U_0 = \int U({\bm r} ) \, d^D{\bm r} $, with units of energy$\times$volume  ($\hat U_0 \propto a^D U_0$). 

\begin{figure}[hc]
\begin{center}
\centerline{\includegraphics[width=10cm]{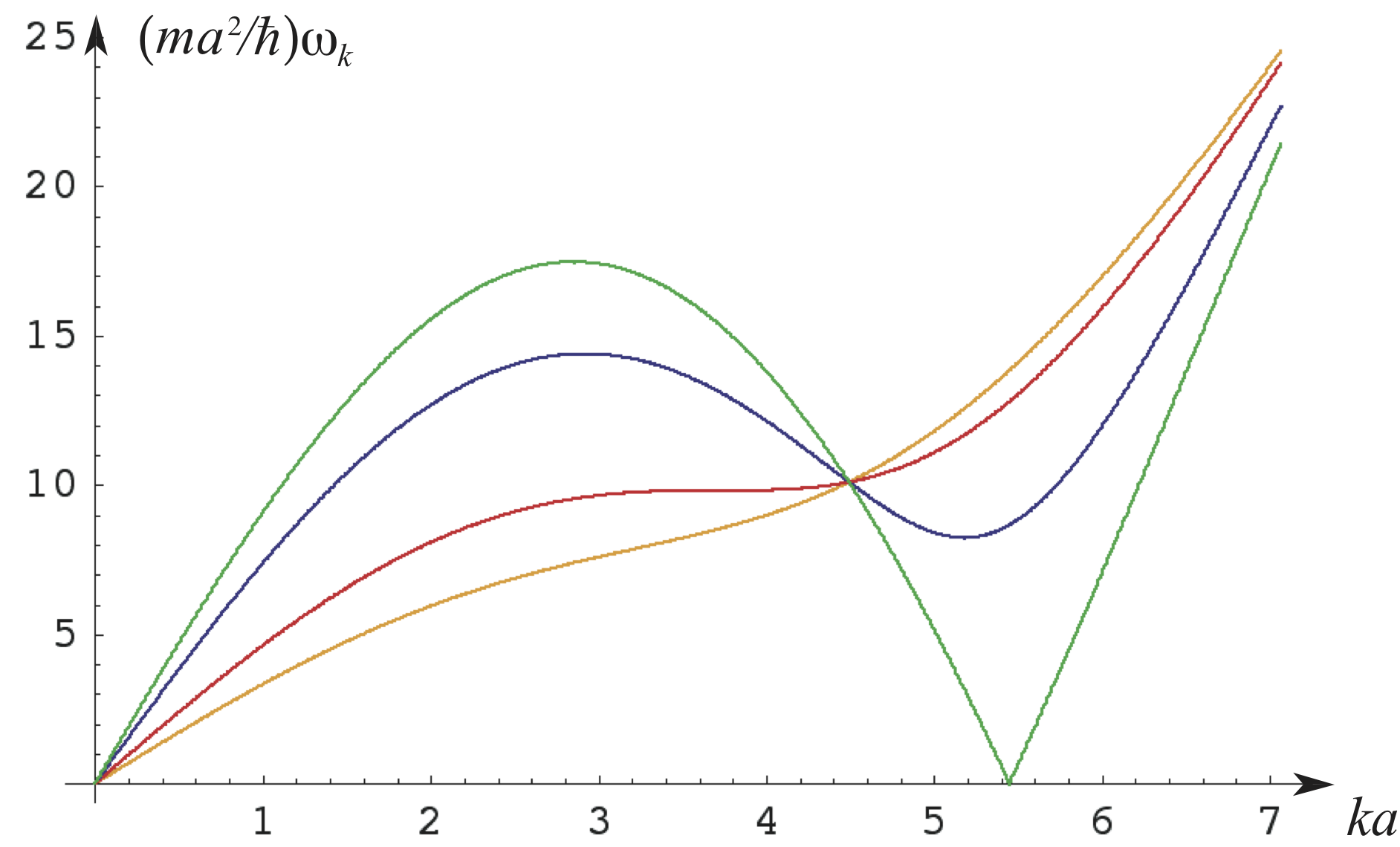} }
\caption{\label{rotons} Various spectrum shapes for different values of $\Lambda=12$, 23.43, 60 and  90.95 in three space dimensions.
}
\end{center}
\end{figure}

The spectrum represents well a phonon part together with a roton one (see figure \ref{rotons}). The long wave fluctuations are phonons that propagate with a speed $c_s =
\frac{\hbar}{ma}\sqrt{\Lambda}$. The short wave regime, the roton part, has  the following characteristics (see figure \ref{rotons}):
\begin{quote}
{\it i)} if $0<\Lambda< \Lambda_s$ the spectrum $\omega_k$ grows monotonically with $k$; therefore 
there is no roton minimum. 
\end{quote}
\begin{quote}
{\it ii)} for $\Lambda= \Lambda_s$ an inflexion point appears at $k_s$ and the energy is given by $\hbar \omega_s =  \frac{\hbar^2 }{ ma^2}   \tilde\omega_s$.
\end{quote}
\begin{quote}
{\it iii)}  If $  \Lambda_s<\Lambda <\Lambda_c$ the spectrum exhibits a roton minimum $k_r \, a = s_r$ which is an implicit function of $\Lambda$:
\begin{equation}\Lambda = - \frac{s_r ^3}{ 2 s_r u_D(s_r) + s_r^2 u'_D(s_r) }.\label{lambdakr}\end{equation}
\end{quote}
\begin{quote}
{\it iv)} for $\Lambda= \Lambda_c$, the spectrum reaches the axis for  $k_c$ as at the edge of the
phonon branch in solids. A reasonable value for $k_c$ is
$k_c = 5.45/a$  ($\approx 2.1 \AA^{-1}$ for
HeII).  
This picture suggest
that the existence of a roton minimum in HeII is, probably, a
reminiscence or a ghost of the solid state as we already suggested~\cite{pomric}.   
\end{quote}
\begin{quote}
{\it v)} if $\Lambda>\Lambda_c$ the spectrum becomes pure imaginary in a finite band-width in $k$, implying the appearence of a linear instability, leading to a periodic pattern of density modulation.
\end{quote}

The parameters considered $  \Lambda_s, \Lambda_c,\Delta_s, \Delta_c , k_s \& k_c$ depend explicitly on the space dimensions.
In the following table are given the values of the listed above for different space dimensions:

\begin{tabular}{|c||c|c|c|}
\hline
  &$ 1 D $  & $2 D$ & $3D$ \\
\hline\hline
$ \Lambda_s $& $9.47\dots$ & $15.81\dots$ & $23.43\dots$ \\ \hline
$k_s a$ &$2.89\dots$ &$3.33\dots$& $3.72\dots$ \\ \hline 
$\tilde\omega_s$ & $4.92\dots$ & $7.26\dots$ & $9.82\dots$ \\ \hline \hline
$ \Lambda_c $& $21.05\dots$ & $46.30\dots$ & $90.95\dots$ \\ \hline
$k_c a$ &$4.08\dots$ & $4.78\dots$ &$5.45\dots$ \\ \hline 
\hline \end{tabular}

Finally, we notice that this phonon-roton spectrum has a meaning only as a linear dispersion relation around an uniform solution.


\subsection{Ground-state of the Gross-Pitaevskii model}
\label{grounds} 

With a convenient value of $\Lambda $ we obtain a Landau spectrum with
rotons. As explained previously, if we increase $\Lambda $ we expect that there is a critical value
$\Lambda_s< \Lambda_{c_1} <\Lambda_c$ for which the system crystallizes, that is has a ground state with a density periodic in space. The increase of $\Lambda$ may be realized, for instance, by keeping constant the range $a$ and the magnitude $U_0$ and
increasing the density $n$. A density increase might be achieved in a physical system by
increasing the pressure and/or by cooling. Crystallization
due to the roton minimum can be expected near the real
solid phase since solid Helium exists only at non zero 
pressure. The transition occurs when the roton minimum is
near the $k$-axis for zero frequency. If we use Landau's notation for rotons: $\hbar\omega_k = \Delta +
\frac{\hbar^2}{2\mu}(|k| - k_r)^2$ for $k\approx k_r$. In
our picture $\Delta, k_r$ and $ \mu$ are non trivial
functions of $\Lambda$. However, $\Delta$ decreases and the roton
minimum $k_r$ increases, as $\Lambda$ increases.
One should also remark that beside the details of the model, the functions $\Delta (\Lambda), \
k_r(\Lambda) \ \& \mu_r(\Lambda)$ are known, and ultimately, must be determined
experimentally.

We shall see next via an energy argument that an uniform ground state cannot be stable for any $\Lambda$. As discussed above, from (\ref{hamilton.polar}), the phase of the ground state is always uniform in 
space, even when this state shows modulations of the density. A physical consequence is that the ground state has zero momentum $\bm P$.  

For low $\Lambda$ the ground state is a homogeneous solution, a superfluid (without positional order).  The uniform ground state  ($\psi= \sqrt{\rho }$), however, cannot be stable for any $\Lambda$, this fact follows from an argument by Enz and Schneider \cite{enz} and by Yu. A. Nepomnyashchii and collaborators \cite{nepo}. Consider  a small perturbation around a uniform solution $\rho(\bm{r}) = \rho_0+\tilde \rho(\bm{r}) $ and $\phi(\bm{r}) = -E_0/\hbar t +\tilde \phi(\bm{r}) $. This  allow us to write the Hamiltonian in  Fourier (\ref{hamilton.polar}) in a simple quadratic form :
$$
  H _2 =  \frac{1}{2} \int  \left[   \left(  \frac{ \hbar^{2} k ^2 }{4 m n}   + \hat U_k  \right)  \left| \tilde \rho_k  \right|^2 +   \frac{  \hbar^{2}  k ^2 }{4 m } \rho  \left| \tilde \phi_k  \right|^2\, \right] \,   {\mathrm{d}}{\bm{k}}    \mathrm{,}
$$
one sees that if $$\frac{ \hbar^{2} k ^2 }{4 m n }   + \hat U_k  <0,$$ then, the uniform solution is no more linearly stable. Therefore a periodic structure is expected at least as the roton minimum reaches the zero frequency axis (that is $\Lambda > \Lambda_{c}$).  In \cite{enz} the possibility of a linear 
instability was only considered, although the transition is subcritical (first-order) in two and three space dimensions \cite{nepo,pomric}. 
Indeed, by decreasing the roton minimum $\Delta$ there is a critical value $\Delta_{c_1}$ \footnote{Similarly, we may say that by increasing $\Lambda$ there is a critical value, $\Lambda_{c_1}, etc. $}   such a that is, if $\Delta<\Delta_c$, then the ground state shows a periodic modulation of density in space.

From now on we shall consider the dimensionless form of the non local Gross-Pitaevskii equation, $\Lambda$ being the only parameter, defined by (\ref{Lambda}). In short, the length scale in the problem will be $a$, the wave function $\psi $ being normalized by the total density $\psi \sim  \sqrt{\rho}$, finally the interaction $U(s)$ will be such a that  $\int U(|{\bm{r}}|) \, \mathrm{d}{\bm{r}} \, \equiv 1$. In those units one can write the Hamiltonian (\ref{nls.hamilton}) like $H/\Omega =   \frac{\hbar^{2}}{ma^2} \, \rho \,   {\mathcal E }$ with $\Omega$ volume of the system and :

\begin{eqnarray}
   {\mathcal E }&= & \frac{1}{\Omega}\left[  \int_\Omega \frac{1}{2} \left| {\bm \nabla} \psi({\bm{r}})  \right|^2  {\mathrm{d}}{\bm{r}} + \frac{\Lambda}{2}\int_\Omega \int_{\Omega}  \tilde U (|{\bm{r}}-{\bm{r}'}|) |\psi({\bm{r}})|^2    |\psi({\bm{r}'})|^2 \,    {\mathrm{d}}{\bm{r}} {\mathrm{d}}{\bm{r}}' \right]
    \mathrm{,}
  \label{densenergy}\\
1&=& \frac{1}{\Omega} \int_\Omega  |\psi(\bm{r},t)|^2 {\mathrm{d}}^D \bm{r} . \label{norm1}
\end{eqnarray}
Here $\Omega$ is the total volume of the system in $D$-space dimension so that the energy density ${\mathcal E }$ converges because the double integral is performed on a compact support (or very localized shape) of the non local interaction $\tilde U (|{\bm{r}}-{\bm{r}'}|)$.

In the following, we shall estimate modulation of periodic  solutions in one, two and three space dimensions based on a variational approach. 

\subsubsection{Weak amplitude periodic modulation in 1D}
In one space dimension the minimization of the energy leads to a supercritical (that is continuous second order transition) from a homogeneous (liquid phase) solution to a periodic (solid phase) solution. The analysis that we follow is the standard perturbation near threshold in the study of pattern formation of lamellar structures \cite{booklen}. 

If $\Lambda \gtrsim \Lambda_c$ a weak amplitude developement with a single wanumber selection is possible, then we consider the wave function that is normalized  in a period $\lambda = 2\pi/k_c$ according to the normalization condition (\ref{norm1})
\begin{equation}
\psi (x) = \frac{1}{\sqrt{1+ 2 |A|^2}}\left(1 + A e^{i k_c x}  + A^*e^{-i k_c x} \right)
\label{trial1}
\end{equation}
Introducing this trial function into 
(\ref{densenergy}) one finds the energy per unit length: 
$${\mathcal E }= \frac{1}{2} \frac{2 k_c^2 |A|^2 } {(1+ 2 |A|^2)}  + \frac{\Lambda}{2} \left( 1+ \frac{8 |A|^2 \hat U_{k_c} } {(1+ 2 |A|^2)^2} + \frac{2 |A|^4 \hat U_{2k_c} } {(1+ 2 |A|^2)^2}\right).
$$
The minimum of this quantity gives the value for the modulus of the complex amplitude $A$:
\begin{equation}|A|^2 =  - \frac{k_c^2 + 4 \Lambda \hat U_{k_c}}{2 ( k^2 + \Lambda(\hat U_{2k_c}-4 \hat U_{k_c})}  
\label{weakamp}\end{equation}
The structure displays a periodic modulation with a wave-number $k_c = 4.078\dots$. Therefore, setting $k=k_c$ into (\ref{weakamp})
one may calculate this amplitude as a function of $\Lambda$:
$$ |A_c|^2 =\frac{-8{\rm sin}(k_c)}{k_c^3(8- {\rm cos}(k_c))} (\Lambda-\Lambda_c)  \approx 0.011 (\Lambda-\Lambda_c) $$ 

\subsubsection{First order transition in two and three space dimensions}
This ground state can be
found  by minimizing the energy functional (\ref{densenergy}).  
We sketch the solution in the case of two
spatial dimensions. As a trial solution we take 
$\psi_0$ the uniform solution plus a modulation, the amplitude of the uniform part being set to one by a convenient choice of units : 
 $$ \psi^2_0 ({\bm x}) = 1 +  \left (\sum_{j=1}^3 A_j
e^{i{\bm k}_j\cdot{\bm x}} + c.c. \right)  \mathrm{,}$$
with the three complex amplitudes $ A_j$ such that $
|A_j|\ll 1$, and the vectors ${\bm k}_j$ form an
equilateral triangle (${\bm k}_1+{\bm k}_2+{\bm k}_3=0$)
with a magnitude $|{\bm k}_j|=k_r$ (in this case the critical wave number will be simply the roton minimum, actually something realistic in solid Helium $\approx 2\pi/k_r \approx 3 \AA$). If we put 
$\psi_0({\bm x})$ in (\ref{densenergy}) and expand in
powers of $A_j$, since $ |A_j|\ll 1$, we obtain: 
\begin{equation}  
{ \mathcal E } = {  \frac{{E }}{\Omega} } = \rho
\frac{k_r^2}{2} \left( 2\mu^2
\sum_{i=1}^3|A_i|^2 - \frac{3}{4}\left( A_1A_2A_3 +
A_1^*A_2^*A_3^* \right)
+ \frac{1}{2}\sum_{i=1}^3|A_i|^4 + 2 \sum_{i<j}|A_i|^2 |A_j|^2 + \dots,
\right)   
\label{energie} 
\end{equation}
with $\mu = \frac{\tilde\omega_\Lambda(k_r) }{k_r^2} $.

\begin{figure}[hc]
\begin{center}
\centerline{\includegraphics[width=10cm]{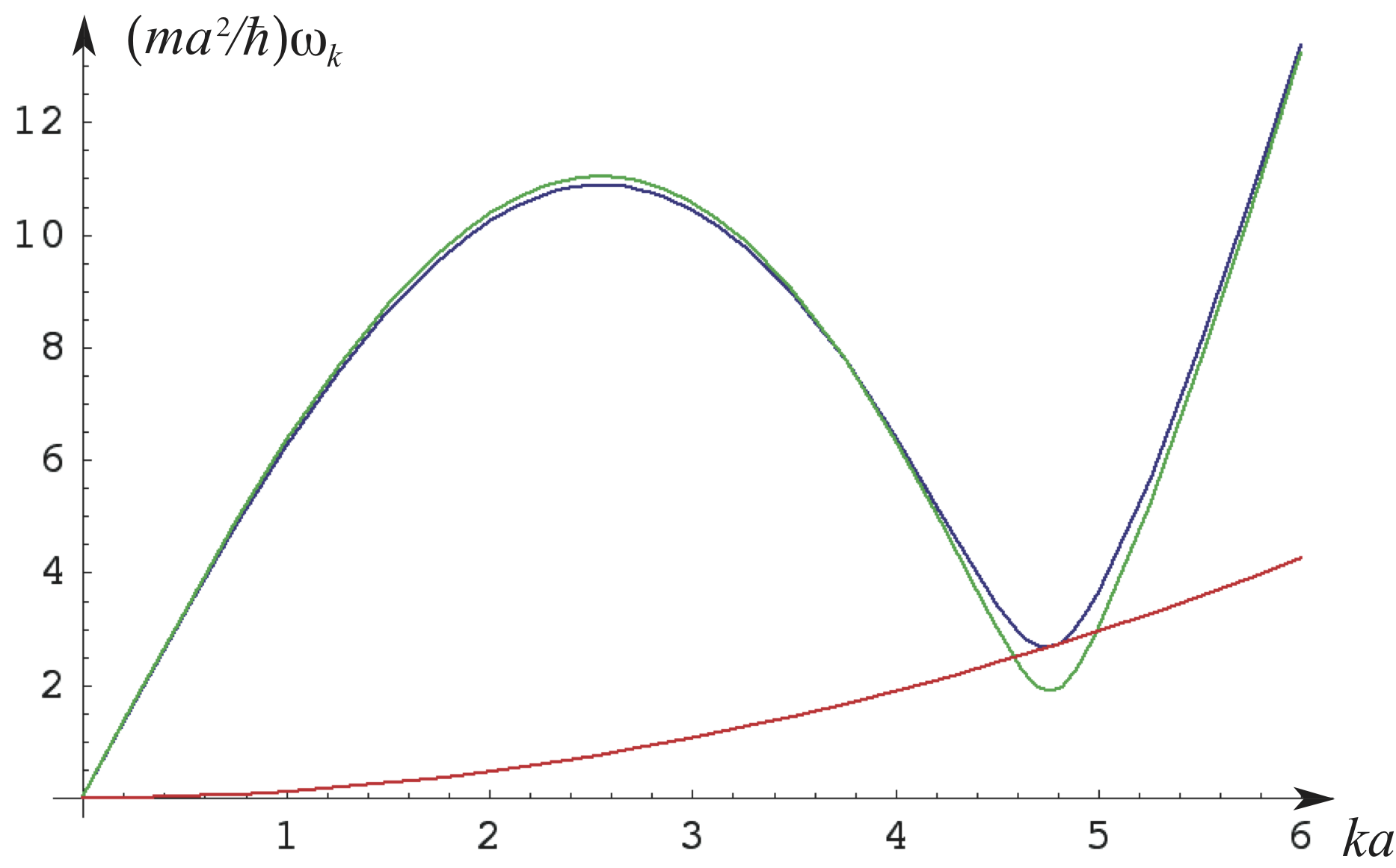}}
\caption{\label{rotonscritic} Various spectrum shapes for different values of $\Lambda=37.36$,  and  $\Lambda=39$ in two space dimensions. The parabola is the critical curve $ \frac{3}{8 \sqrt{10}} k^2$, when it touches the roton spectrum at $\Lambda_{c_1} =  37.36...$ sub-critical transition towards an hexagonal pattern is possible.
}
\end{center}
\end{figure}

The ground state is given by the minimum of the energy,
i.e., $\frac{\delta E}{\delta A_i} = 0$.
The hexagonal crystal solution is
($A_j=R_je^{i\varphi_j}$) such that each of the three
amplitudes is equal to:   \begin{equation} 
R_j = \frac{3}{40}\left(1\pm
\sqrt{1-\frac{640}{9}\mu^2 } \ \right),
\ \ j=1,2,3, 
\label{hex1}
\end{equation}
and the phases satisfy
$\varphi_1+\varphi_2+\varphi_3=0$ (mod. $2\pi$). The positive
sign in (\ref{hex1}) gives the stable solution and the
negative sign gives an unstable solution. The existence of
this kind of solution constrained by the condition $\mu^2<9/640$; or better  as represented in Fig. \ref{rotonscritic} :
$$\frac{\tilde\omega_{\Lambda_{c_1}}(k_r) }{k_r^2}  = \frac{3}{8 \sqrt{10}}$$ has a single (degenerate) root in $k_r$.
This  gives a critical wave number $k_r a = 4.55...$ and a subcritical threshold for the control parameter $\Lambda_{c_1} =  37.36...$

In physical terms the roton minimum is a more natural parameter than $\Lambda$.  In physical units  $\Delta_{c_1}=\frac{3}{\sqrt{40}}
\frac{\hbar^2 k_r^2}{4m}$.
For superfluid liquid Helium
($k_r = 1.95 \AA^{-1}$) this condition gives the value\footnote{All the estimations concern Helium, the
energies being in Kelvin ($K$). It is useful to note
that $\frac{\hbar^2}{m} = 12 \ K\AA^{2}$, the quantum of
circulation is $\frac{\hbar}{m} = 158 \ m/s \AA$ and $a =
2.57 \AA$.}:
$\Delta_{c_1}= 5.4 K$ 
and the corresponding energy by particle
is about $E/N \approx 10^{-3} K$. We note that this energy
is reduced, probably by hybridization, by three orders of
magnitude with respect to all of the energies in the
theory. $\Delta_{c_1}$ is bigger in the three dimensional
case, since there are more possible stable crystalline
configurations, such as $bcc$ or $hcp$. In Ref. \cite{pomric} we studied the
$bcc$ configuration, in this case we have six amplitudes and the waves
vectors form an octahedron \cite{anne}, each vector
participates in  two equilateral triangles, which
produces a stable configuration. A $bcc$
lattice is stable if $\Delta <
\Delta_{c_1}=\frac{3}{\sqrt{22}} \frac{\hbar^2 k_r^2}{4m}$,
which gives $\Delta_{c_1}= 7.3 K$ and we have an
energy by particle about $E/N \approx 5\times 10^{-3} K$.
We suggest that the supersolid phase occurs by increasing the
density, therefore an estimate of the critical density $\rho_{c_1}$ is
possible, in principle, using an empirical law, $\Delta(\Lambda\sim \rho)$,
together with the critical gap $\Delta_{c_1}$. Unfortunately we do not
have adequate experimental data ($\Delta$ as a function of $n$).
However, from \cite{varoco} we can fit
$\Delta$ as a function of the pressure ($p$). The fit gives $\Delta =
\Delta(0) (1-p/p_0)$, with $\Delta(0) = 8.74 K$ and $p_0 = 157 bar$.
We find a critical pressure $p_c = 26 bar$ for the transition to a $bcc$ lattice, in good agreement with the experimental value. 

\begin{figure}[hc]
\begin{center}
\centerline{a) \includegraphics[width=8cm]{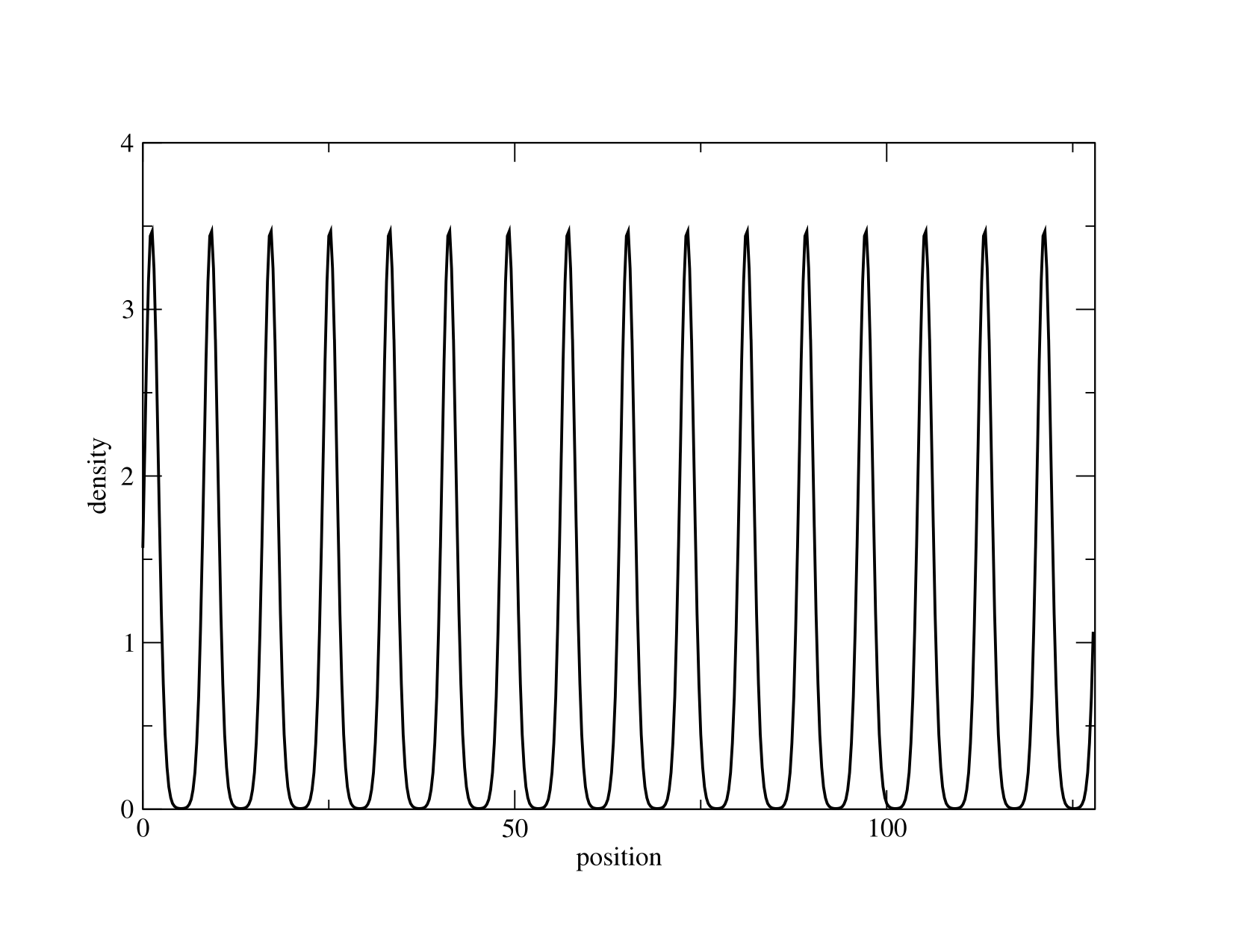} \quad b)  \includegraphics[width=8cm]{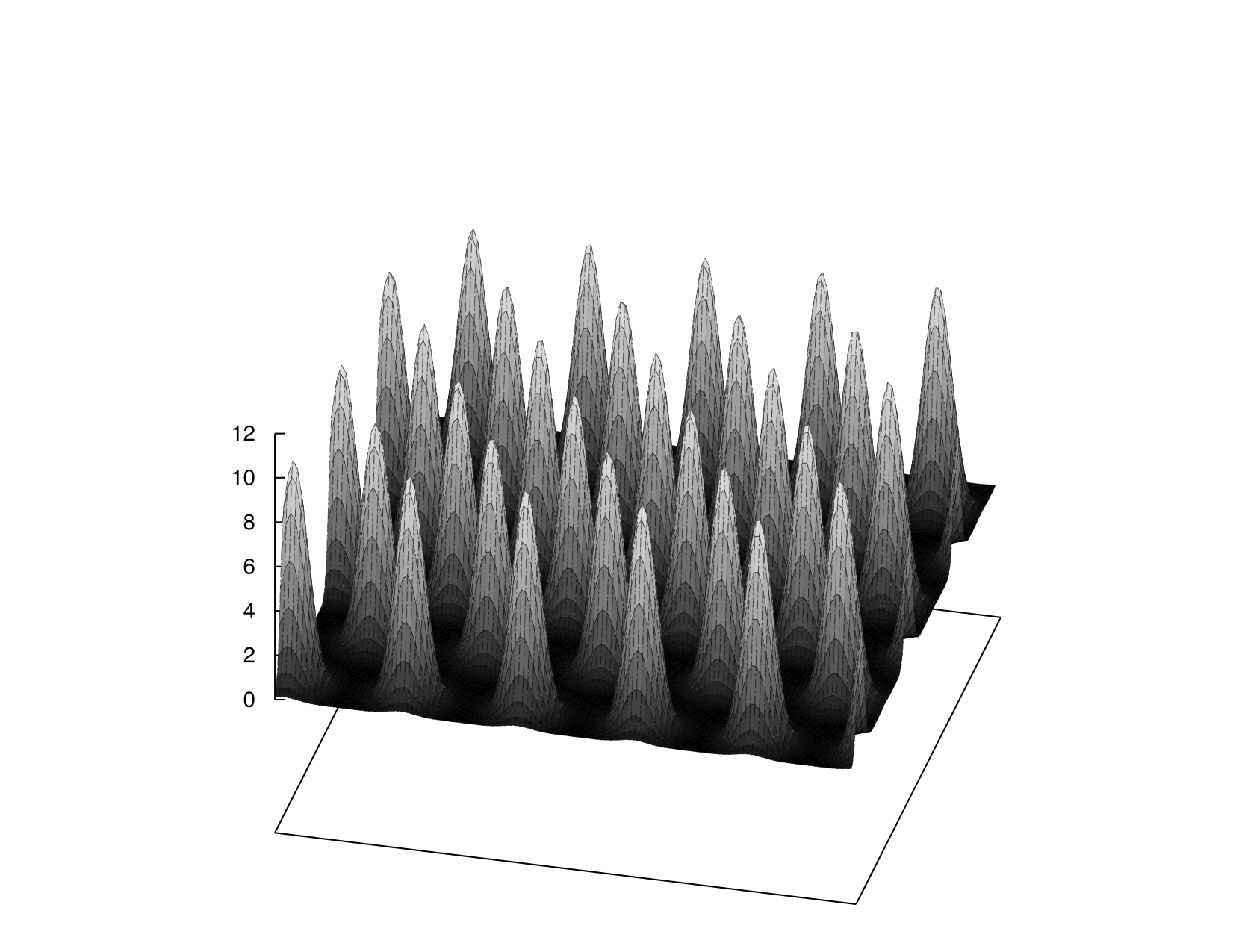} } 
\caption{\label{fig:ground} Ground state of the G-P equation obtained numerically using the dissipative G-L equation a) in 1D, where a periodic crystal of density peaks is formed (here $\Lambda=43.2$); b) in 2D, it leads to a periodic hexagonal pattern ($\Lambda=107$).
}
\end{center}
\end{figure}

Numerically, the ground state can be reached by considering the dissipative version of the G-P equation, called the Ginzburg-Landau (G-L) equation  that can be understood as a imaginary time evolution of the G-P equation. It writes in dimensionless form:

\begin{equation}
\frac{\partial \psi}{\partial t} = 
    \frac{1}{2}{\bm{\nabla}}^{2}\psi -\Lambda \psi({\bm{r}})\int U(|{\bm{r}} - 
{\bm{r}}'|) |\psi({\bm{r}}')|^2 {{d}}{\bm{r}}'+\mu \psi
    {,}
\label{eq:GL}
\end{equation}

where $\mu$ is the chemical potential to impose the mean density (or total mass) of the system. The G-P and G-L equations have the same stationary solutions and ground states and the dissipative G-L dynamics converges to a local minimum of the free energy that can be deduced from the Hamiltonian. Thus, starting with a noisy initial conditions and running numerically the G-L dynamics, one can achieve a ground state. We use two types of numerical methods, a pseudo-spectral one when periodic boundary conditions can be considered and a finite difference one based on a Crank-Nicholson scheme otherwise. Figure (\ref{fig:ground}) shows the ground state in 1 and 2 dimensions with periodic boundary conditions. 
A regular 1D crystal is observed and in 2D an hexagonal pattern is formed, minimizing the free energy of the system.

\begin{figure}[hc]
\begin{center}
\centerline{\includegraphics[width=12cm]{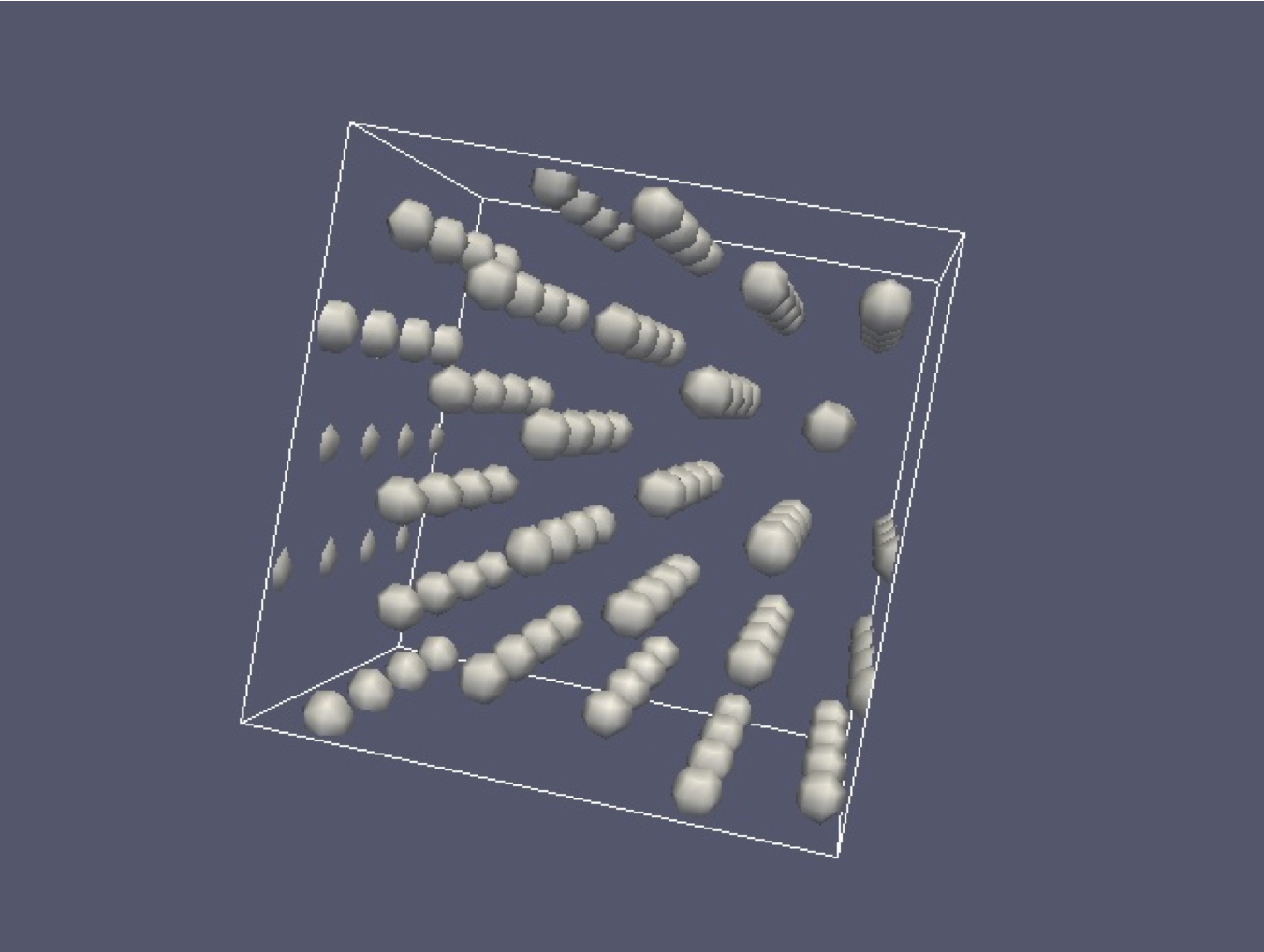} } 
\caption{\label{fig:g3D} Ground state of the G-P equation obtained numerically using the dissipative G-L equation in 3D, exhibiting a regular $hcp$ crystal ($\Lambda=429$).
}
\end{center}
\end{figure}

In three space dimensions the most stable configuration is the  $hcp$ crystalline structure as it can be seen on figure (\ref{fig:g3D}). In this case the critical wave vectors belong to a tetrahedron, whereas in the $bcc$ structure each vector belongs two resonant equilateral triangle.

\subsubsection{Ground-state in the large $\Lambda$ limit.}
The  crystal structures with a weak modulation of density have been described in 1-D for moderate values of $\Lambda$, say $\Lambda \gtrsim \Lambda_{c_1}$. In fact, the ground state defined as the minimum of the energy functional (\ref{densenergy}) exists and is unique, because of the convexity, for any $\Lambda$. For larger values of $\Lambda $ the weakly nonlinear approach cannot be used anymore to derive the amplitude of the density modulation.  

For large $\Lambda$, the potential energy in (\ref{densenergy}) requires small $\psi$ while the mass 
normalization (\ref{norm1}) forbids $\psi$ to be small everywhere. Therefore the energy 
minimization leads to a periodic structure with zones where $\psi \approx 0$ balancing zones where 
$\psi \gg 1.$ Recently it has been proven by Aftalion et al.~\cite{mamandine} 
a remarkable, in the opinion of the authors, theorem: in the limit $\Lambda \rightarrow \infty$, the minimization of  (\ref{densenergy}) with the restriction (\ref{norm1}) is equivalent to a close packing arrangement of rods, disk and spheres in one, two and three space dimensions respectively.

In the particular case of 1-D,  it is shown, in Ref. \cite{mamandine},  that in the large $\Lambda$ limit $\psi \neq 0 $ only in a small zone : $x\in (-\delta,\delta)$ 
of the whole period: $(-\lambda/2,\lambda/2)$, where the Euler-Lagrange condition deduced from (\ref{densenergy}) together with  (\ref{norm1}) leads to the Hemholtz equation in the domain $(-\delta,\delta)$ : $-\psi''(x) =  \mu \psi$. Finally the minimization of the energy gives $\delta$ and the wave number $\lambda$ of the periodic structure. 

Following this approach, with Sep\'ulveda~\cite{nestor} we estimated such a ground state for $\Lambda \gg 1$ (the extension to higher dimensions seems natural but the computations are harder). We sketch below the corresponding results. 
We consider the energy and normalization (\ref{densenergy},\ref{norm1}) in a single period with the trial function in the unit cell\footnote{A different trial function that does not satisfy the equation $-\psi''(x) =  \left(\frac{\pi }{2\delta}\right)^2 \psi$ could be used: $\psi(x) = 
		\sqrt{\frac{15\lambda}{16\delta}} \left(1- \left(\frac{x}{\delta}\right)^2\right) $ in $ x \in[-\delta,\delta]$ and zero elsewhere. It provides  however similar asymptotic results.}
\begin{equation}
\psi(x) =  \left\{ 
		\begin{array}{ll}
  0& \quad x\in[-\lambda/2,-\delta]\\
		\sqrt{\frac{\lambda}{\delta}} \cos\left(\frac{\pi x}{2\delta}\right) & \quad x \in[-\delta,\delta]\\
0& \quad x\in[\delta,\lambda/2]\\
		\end{array}
\right.
\label{trial2}
\end{equation}
The integrals maybe computed more or less easily because of the interaction term. The minimization yields a relation between $\delta$, the wavelength $\lambda$ and the dimensionless $\Lambda$.: 
\begin{equation}1 + 2\,\delta  - \lambda  = 2\left( \frac{15}{\pi^2} \right)^\frac{1}{5} \frac{ \delta}{ \left(\lambda ( \lambda -1)^2 \, \Lambda  \right)  ^\frac{1}{5}}. \label{asympl2} \end{equation}
This variational result is in complete agreement with the numerics, a more complete discussion on this problem may be found in Reference \cite{nestor}.

\subsubsection{The question on the commensurability in the Gross-Pitaevskii model.}

The question of the number of  ``atoms'' $N$ versus the number of sites $N_s$ and how are they related, if they are a relation, needs some discussion in the context of the present model. 
The problem if the solid is commensurable or incommesurable, that is if $N_s/N\equiv 1$ or $N_s/N\neq 1$ is, in the context of this model, rather different from a commensurability problem in a classical solid. More precisely, let us consider a initial wavefunction with initial number of atoms (or mass) $\int |\psi|^2 {\rm d} {\bm r} = N$, being an integer in a periodic box of size $L$, $L\times L$, $L\times L \times  L$ in one, two and three space dimensions respectively.  The number density is, therefore, $N/L^D$.

In the present model the crystal appears by a spontaneous breaking of the symmetry under translations of the uniform state. Neither the sites nor the peaks (or ``atoms'' ?) were there before. Moreover, as in every spontaneous symmetry breaking, the appearance of the structure is a dynamical process governed by transient structures full of defects, vacancies, interstitials, grain boundaries, {\it etc.} In infinite length domain, it is expected that these defects, vacancies, interstitials, grain boundaries, {\it etc.}  move away the system leading to a defect-free  crystal. This process may take, however, a very long time. On the other hand, for finite length system, which is the particular case of numerical simulations, the  formation of a periodic  structure needs to match the boundary conditions that makes difficult the realization of a free defect system.  

As an example, the ``pattern formation'' process in one space dimension is such a that the wavelength cannot be selected via the most unstable mode $\lambda_c = \frac{2\pi}{k_c}$ because the wavelength should match the boundary conditions, something that becomes more or less irrelevant as the system size grows $L\gg \lambda_c$.  The number of sites is related to the selected wavelength via  $\lambda N_s =L $. So in practice the ratio of the number of sites to the number of ``atoms'' (the initial normalization) $N$ are not the unity $N/N_s =  N\lambda/ L = \rho \lambda $ except for very particular cases.  In higher dimensions the mismatch of the crystal structure-boundary conditions becomes more complex.

One of us (SR) has performed 2D numerical simulations of equation (\ref{nls.org}) in a $64\times64$ periodic box. The interaction range was chosen to be $a=8$, the mesh size being $dx = 1$. The normalization condition is for $\int |\psi|^2 {\rm d} {\bm r} = N= 36$ ``atoms''. Although the simulations of the original Gross-Pitaevskii equation (\ref{nls.org}) are formally reversible, an irreversible behavior appears naturally that takes the energy from long scales to the smallest ones $\approx dx$ driving the system finally to a macroscopic state that minimizes the energy \cite{hdr}. It is important that the relevant structures are much larger than this size. Indeed in the present case the roton minimum is placed at $k_r \approx 4.55/8 = 0.57$ which is much smaller than the wave number associated to the mesh size: $ \pi.$

\begin{figure}[hc]
\begin{center}
\centerline{a) \includegraphics[width=6cm]{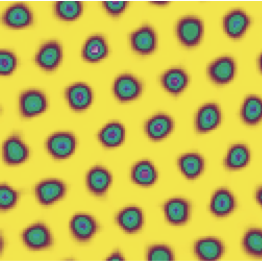} \quad b)  \includegraphics[width=6cm]{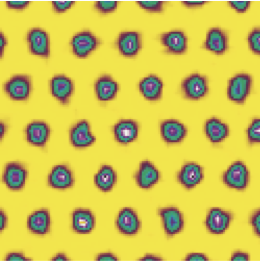} } 
\centerline{c) \includegraphics[width=6cm]{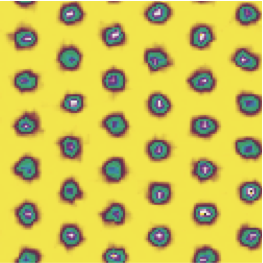} \quad d)  \includegraphics[width=6cm]{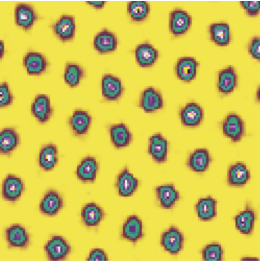} } 
\caption{\label{commens} Various density plot for long-time (1000 time units) numerical simulation of (\ref{nls.org}) with $a=8$ in a $64\time 64$ system size with periodic boundary conditions. In all simulations the initial conditions are identical and  $N=36$. The values of the interaction parameter $U_0 $ are: a) $U_0 = 0.5$ and $\Lambda=56.55$ ; b)  $U_0 = 0.7$ and $\Lambda=79.17$; c) $U_0 = 0.75$ and $\Lambda=84.82$ and d) $U_0 = 0.85$ and $\Lambda=96.13$. The number of sites are, respectively, a) $N_s = 34$; b)  $N_s =36$; c)  $N_s =36 $ and d)  $N_s =40$.
}
\end{center}
\end{figure}

One sees that despite the same initial condition, for the four cases presented in the figure \ref{commens}, the system reaches different commensurability fraction $N/N_s$ both larger and smaller than the unit depending on $\Lambda$. In fact, for larger $\Lambda$ (that is larger compressibility because $\Lambda\sim \rho$) the system makes in average more sites of lower mass each. And conversely for small $\Lambda $ (small compressibility) the system makes a smaller number of sites. 
We can conclude that in the present model the commensurability is not a relevant parameter, as suggested by the mean field origin of the model.
In fact, despite the different commensurability, figures a), b) or c) and d) are qualitatively the same.

This question of commensurability could be investigated experimentally. If there is in average a non integer number of atoms per lattice site, by neutron (or X ray) scattering, one could 
see the mismatch between one and the actual number of atoms per site. This is surely not an easy experiment, because the mismatch is likely small and it requires an accurate measurement of the mass of the Helium crystal unit for diffraction. Perhaps a differential experiment could be done by changing the volume under applied pressure.

\subsection{A model combining elastic and superfluid properties}

\subsubsection{Non-Classical Rotational of Inertia}
As a built-in superfluid model, it is expected that the Gross-Pitaevski\u{\i} equation exhibits superfluid properties in presence of a crystal lattice. This has been shown first in~\cite{pomric} where the superflow of such supersolid model was observed around a cylindrical obstacle with a critical velocity above which vortices where nucleated, similarly than for the
superfluid model~\cite{fpr}.

\begin{figure}[hc]
\begin{center}
\centerline{{\it a} \includegraphics[width=8cm]{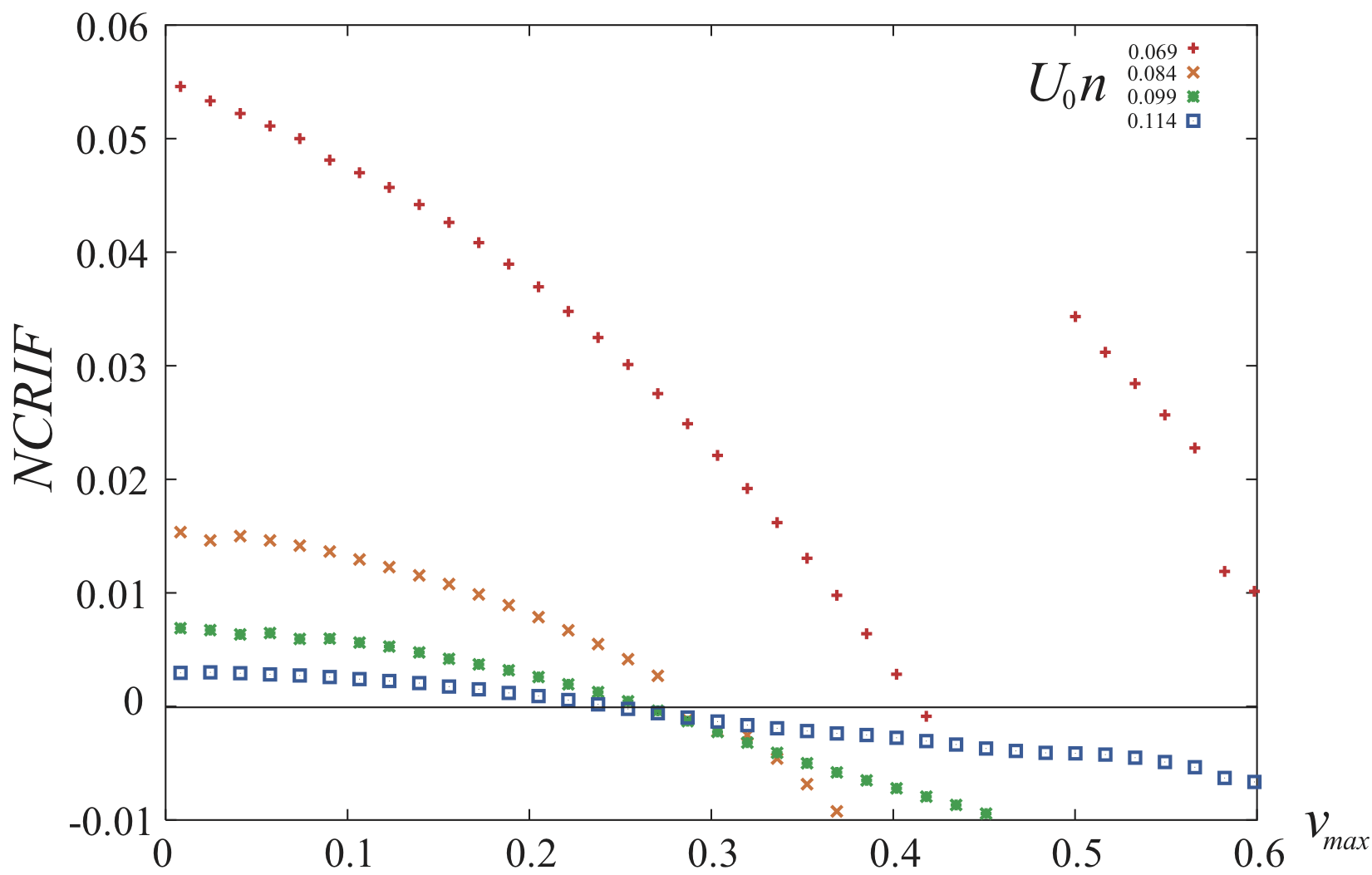} {\it b} \includegraphics[width=8cm]{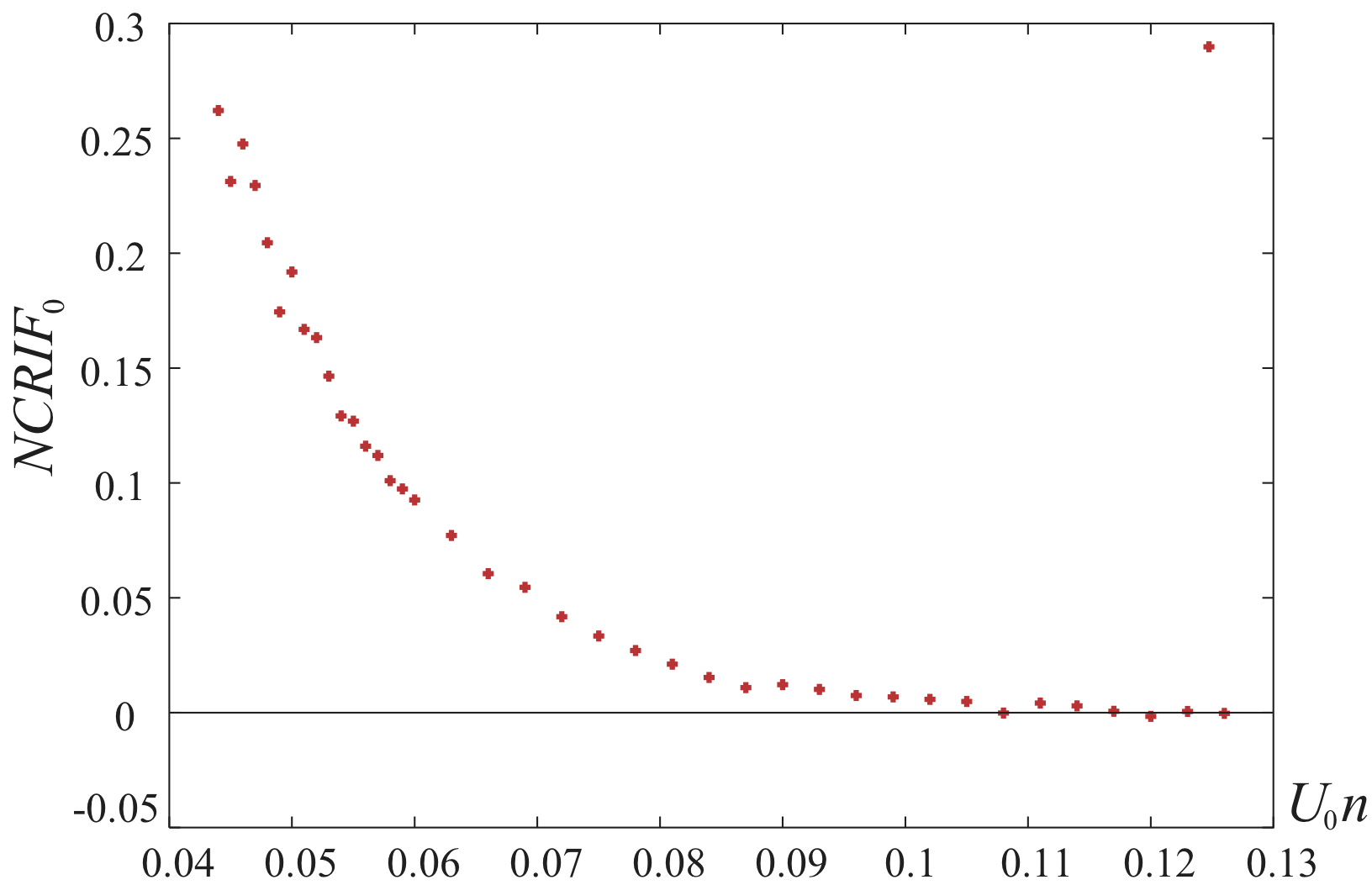}}
\caption{ \label{ncri2D} 2D numerical simulation of the dimensionless equation with $128\times 128$ modes in a square cell of $96 \times 96$ units for different values of $\rho U_0 $; the potential range is $a = 4.3$. {\it (a)} The $NCRIF \equiv 1-L_z'(\omega)/\left<I_{rb}\right>$ {\it vs.} the local maximum speed $v_{max} = \omega L/\sqrt{2}$ for  $\rho U_0= 0.069$, $0.084$, $0.099$ $\&$ $0.114$ (res. $\Lambda = 74.1$, $90.2$, $106.3$, \& $122.4$). Here $\left<I_{rb}\right>$ is the converging inertia moment computed numerically for large $n U_0$ at $\omega=0$. Note that the jump in $NCRIF$ for $\rho U_0 = 0.069$ ($\Lambda = 74.1$) corresponds to the nucleation of a vortex in the system. 
 {\it b)} NCRIF at $\omega=0$ as a function of $\rho U_0$. We have verified that {\it (a)} and {\it( b)} are 
 almost independant of the box size. This figure is similar than that in~\cite{ss1}}
\end{center}
\end{figure}

In fact, it is possible to use this model to mimic the torsional oscillator experiment where NCRI has been experimentally observed~\cite{chan04a}. In~\cite{ss1}, we have observed the NCRI effect using a 2D numerical simulation where a square sample is put under rotation. Measuring the rotational inertia, we have demonstrated that a part of the total mass was decoupled from the rotational motion, as shown on figure (\ref{ncri2D}a)). We use a relaxation algorithm (based on the Ginzburg-Landau equation which consist on a imaginary time evolution of the G-P equation as explained above) to converge towards an equilibrium state of the system (close to ground state). The numerical simulation are performed using a pseudo-spectral method and the system is put under rotation by considering the equation in a constant rotating frame. Figure (\ref{ncri2D}b)) shows the evolution of the NCRIF in the limit of zero angular valocity as function of $\Lambda$ (here by varying $\rho U_0$), indicating that the superfluid fraction decreases as the pressure increases in agreement with Leggett's argument that the superfluid fraction should decrease with the minimal value of the wave function~\cite{leggett,mamandine,nestor}.

\begin{figure}[hc]
\begin{center}
\centerline{{\it \includegraphics[width=8cm]{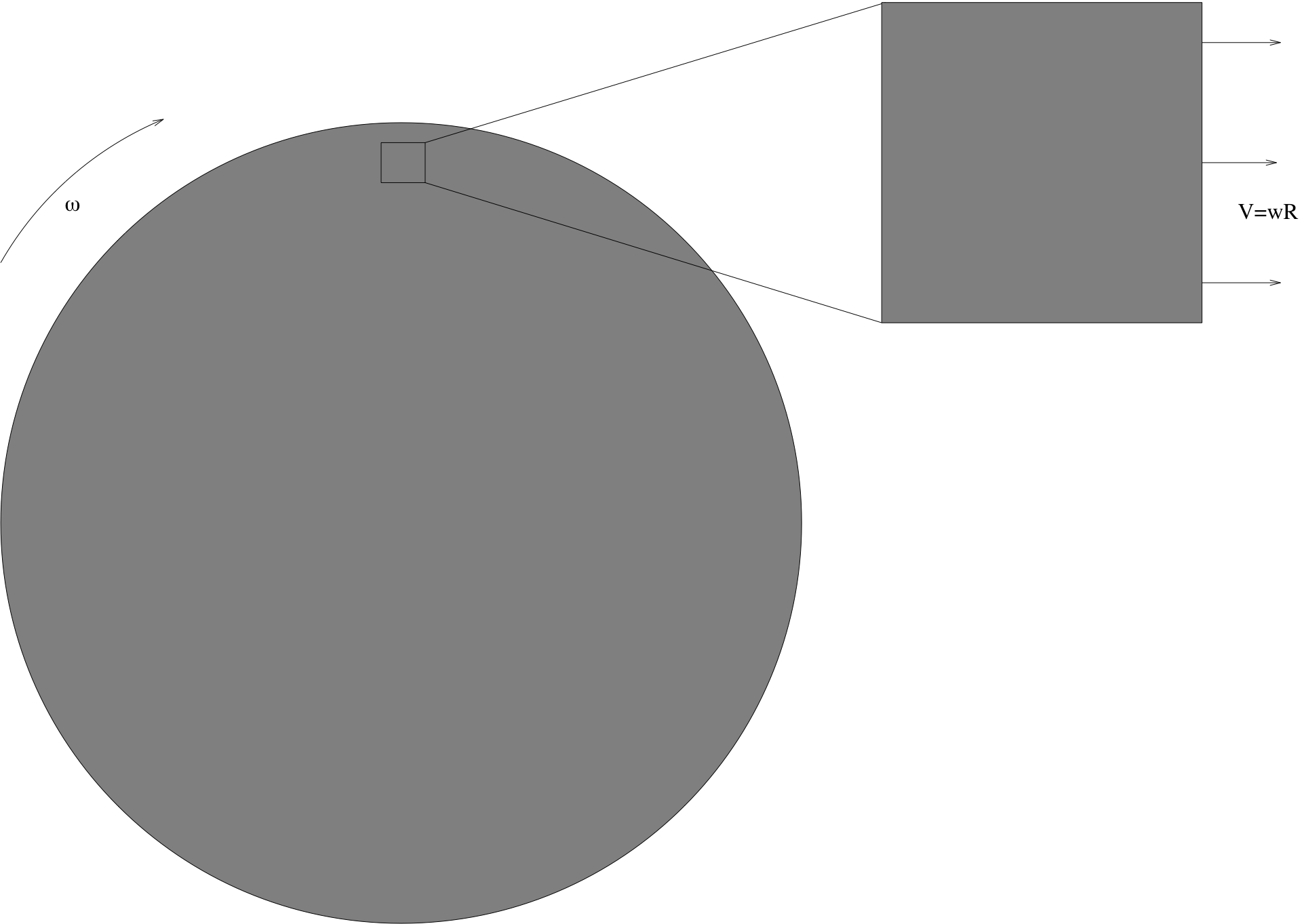}}}
\caption{ \label{sketch-boost}  Sketch illutrating how a rotating cylinder can be analysed {\it via} a cubic sample.}
\end{center}
\end{figure}

A more accurate way to compute the NCRI can be obtained by considering a small fraction of the sample only (see figure (\ref{sketch-boost})). In such case, neglecting the centrifugal acceleration term (which would induced a correction in $\omega^2$ on the solution of the wave-function), the rotation can be approximated by a Galilean boost in the azimuthal direction. The limit case of a Galilean boost of a 1D system can be understood in this context as the rotation of a very thin torus of the crystal. With N. Sepulveda, we have numerically 
computed the NCRI in 1 and 2 spatial dimension using this method, allowing an accurate measure of the superfluid fraction in this model of supersolid~\cite{nestor,Nestor2D}.

\begin{figure}[hc]
\begin{center}
\centerline{{\it a} \includegraphics[width=8cm]{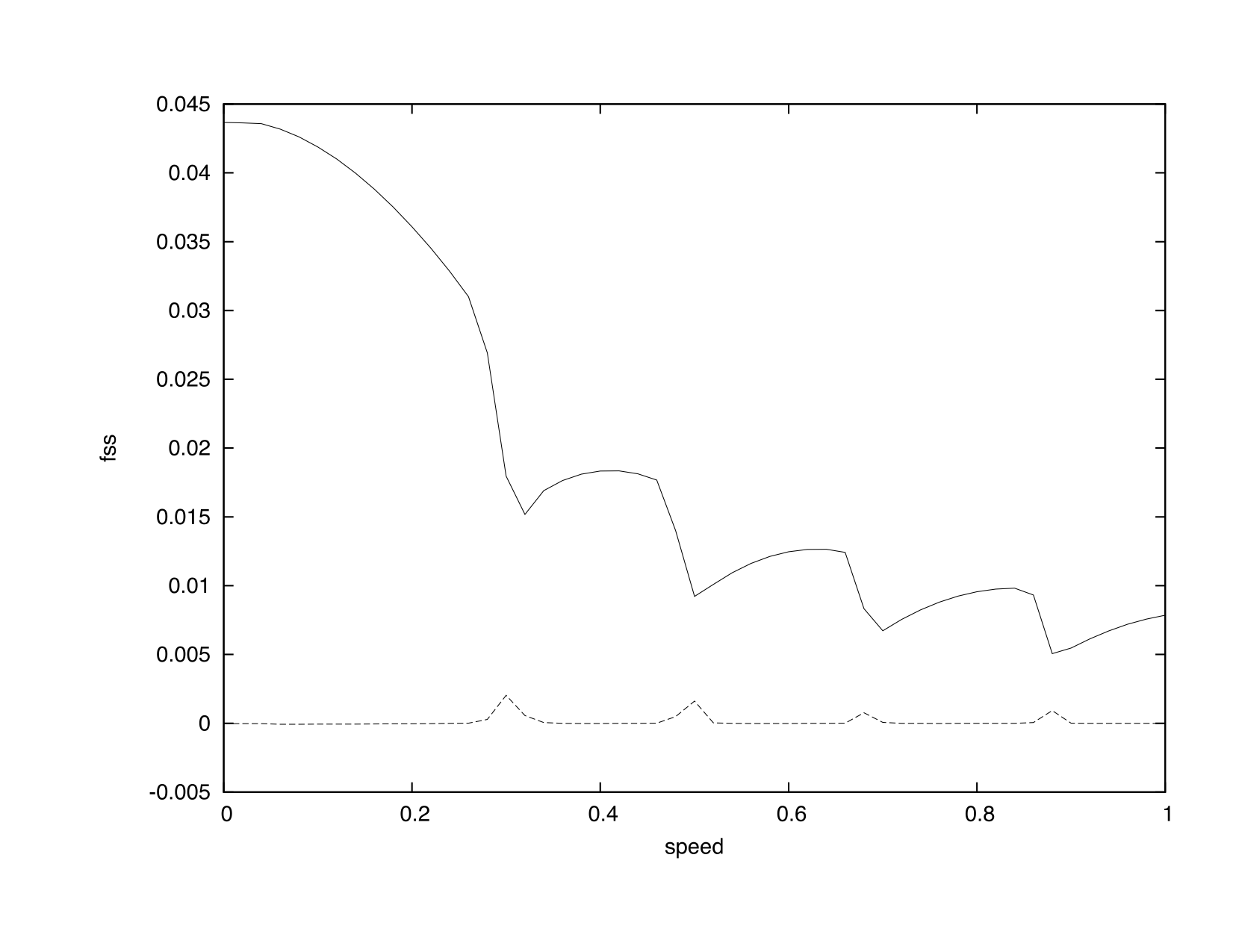} {\it b} \includegraphics[width=8cm]{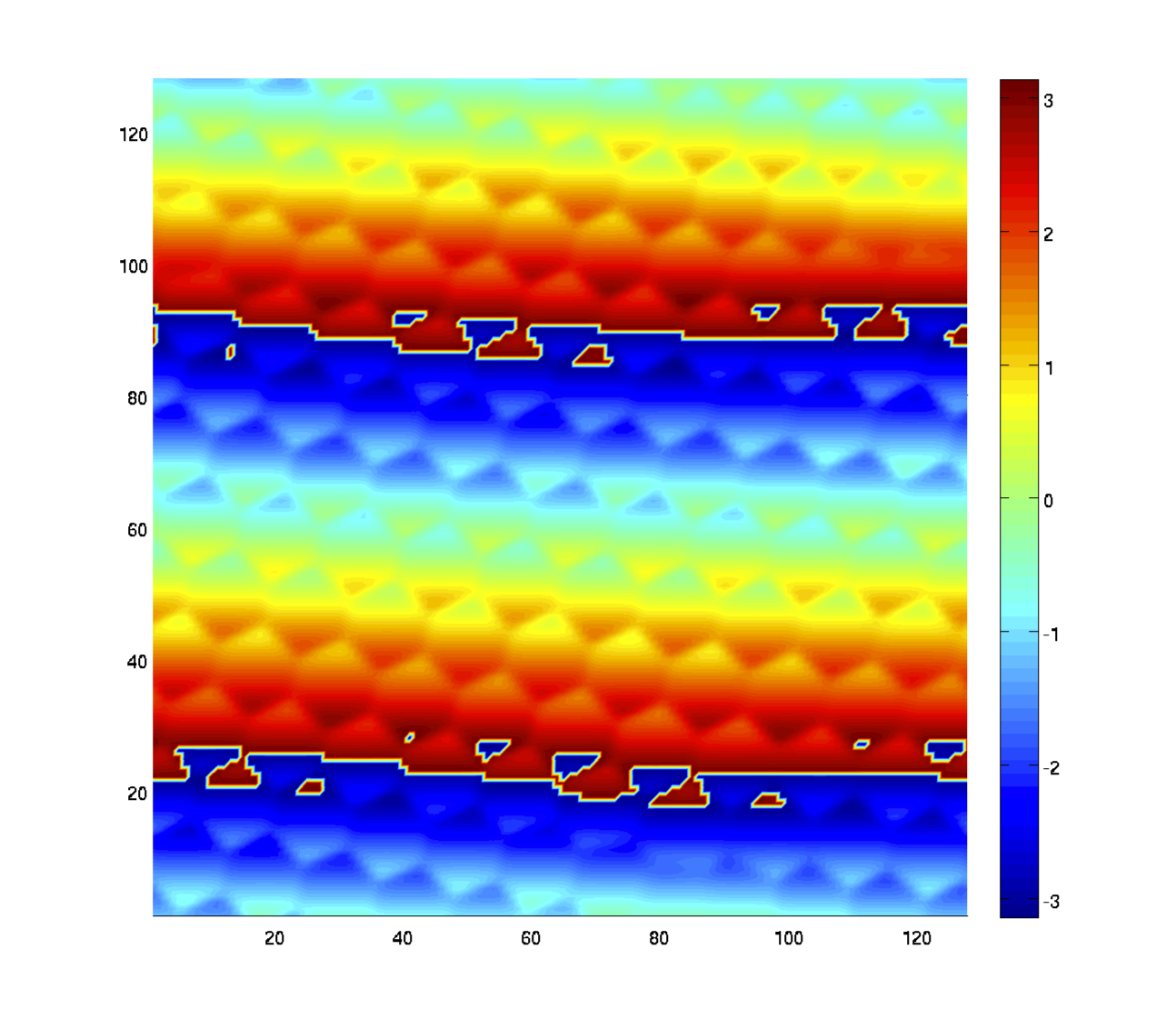}}
\caption{ \label{ncri-boost} {\it (a)} Supersolid fraction as a function of the boost velocity when the crystal is submitted to a Galilean boost. The continuous line shows the component of the supersolid tensor parallel to the velocity while the dashed line is for the orthogonal direction.  
 {\it b)} Phase of the wavefunction for the boost velocity $v_{b}=0.4$. It shows twice a $2\pi$ phase jump (the phase is defined with a multiplier of $2 \pi$) indicating that vortices are
 present in the sample (see figure (\ref{sketch-boost})).}
 \end{center}
\end{figure}

Figure (\ref{ncri-boost}-a) shows the NCRI as function of the Galilean boost velocity in 2D, showing a decrease of the NCRI as the velocity increases. Moreover, in this configuration rapid variations of the NCRI appear regularly and can be. As shown on figure 
(\ref{ncri-boost}-b), these sudden decrease of the superfluid fraction is related to the nucleation of quantum vortices which are known to lead to the loss of superfluidity in superfluids~\cite{peter}.

\subsubsection{Quantized vortices and permanent currents}
\begin{figure}[hc]
\begin{center}
\centerline{a) \includegraphics[width=6cm]{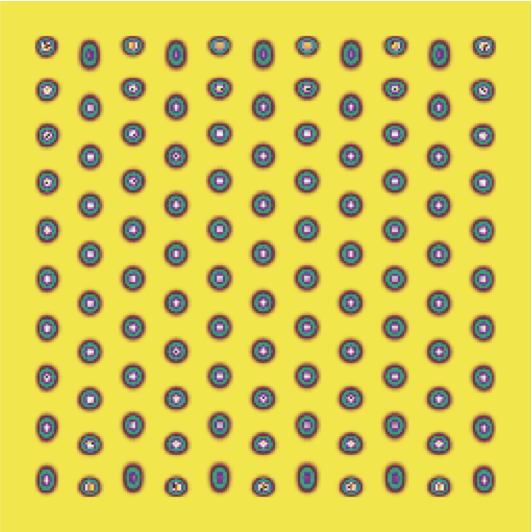} \quad b)  \includegraphics[width=6cm]{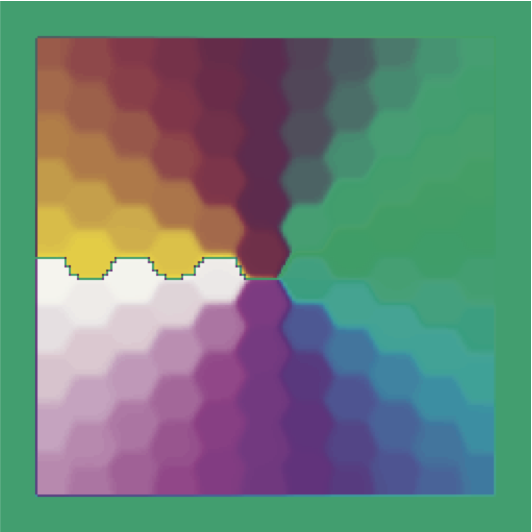} } 
\caption{\label{vortex} Numerical finding of an energy minimum with a topological vortex in the middle of the system in the frame of the energy (\ref{nls.hamilton}) with a constant total number of particles $N=121$, the range of interaction is  $a=8$ and the size system is $112 \times 112$ with Neumann boundary conditions, that is $\psi(x,y)\equiv 0$ if $(x,y)$ are outside the domain. The  interaction parameter is $U_0 = 0.75$, that is  $\Lambda=82.94$. The right-hand-side figure a) represents the density of the crystal, that is  $|\psi|^2$ with $N_s=110$ sites, while the left-hand-side b) represents the phase of the complex wave function $\psi$. A phase jump is clearly visible on the left side of the vortex which is located at the center of the system at the end of phase jump. It is interesting to note that the phase maybe split in two parts, a slowly varying part that change over the system size and a periodic perturbation with a well defined relief of the crystalline pattern as is guessed in the homogenization theory. See section \ref{Approach1} for more details on the model.
}
\end{center}
\end{figure}

In general, the existence of a macroscopic quantum phase suggests that quantized vortices and permanent currents should be present. Such "{\it supersolid vortices}" consists of
a $\pm 2\pi$ jump of the quantum phase around the so-called vortex core. It corresponds thus to topological defects that cannot be removed by infinitesimal perturbations of the system.
Actually, quantum vortices can be observed within a mean-field approach of a quantum solid  proposed here, as illustrated in figure \ref{vortex}. Using an energy minimization argument, one shows that the core of the vortex is located at a minimum of density of the crystal lattice.

How could these vortices be observed in real supersolids? Imagine a number of superfluid vortices in a HeII superfluid created by a rotation at very low temperature, (typically below 
$0.1K$). Then increase the pressure in order to solidify Helium. If it is a normal solid phase, no quantum phase is anymore present and the vortices disappear. But, if it is a supersolid, then the evolution of the vortices during the crystallization process should be as follows:

1) The vortices in the fluid are pushed by interaction with the crystal
boundary and driven to the container boundary; or, 

2) The crystal grows around the vortex to form a ``supersolid vortex" (numerical simulations of (\ref{nls.org})
seem to confirm this second possibility). 

Finally to see experimentally the existence of supersolid vortices, we again melt the
supersolid. The vortex cannot disappear and thus we obtain a
vortex in liquid HeII which could be detected as done in
superfluids.  Therefore, if one could show experimentally that a vortex survives the freezing-melting process, we would 
have a clear proof of the existence of a supersolid phase!

However, the first possibility might happen and no vortex could be measured while the supersolid phase still exist! A way to avoid this problem could be by
freezing a circulating superfluid in a multi-connected
domain\footnote{This idea was suggested by Bernard Castaing to Sergio Rica in 1993.}.
Similar {\it Gedanken} experiments could be considered using permanent currents. A permanent current may be obtained in a supersolid filling a multi-connected domain. Indeed, imposing a phase jump of $2\pi$ as one turns along a non collapsible closed curve in a multi-connected domain we assure the existence of a non-uniform phase as in the case of a vortex. 
\begin{figure}[hc]
\begin{center}
\centerline{a) \includegraphics[width=6cm]{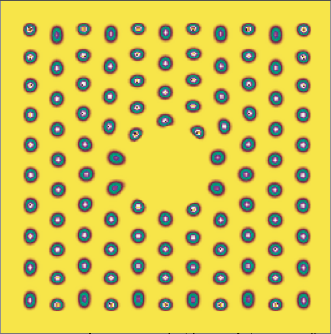} \quad b)  \includegraphics[width=6cm]{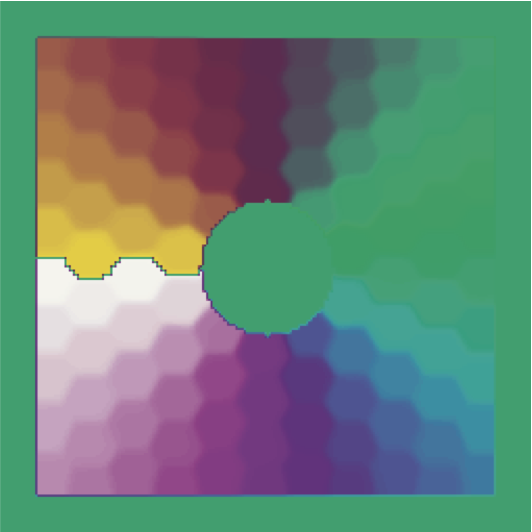} } 
\caption{\label{permanent} Numerical finding of a energy minimum with a permanent current in the frame of the non local Gross-Pitaevskii energy (\ref{nls.hamilton}) with a constant total number of particles $N=96$ the size system is a multi-connected domain composed of a $112 \times 112$ square box with a circular hole of a radii of 15 units in the middle, and we use Neumann boundary conditions, that is $\psi(x,y)\equiv 0$ if $(x,y)$ outside the domain.  The range of interaction is  $a=8$ and the  interaction strength is $U_0 = 0.75$, that is  $\Lambda=82.94$. The right-hand-side figure a) represents the density of the crystal that is  $|\psi|^2$ with $N_s=106$ sites, while the left-hand-side b) represents the phase of the wave function. A phase jump is clearly visible on the left side of the hole.
}
\end{center}
\end{figure}

Vortices and permanent currents are a non-ambiguous property that would confirm the existence of a supersolid coherent state of solid Helium.  To date there is no direct experimental evidence of such properties in solid Helium at conditions where NCRI exists.

\subsubsection{Elastic properties}
The crystal structure that forms naturally in this model exhibit also classical elastic properties. Obviously, the elastic response of the system is due to the rapid density variation of the
mean field in the crystal. Using a homogenization technique~\cite{homogenization} that separates the small scale short time variation at the level of a density peak with the long wavelength slow macroscopic modes, we have been able to deduce from this G-P equation the macroscopic model described in section \ref{macro}-B~\cite{ss1,ss2}. In addition, this
technique provides an explicit protocol to compute the superfluid tensor as it has been checked in details in~\cite{Nestor2D}.

\subsubsection{The U-tube experiments visited in the numerics}\label{Utube}

In Ref.~\cite{ss1} we studied a 
gravity driven supersolid flow. As early suggested by Andreev {\it et al.} \cite{andreev2} an experiment of an obstacle pulled by gravity in solid helium could be a proof of supersolidity. Different versions of this experiment failed to show any motion \cite{Bali06}, therefore a natural question arises: How we can reconcile the NCRI experiment by Kim and Chan and the absence of pressure or gravity driven flows? 

As already said in section \ref{Examples}, our supersolid model (and it seems that supersolid helium too) reacts in different ways under a small external constrain such as stress, bulk force or rotation in order to satisfies the equation of motion and the boundary conditions. For instance, if gravity (or pressure gradient) is added then the pressure ${\mathcal E}'(\rho)$ balances the external  ``hydrostatic'' pressure $mg z$ in equation (\ref{eq:lagrangtimedeptotalBernoulli}) while the elastic behavior of the solid of equation (\ref{eq:lagrangtimedeptotalcauchy}) balances the external uniform force per unit volume $m \rho g$. No ${\bm \nabla} \Phi$ nor $\dot { \bm u}$ are needed to satisfy the mechanical equilibria. Moreover, a flow is possible only if the stresses are large enough to display a plastic flow as it happens in ordinary solids. 
In \cite{pomric} we showed that a flow around an obstacle is possible only if defects are created in the crystal, in this sense we did observe a plastic flow, however in the same model  we observe a ``superfluid-like''  behaviour under rotation without defects in the crystal structure. Indeed for a small angular rotation the elastic deformations come to order $\omega^2$ while ${\bm \nabla}\Phi$ or $ \dot { \bm u}$ are of order $\omega$, the equations of motion together with the boundary conditions leads to a NCRIF different from zero.

\begin{figure}[hc]
\begin{center}
\centerline{{\it a} \includegraphics[width=6cm]{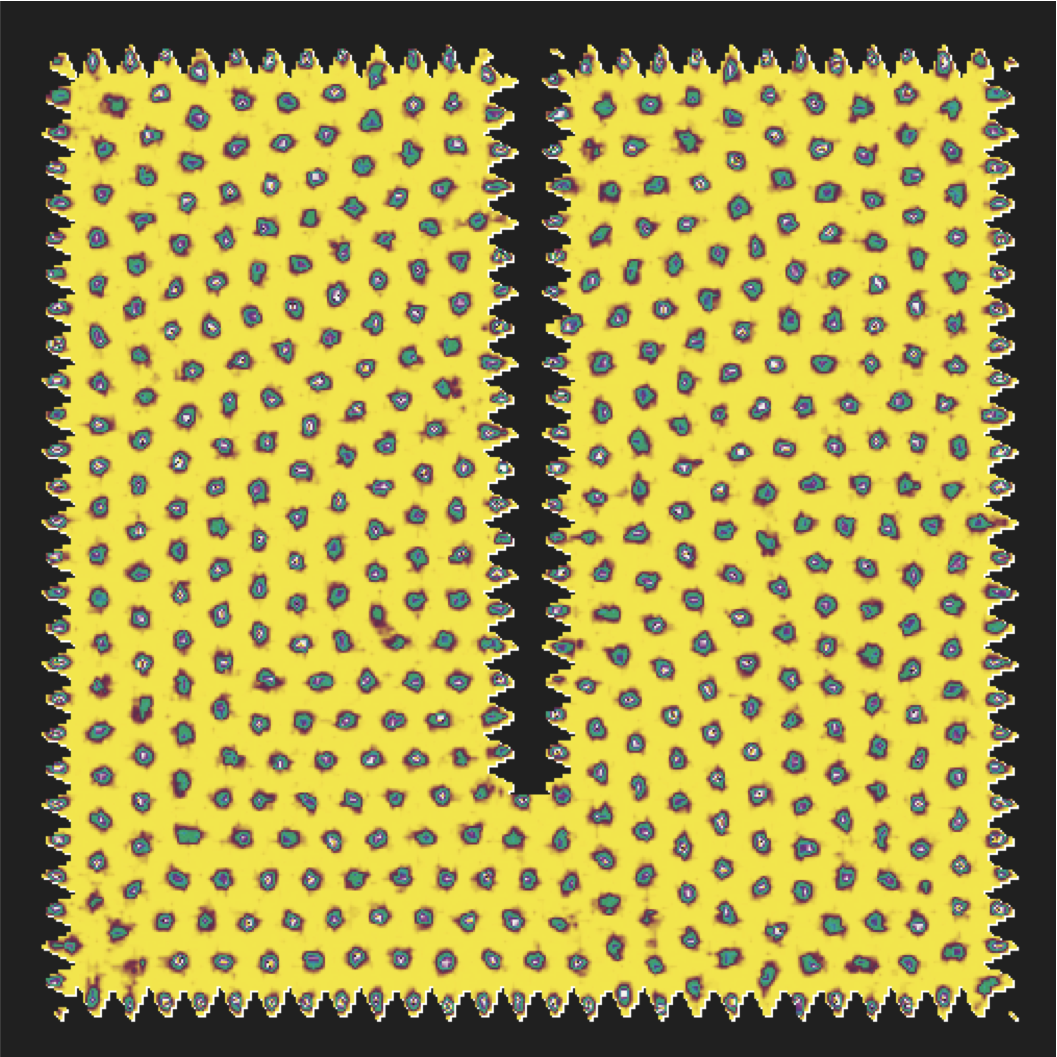} {\it b} \includegraphics[width=6cm]{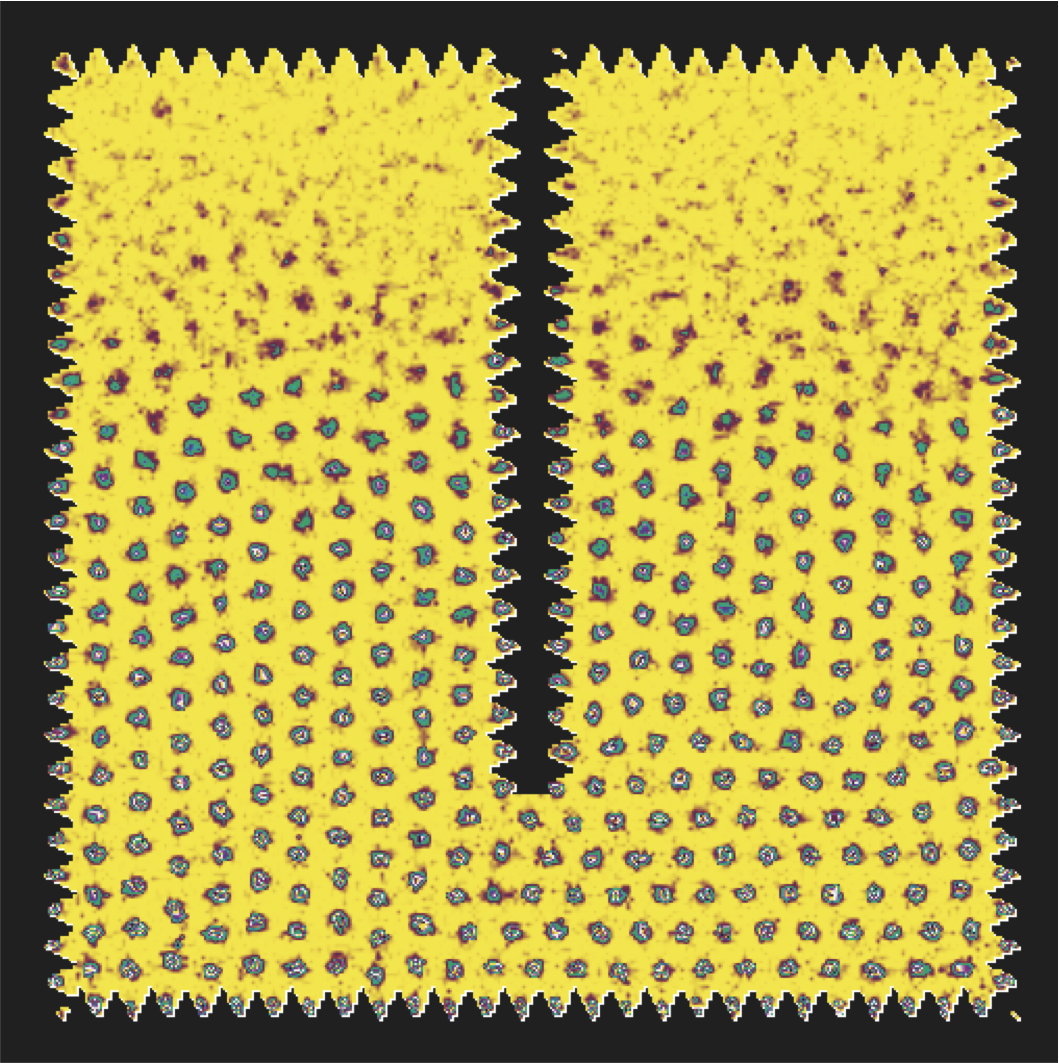}}
\centerline{{\it c} \includegraphics[width=6cm]{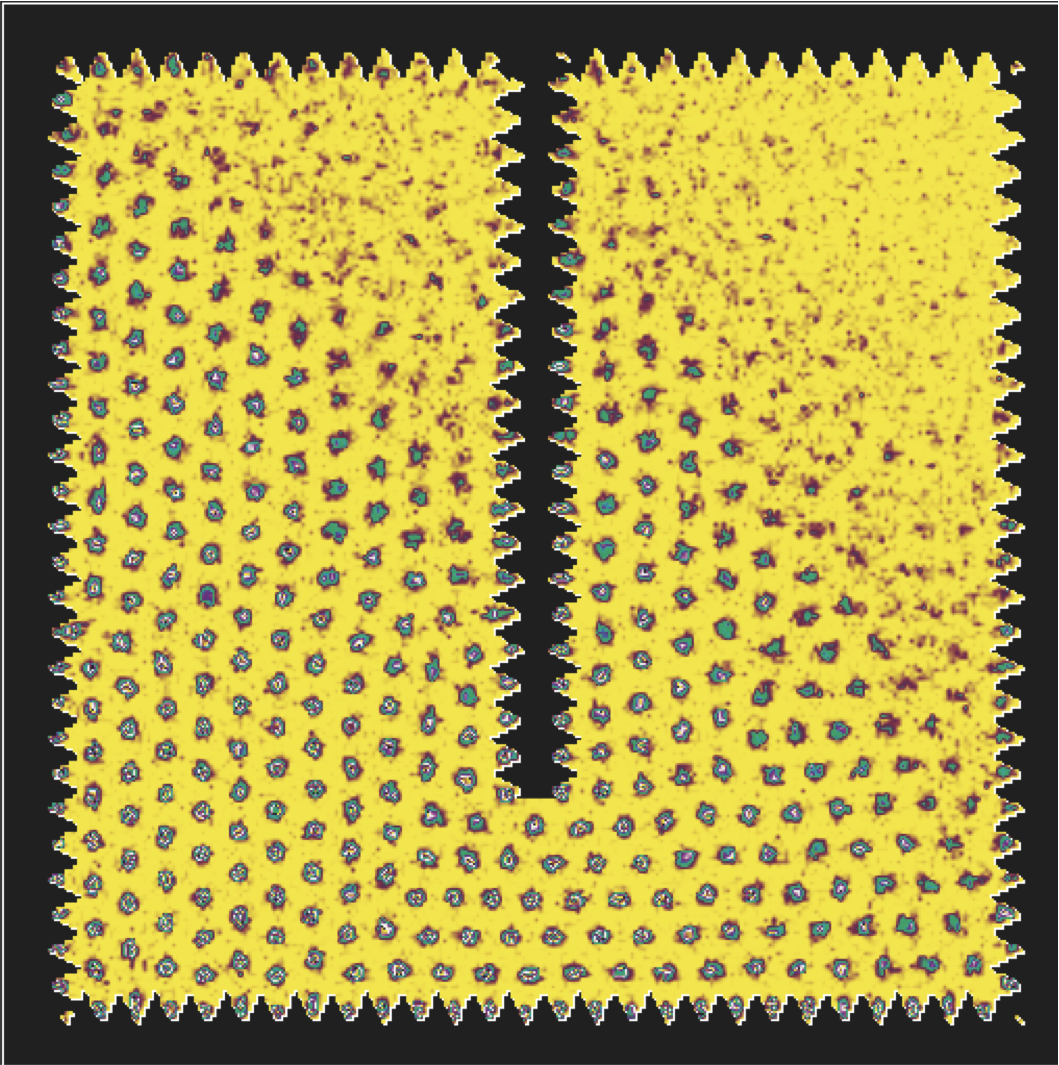} {\it d} \includegraphics[width=6cm]{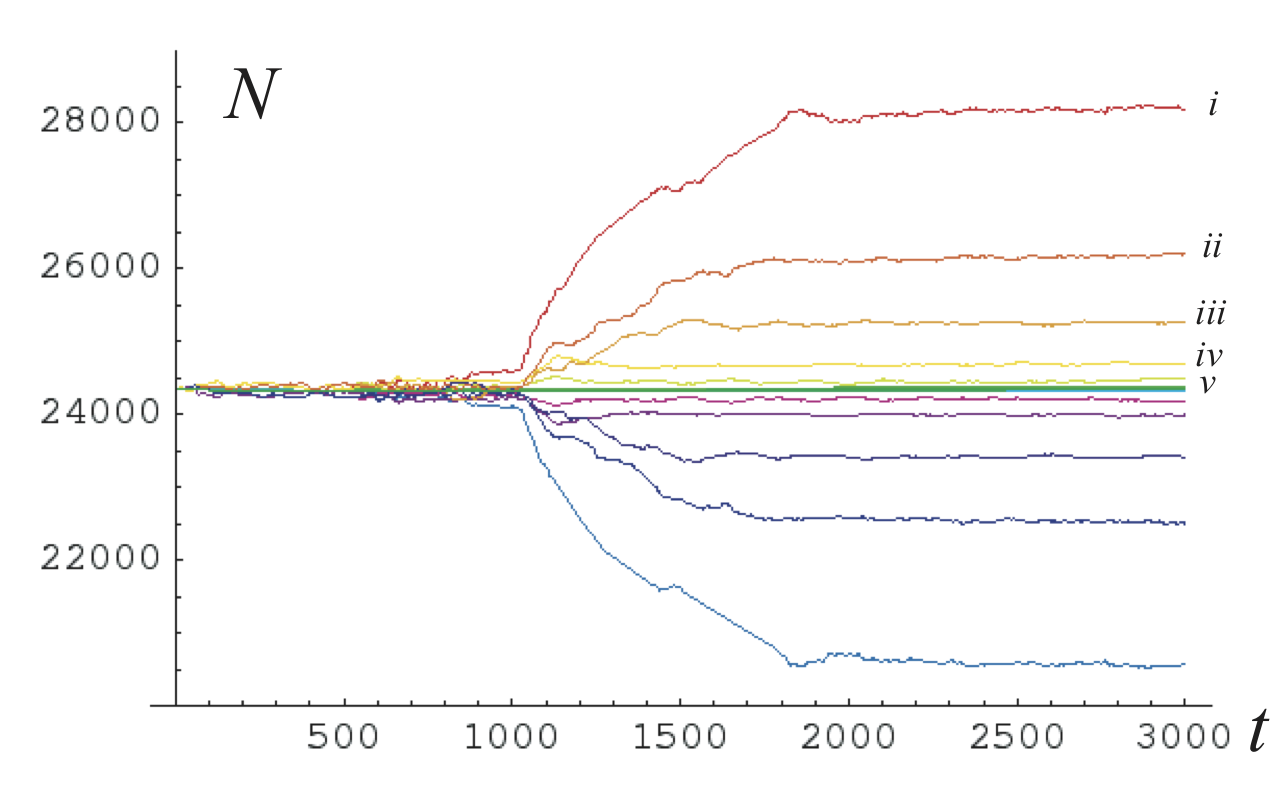}}
\caption{ \label{grav} We plot the three snapshots of density modulations $|\psi|^2$ (the dark points means a large mass concentration) of a numerical Simulation of eqn. (\ref{nls.org}) with Dirichlet boundary conditions with the shape of an u-tube as in the figure. We use a Crank-Nicholson scheme that conserves  the total energy and mass. The mesh size is $dx=1$, the nonlocal interaction parameters are chosen as $U_0 = 0.01$ and $a=8$ (physical constants $\hbar$ and $m$ are 1), finally the initial condition is an uniform solution $\psi= 1$ plus small fluctuations.  {\it a)} A crystalline state is obtained after $500$ units of  time;  {\it b)} then a vertical gravity of magnitude ${\mathcal G} =0.01$ is switched-on, and the system evolves for 500 time unites up to {\it b}: the solid makes a neat interface;  {\it c)}  Finally,  the gravity orientation is tilted in 45$^\circ$. After 2000 time units the system evolves to a stationary situation {\it c} showing that the mass flow is only a transient. Moreover,  in {\it d)} it is observed, after 1000 time units, a stationary situation showing that the mass flow  is only a transient. 
}
\end{center}
\end{figure}

  We have realized a numerical simulation to test the possibility of a permanent gravity flow for different values of the dimensionless gravity ${\mathcal G}=\frac{m^2ga^3}{\hbar^2}$.
 Let us consider an U-tube as in Fig.\ref{grav}.
The system is prepared for $500$ time units into a good quality (but not perfect) crystalline state. A vertical gravity of magnitude $\mathcal G$ is switched-on and the system evolves for 500 time units more up to a new equilibrium state (see Fig. \ref{grav}-{\it a}).  The gravity is then tilted (with the same magnitude) at a given angle. 
A mass flow is observed at the begining from one reservoir into the other, but both vessels do not reach the same level eventually (see Fig. \ref{grav}-{\it b}). 
There is some dependence of the transferred mass on $\mathcal G$
till $\mathcal G \approx 0.0005$  and the mass transfer becomes negligible from fluctuations for $\mathcal G < 0.00025 $ indicating the existence of a yield-stress. The flow is allowed by dislocations and grain boundaries and it is a precursor of a microscopic plastic flow as in ordinary solids ({\it e.g.} ice) and as it is
probably observed in Ref. \cite{Bali06}. A microscopic  yield-stress could be defined by the smallest gravity $\mathcal G$ such that no dislocations, defects nor grain boundaries appear. In the present model this is for  $\mathcal G <  10^{-4} $.

\noindent
In conclusion, we have shown a fully explicit model of supersolid that display either solid-like behavior or superflow depending on the external constrain and on the boundary conditions on the reservoir wall.
Our numerical simulations clearly show that, within the same model a nonclassical rotational inertia is observed as well a regular elastic response to external stress or forces without any flow of matter similarly than in the macroscopic model and in agreement with the experiments \cite{chan04b,Bali06}.

\subsubsection{Sound as an alternative test for supersolidity}
 
As explained in~\cite{ss1,ss2}, this model, beside the NCRI property, has an intrinsic elastic behavior and follows the macroscopic equations discussed in section \ref{macro}-B. In particular, it has been shown numerically there that the quantum solid of the model can present a pure elastic response to an imposed external strain due to gravity. Let us also recall that such solid exhibits different perturbation modes (sound waves), a pure shear mode on one hand, and compression-phase coupled modes on the other hand. In particular, in the limit of small superfluid fraction, a low-velocity mode exist which has the same characteristics than the Bogoliubov mode for superfluid or Bose-Einstein condensate. In this 
limit, the velocity of this phase mode is proportional to the square-root of the superfluid fraction of the supersolid ($v_2^2 =\frac{ \varrho^{ss} {\mathcal E}''(\bar\rho)}{m}$). The existence of such mode suggests an alternative way of determining the superfluid fraction in a supersolid (and by consequence an alternative test for showing the existence of the supersolid state), by looking to the resonant frequencies of a "supersolid" cavity. In this paragraph, we will show numerically in the present mean field model that the lower frequency eigen-mode of a cavity provides a correct (and thus alternative) measure of the superfluid fraction.

\begin{figure}[hc]
\begin{center}
\centerline{ \includegraphics[width=10cm]{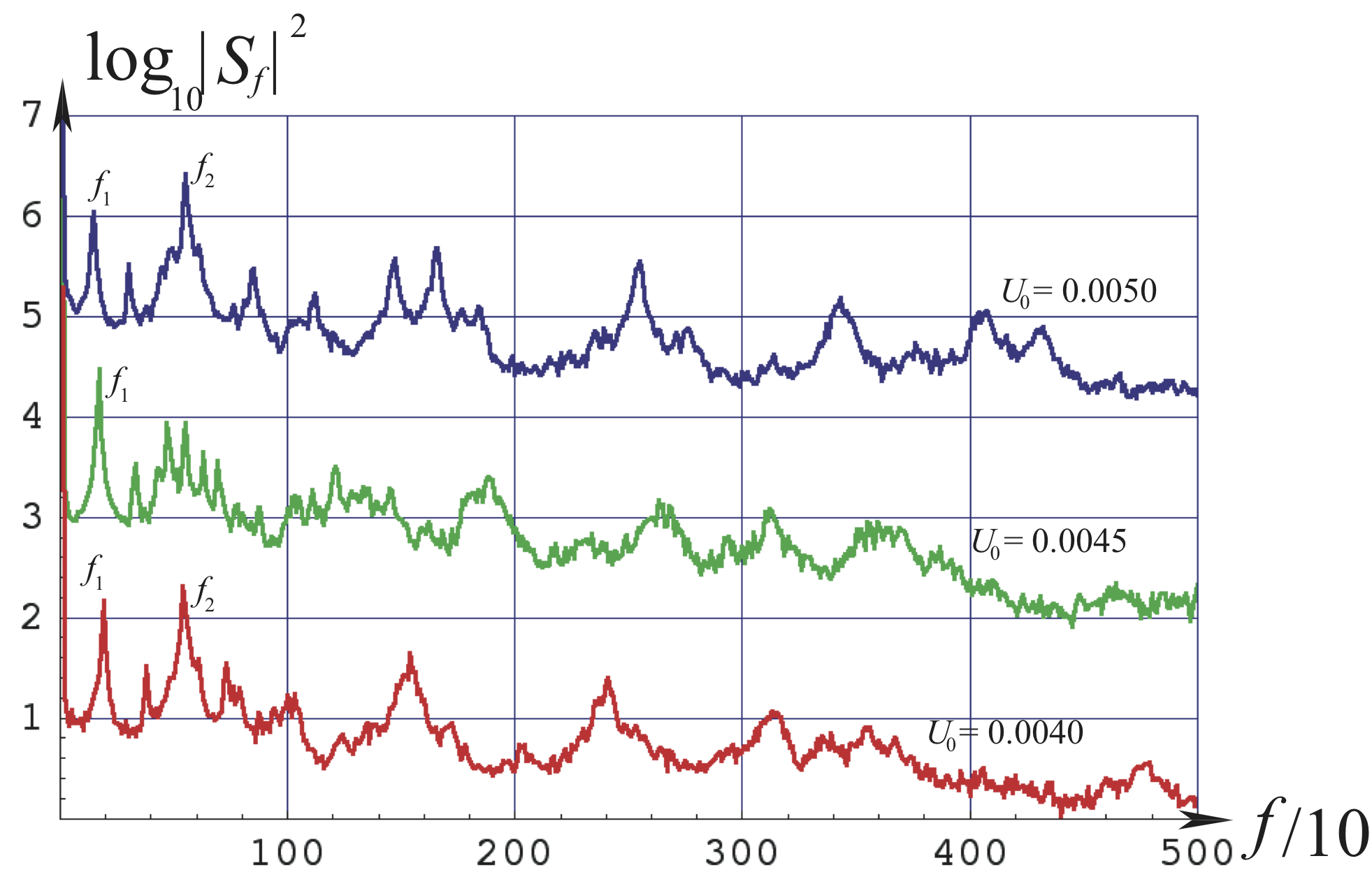}} 
\caption{\label{spectre}  Fourier transform of the signal $N_1(t)$. 
}
\end{center}
\end{figure}

In order to measure numerically the cavity modes we are interested in, we create a low frequency oscillation of the system by starting the numerical simulation of eq. (\ref{nls.org}) 
in a closed cavity of size $LxL$ with a small initial mass difference between the left side ($0<x<L/2$) and the right side ($L/2<x<L$). More precisely, we use as initial condition: 
$$\psi(x,y,t) = \left\{ \begin{array}{cc}
1 -\epsilon + {\rm noise}&0< x < L/2 \\
1 +\epsilon + {\rm noise}& L/2<x<L
\end{array}
\right.
$$
with typically $\epsilon = 0.1 $ and a zero mean noise of amplitude $0.05$. We will vary the dimensionless parameter $\Lambda$ by varying in the numerics the value of $U_0$ for $a$ fixed. By measuring the time evolution of the mass in the left side of the cavity ($N_1(t) = \int_0^L\, dy \int_{L/2}^L\, dx  |\psi(x,y)|^ 2$), we can extract the vibration modes of the cavity using Fourier transform. Figure (\ref{spectre}) shows such frequency spectrum and one can identify the eigen-modes of the cavity as the different peaks of the spectrum.

\begin{figure}[hc]
\begin{center}
\centerline{{\it a} \includegraphics[width=6cm]{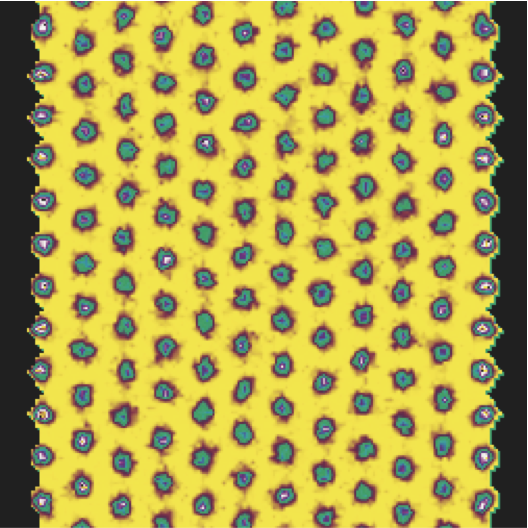} {\it b} \includegraphics[width=6cm]{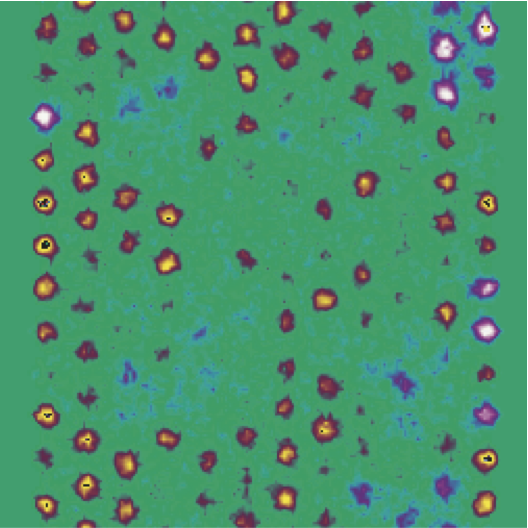}}
\caption{ \label{sound} Snapshot of the density modulations $|\psi|^2$, {\it a)}, and the real part of the wave function ${\rm Re} \psi$,  {\it b)} of numerical Simulation of eqn. (\ref{nls.org}) with in an annular domain with Dirichlet boundary conditions in the horizontal axis and periodic boundary conditions on the vertical axis.   We use a Crank-Nicholson scheme that conserves  the total energy and mass. The mesh size is $dx=1$, the nonlocal interaction parameters are chosen as $U_0 = 0.005$ and $a=8$ (physical constants $\hbar$ and $m$ are 1), both snapshots are at the same instant of time $t= 125,000 $ time units. Though the modulus seems more or less uniform in the large scale one sees the long wave oscillations of the phase of the wave function in the figure {\it b}.}
\end{center}
\end{figure}

Seeking to identify the low velocity mode with the low frequency cavity, we exhibit the two smallest frequency peaks as we vary $\Lambda$, by changing $U_0$ and the initial density, and for different boundary conditions (usual conditions is no split at the boundary) in table \ref{table1}. We also compute for these values of the parameters the superfluid fraction bounds using the Leggett theory~\cite{leggett}. Leggett deduces from general argument an upper and lower bound for the superfluid density which is easy to compute when the wave function is known (see ~\cite{nestor,Nestor2D} for a discussion on Leggett's formulae). As it can be seen in the table, the lower and the upper superfluid fractions are very close one from each other and we will take the superfluid fraction  as the average of both bounds $f^{ss} =  (f^{ss}_-+f^{ss}_+)/2 \pm (  (f^{ss}_+-f^{ss}_-)/2 $.

\begin{table}[!h]
\begin{tabular}{|c|c||c|c|c|c|c|}
\hline
$ U_0$ &$  \Lambda $  & $f_1/10$ & $\pm \Delta f_1 \times 10^{-5}$ & $f_2/10$ &  $f^{ss}_{-}$ & $f^{ss}_{+}$ \\
\hline\hline
$\bar\rho=1$\\ \hline \hline
$0.00300  \, \gr \blacksquare$ & $ 38.6039 $ & $1/39$ & $6.58$  & $1/16$ & $0.984752 $  & $0.9983 $   \\ \hline
$0.00325  \, \gr \blacksquare $ & $ 41.8209 $ & $1/41$ & $5.95$  & $1/17$ & $0.913244 $  & $0.920613 $   \\ \hline
$0.003375  \, \red \blacktriangle$ & $ 43.4294 $ & $1/42$ & $5.67$  & $1/17.7$   & $0.559121 $& $0.607666 $   \\ \hline
$0.00350  \, \red \blacktriangle$ & $ 45.0379 $ & $1/43.5$ & $  2.64    $  & $1/18.5$   & $0.481871 $ & $0.531268 $  \\ \hline
$0.003625  \, \red \blacktriangle$ & $ 46.6464 $ & $1/44.5$ & $  5.17    $  & $1/14.7 ??$   & $0.422077 $ & $0.490596 $  \\ \hline
$0.00375  \, \red \blacktriangle$ & $ 48.2549 $ & $1/47$ & $4.53$  & $1/15$  & $0.376049 $  & $0.432442 $  \\ \hline
$0.003875  \, \red \blacktriangle$ & $ 49.8634 $ & $1/61$ & $  3.97    $  & $1/17.5$   & $0.33623 $ & $0.408770 $  \\ \hline
$0.00400  \, \red \blacktriangle$ & $ 51.4719 $ & $1/57$ & $3.08$  & $1/19.5$ & $0.303137 $  & $0.365992 $   \\ \hline
$0.00425  \, \red \blacktriangle$ & $ 54.6888 $ & $1/62$ & $2.69$  & $1/18$ & $0.247761 $  & $0.311837 $   \\ \hline
$0.00450 \, \red \blacktriangle$ & $ 57.9058 $ & $1/65$ & $2.36$  & $1/19-1/15$ & $0.203728 $  & $0.269468 $   \\ \hline
$0.005  \, \red \blacktriangle$ & $ 64.3398 $ & $1/74$ & $1.82$  &  $1/18.5$ & $0.141113 $  & $0.206708 $   \\ \hline
$0.006 \, \red \blacktriangle$ & $ 77.2078 $ & $1/89$ & $1.26$  & $1/18.5$ & $0.0723346 $  & $0.131552 $   \\ \hline
$0.0065 \, \red \blacktriangle$ & $ 83.6418 $ & $1/108$ & $5.91$  & $1/15.3?$ & $0.05296 $  & $0.10862 $   \\ \hline
$0.007 \, \red \blacktriangle $ & $ 90.0757 $ & $1/130$ & $5.95$  & $1/18.5$& $0.0395829 $  & $0.0911309 $    \\ \hline
$0.0075  \, \red \blacktriangle$ & $ 96.5097 $ & $1/115$ & $7.62$  & $1/16.5$ & $0.0303889 $  & $0.0780576 $   \\ \hline
$0.007875  \, \red \blacktriangle$ & $ ?? $ & $1/115$ & $7.62$  & $1/16$ & $ 0.02496 $  & $0.07125 $   \\ \hline
$0.008  \, \red \blacktriangle$ & $ 102.944 $ & $1/220$ & $2.07$  & $1/16.5$ & $0.0233344 $  & $0.0673922 $   \\ \hline
$0.010  \, \red \blacktriangle$ & $ 128.68 $ & $--$ & $--$  & $??$ & $0.0101029 $  & $0.0440862 $   \\ \hline
\hline
Slidding BC\\ \hline \hline
$0.0035\, \blue\blacklozenge   $ & $ 45.0379 $ & $1/43.5$ & $5.2$  & $1/15.5$ & $0.913244 $  & $0.920613 $   \\ \hline
$0.0050\,  \blue\blacklozenge   $ & $ 64.3398 $ & $1/74$ & $5.5$  & $1/18.5$ & $0.2067 $  & $0.1411 $   \\ \hline
$0.0055\,  \blue\blacklozenge   $ & $ 70.7738 $ & $1/80$ & $3.12$  & $1/18.5$ & $0.162815 $  & $0.09994 $   \\ \hline
$0.0075  \, \red \blacktriangle$ & $ 96.5097 $ & $1/123$ & $1.07?$  & $1/17$ & $0.0303889 $  & $0.0780576 $   \\ \hline
\hline
$\bar\rho=2$\\ \hline \hline
$0.001625\,  \purple\star  $ & $ 41.8209 $ & $1/41$ & $2.97 $  & $ 1/17$ & $0.92876 $ & $0.9317293 $     \\ \hline
$0.00175\,  \purple\star  $ & $ 45.0379 $ & $1/43$ & $2.77 $  & $ 1/13$ & $0.4751098 $  & $0.567067 $   \\ \hline
$0.002\,  \purple\star $ & $ 51.4719 $ & $1/60$ & $2.78 $  & $ 1/17$ & $0.39088 $  & $0.30088 $   \\ \hline
$0.0025\,  \purple\star  $ & $ 64.3398 $ & $1/75$ & $2.7$  & $1/19$ & $0.139155 $  & $0.22566 $   \\ \hline
$0.0035\,  \purple\star  $ & $ 90.0757 $ & $1/104$ & $4.72$  & $1/19$ & $0.0383266 $  & $0.10147648 $   \\ \hline
$0.0045\,  \purple\star $ & $ 115.812 $ & $1/190$ & $2.78$  & $1/14$ & $0.0173359 $  & $0.09374663 $   \\ \hline
\hline
\end{tabular}
\caption{\label{table1} Table showing the first two frequency peaks observed in the numerical simulations. All numerical simulations ($\red \blacktriangle$) are for $a=8$ and a mean density $\bar\rho = \frac{1}{L^2} \int |\psi|^2 dx\, dy = 1$. $\gr \blacksquare$ represents a situation where the density is more or less homogeneous in space, that is a liquid. $\blue\blacklozenge  $ are numerical simulations with a sliding boundary conditions and $\purple\star$ are for $\bar\rho = 2$. The two Legget's bounds for the superfluid fraction are also computed for each numerical simulation.} 
\end{table}

\begin{figure}[hc]
\begin{center}
\centerline{ a) \includegraphics[width=8cm]{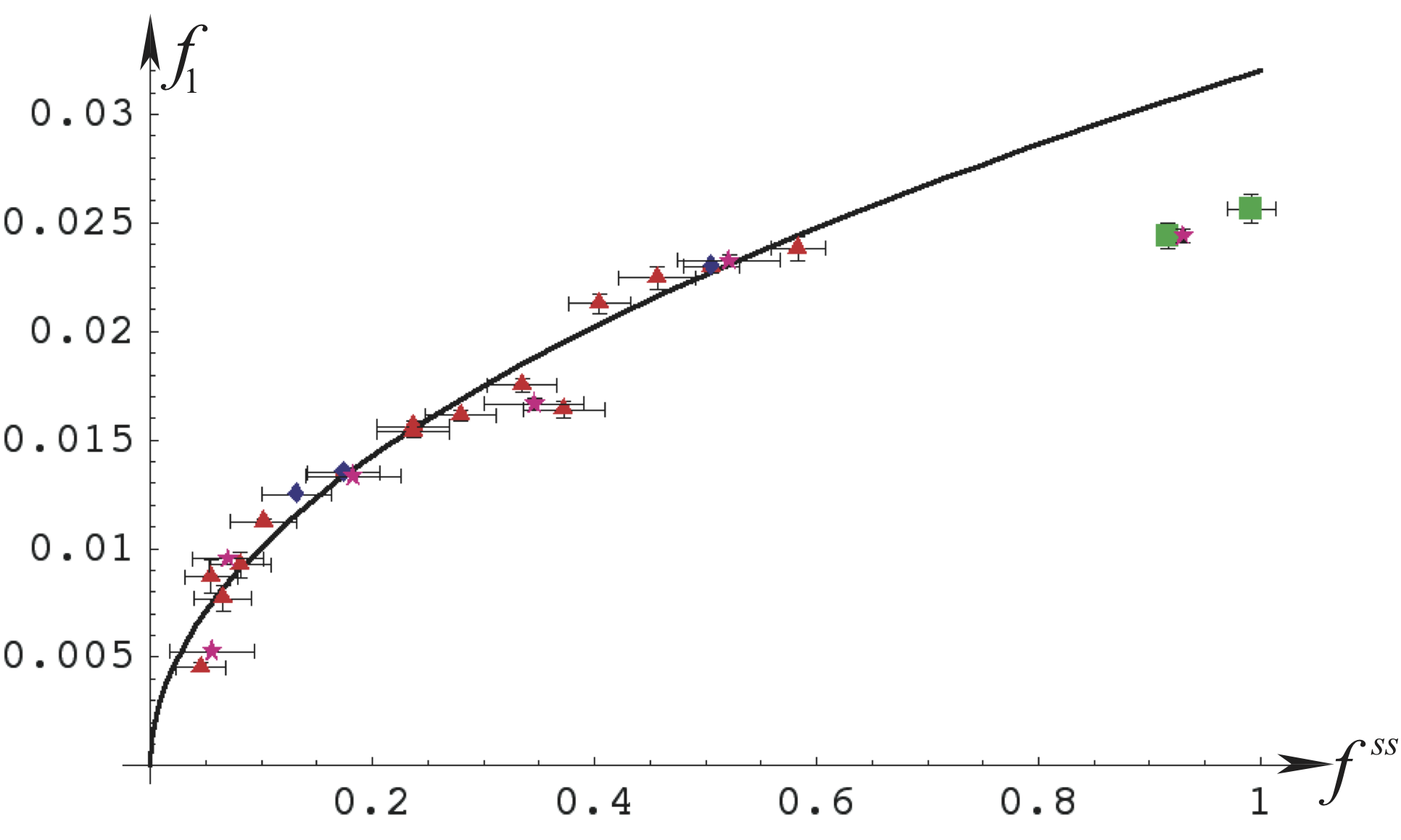}\quad b)  \includegraphics[width=8cm]{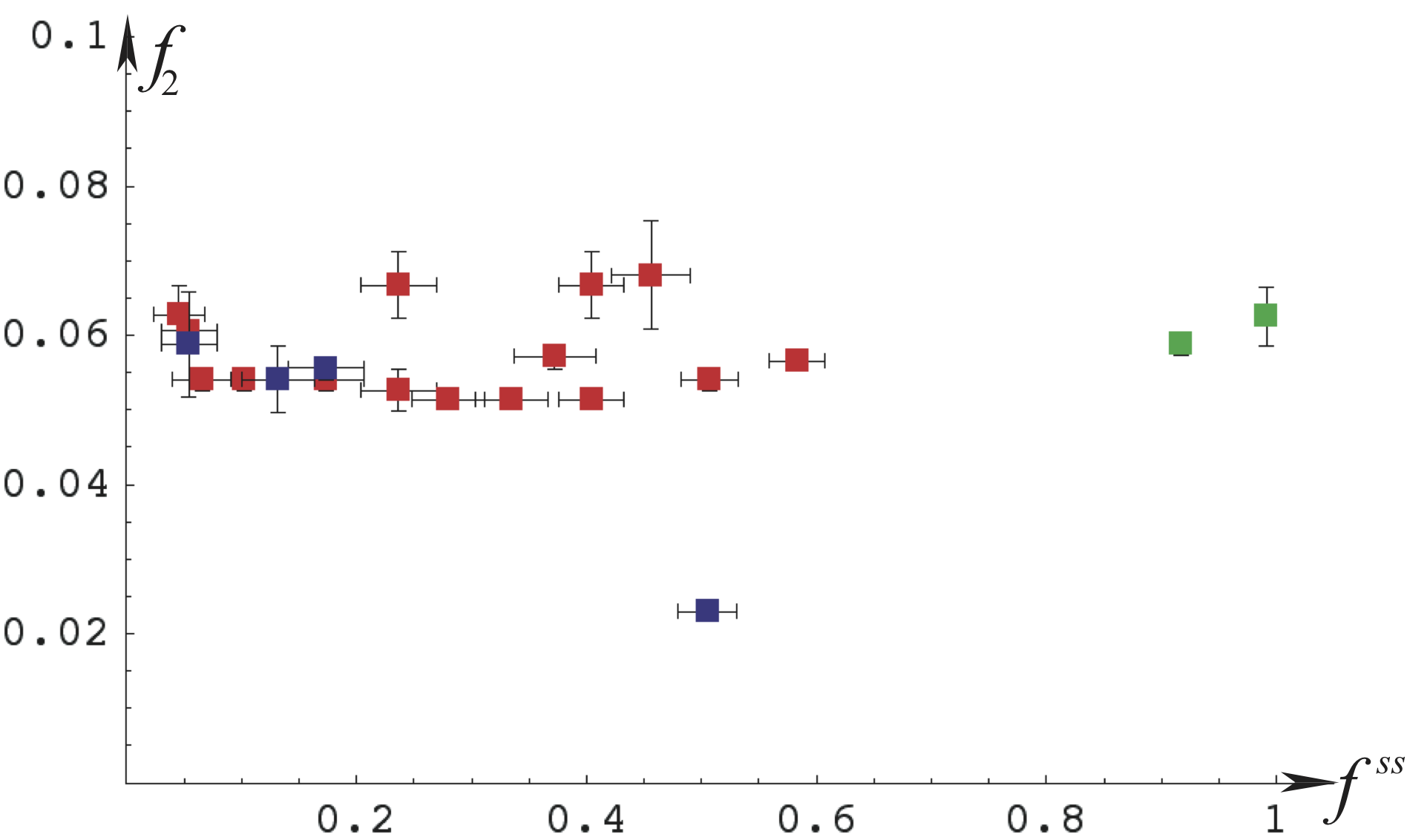}} 
\caption{\label{freqvsfss} The first resonance $f_1$ (a) and second resonance (b) versus the supersolid fraction $f^{ss}$ measured thanks to Leggett's iformulas applied to the model as explained in the text. In a) The line represents the curve $0.032 \sqrt{f^{ss}}$ for guideline to the eyes. The two points represented by a $\gr\blacksquare$ are low values $U_0$ a situation where the ground state is homogeneous in space and represents better a liquid than a solid.
}
\end{center}
\end{figure}

Fig. \ref{freqvsfss} shows these two lowest frequency modes as function of this $f^{ss}$. We observe that the first mode satisfies a Bogoliubov-like dependance on the superfluid fraction as expected from the macroscopic theory while the second mode does not depend on the superfluid fraction and should be associated to a pure elastic mode.

From this preliminary numerical study, we suggest that an alternative (and complementary) way to measure (and somehow to exhibit) supersolidity could be by identifying pressure and/or temperature dependence of the frequency of the lowest eigen-mode of vibration of solid Helium.
\section{Conclusions and Perspecives}
We have presented in this review various aspects of the theory of
supersolids. A discussion based on fundamental physical concepts has shown
that a regular lattice with elastic properties may coexist with a
superfluid component. We derived from the equation of motion of the full
system (elastic lattice + superfluid component) the observed property that
{\underline{rotation} does induce a superflow not implying the whole mass
but that a pressure difference does induce a purely elastic response
without superflow. In this respect the often neglected question of the
boundary conditions play a central role in the understanding of the
properties of the full system. Later on we discussed mathematical models
of supersolidity, with various degree of realism as far as the detailed
microscopic interactions are taken into account. Because of the major
difficulty of solving honestly and accurately the Schr\"{o}dinger equation for
a sizable number of interacting atoms, even with idealized two-body
potentials, it is likely that no other fully microscopic theory will be
available soon for this class of many body systems. We also addressed the
connection between the "ideal model" leading to the macroscopic equations
and the wide variations in superfluid density from one experiment to the
other in nominally similar thermodynamical conditions. Almost ten years after the
experiments by Chan and collaborators and many more years after the
theoretical predictions by Onsager and by Leggett, one can say that a fair
understanding of the physics of supersolidity has been reached and likely
will be confirmed by experiments now in design or to come.

The authors acknowledge Peter Mason and Nestor Sep\'ulveda for fruitful discussions. CJ and SR thank the financial support of the Agence Nationale de la Recherche through the grant SYSCOM COSTUME ANR-08-SYSC-004 (France). SR thanks the Fondecyt Grant  N$^0$ 1100289 (Chile).

	  \end{document}